\documentclass[a4paper,fleqn,usenatbib,useAMS]{mnras}

\usepackage{graphicx}	
\usepackage{amsmath}	
\usepackage{amssymb}	
\usepackage{multicol}        
\usepackage{bm}	
\usepackage{pdflscape}	
\usepackage{subfigure}
\usepackage{longtable}
\usepackage{array}
\usepackage{booktabs}

\usepackage{xcolor}
\usepackage{tabularray}
\usepackage{lipsum}
\usepackage{multicol}

\usepackage[T1]{fontenc}
\usepackage{ae,aecompl}

\title[Cocoon cooling emission]{Cocoon cooling emission in neutron star mergers}

\author[Hamidani \& Ioka]{Hamid Hamidani$^{1,2}$\thanks{E-mail: hamidani.hamid@yukawa.kyoto-u.ac.jp} and Kunihito Ioka$^1$
\\
$^{1}$ Yukawa Institute for Theoretical Physics, Kyoto University, Kyoto 606-8502, Japan\\
$^{2}$ Astronomical Institute, Tohoku University, Aoba, Sendai 980-8578, Japan\\
}

\date{Accepted 2023 June 23. Received 2023 May 25; in original form 2022 October 24}

\pubyear{2023}

\begin{document}
\label{firstpage}
\pagerange{\pageref{firstpage}--\pageref{lastpage}}
\maketitle

\begin{abstract}
In the gravitational wave event GW170817, there was a $\sim 10$ hours gap before electromagnetic (EM) observations, without detection of the cocoon.
The cocoon is heated by a \textit{short} gamma-ray burst (\textit{s}GRB) jet propagating through the ejecta of a Neutron Star (NS) merger, and a part of the cocoon escapes the ejecta with an opening angle of $20^{\circ}$--$30^{\circ}$. 
Here we model the cocoon and calculate its EM emission. 
Our 2D hydrodynamic simulations suggest that the density and energy distributions, after entering homologous expansion, are well-fitted with power-law functions, in each of the relativistic and non-relativistic parts of the escaped cocoon. 
Modeling these features, we calculate the cooling emission analytically. 
We find that the cocoon outshines the r-process kilonova/macronova at early times (10--10$^{3}$ s), peaking at UV bands. 
The relativistic velocity of the cocoon's photosphere is measurable with instruments such as Swift, ULTRASAT and LSST. 
We also imply that energetic cocoons, including failed jets, might be detected as X-ray flashes. 
Our model clarifies the physics and parameter dependence, covering a wide variety of central engines and ejecta of NS mergers and \textit{s}GRBs in the multi-messenger era.
\end{abstract}



\section{Introduction}
\label{sec:1}
High energy, short timescale, and extreme luminosities of Gamma-Ray Bursts (GRBs) (in comparison to other astrophysical transients, e.g., supernova, nova, etc.) can only be explained by highly relativistic jets (\citealt{1975NYASA.262..164R}; \citealt{1978Natur.271..525S}; \citealt{1986ApJ...308L..43P}; \citealt{1986ApJ...308L..47G}; \citealt{1991ApJ...373..277K}; \citealt{1993ApJ...405..273W}).
GRB jets are short lived, powered by a central compact object surrounded by an accretion disk (a system often referred to as the central engine) formed after a cataclysmic stellar event.
GRBs consist of two distinct classes with different astrophysical origin.
\textit{Short} GRBs (hereafter \textit{s}GRBs) have a duration of $\lesssim 2$ s, and \textit{long} GRBs (hereafter \textit{l}GRBs) are $\gtrsim 2$ s (\citealt{1993ApJ...413L.101K}; \citealt{2009ApJ...691..182S}). 

\begin{figure}
    \centering
    \includegraphics[width=0.999\linewidth]{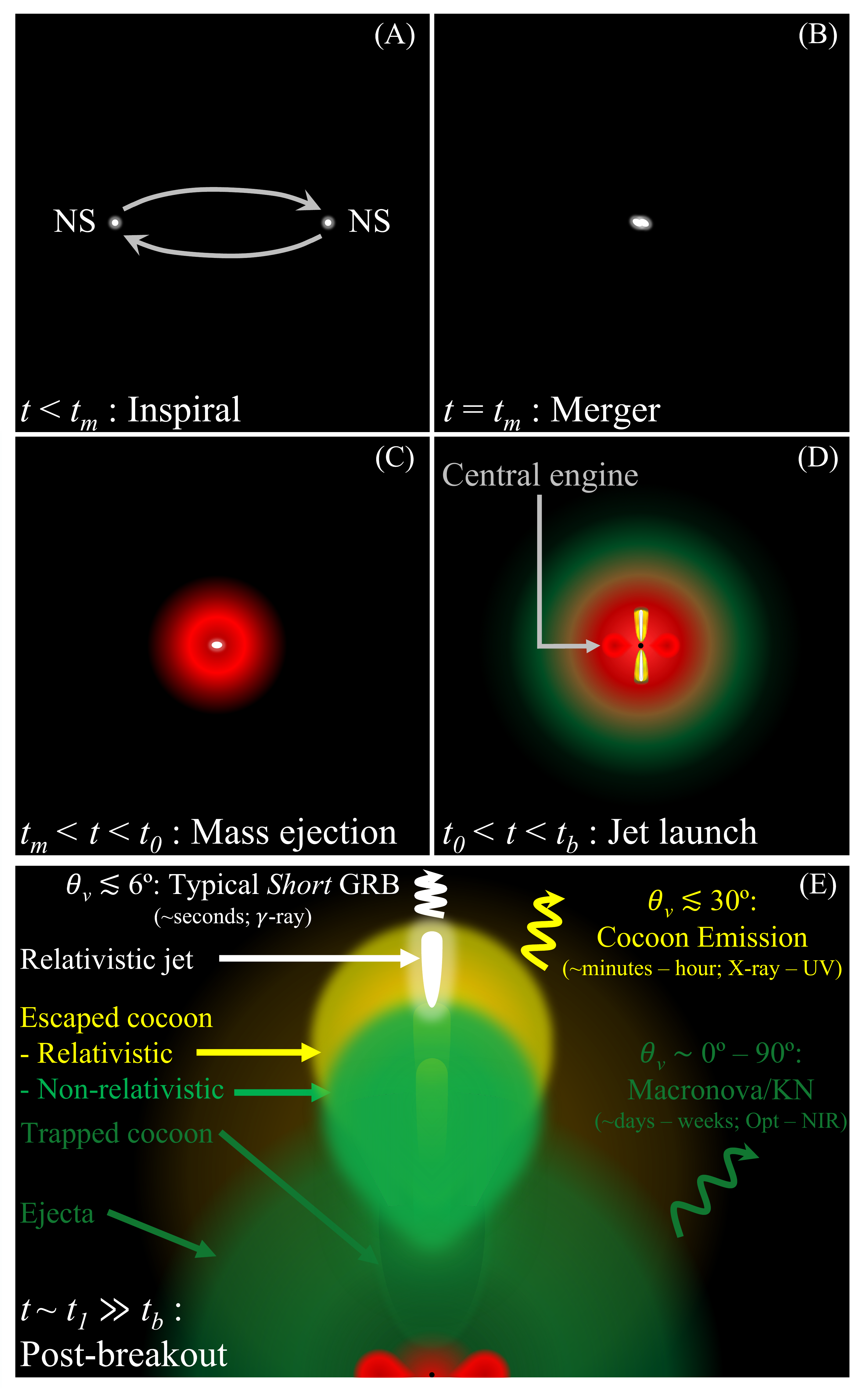} 
  \caption{
  Schematic illustration of the timeline and key phases in NS mergers, with the observational perspective. 
  Initially, a pair of compact objects in the inspiral phase [NS-NS here; applies also for a NS-BH system] (A) . 
  The two objects merge into one compact object (B). 
  This interaction triggers mass ejection [$\sim 0.01 M_{\odot}$ expanding at $\langle{\beta}\rangle \sim 0.2c$; \citealt{2013PhRvD..87b4001H}] (C). 
  Soon after ($\sim0.1-1$ s in the case of GW170817; see \citealt{2021MNRAS.500..627H}), a system of a central compact object with an accretion disk is formed (D).
  This system (i.e., central engine) powers two polar jets (D) [white]. 
  Each jet propagates through the surrounding dense ejecta (D) [red and dark green].
  This forms a bubble of hot gas, ``cocoon", around the jet (D) [yellow].
  Soon after the jet/cocoon breaks out of the ejecta,
  the system enters the free-expansion phase (E). 
  Only a small fraction of the cocoon escapes from the ejecta and expands in the conical manner [with an opening angle $\theta_c^{es}\sim 20^\circ - 30^\circ$] (E) [see \citealt{2023MNRAS.520.1111H}].
  This escaped cocoon contains a relativistic component, and a non-relativistic component (E) [yellow (mostly shocked jet cocoon), and light green (mostly shocked ejecta cocoon), respectively].
  Three EM transients are highlighted;
  from hard to soft, short to long, and narrow to wide emission's opening angle:
  \textit{s}GRB [white], cocoon emission [yellow], and KN [dark green] (E).
  }
  \label{fig:key1} 
\end{figure}

While \textit{l}GRBs are explained by the death of massive stars in the collapsar model (\citealt{1993ApJ...405..273W}; \citealt{1999ApJ...524..262M}) and confirmed by observations (supernova explosions; \citealt{1998Natur.395..672I}; \citealt{2003Natur.423..847H}; and star forming regions; \citealt{1998ApJ...494L..45P}),
\textit{s}GRBs have been linked to the merger of two compact objects, NS-NS 
(binary neutron star mergers)
or BH-NS (black hole-NS mergers)
(\citealt{1986ApJ...308L..43P}; \citealt{1986ApJ...308L..47G}; \citealt{1989Natur.340..126E}) (see Figure \ref{fig:key1}).
In both scenarios, the cataclysmic event generates the necessary ingredients for the launch of the relativistic jet, and for the prompt emission. 

NS mergers are studied with numerical relativity simulations (\citealt{1999PhRvD..60j4052S}; \citealt{2000PhRvD..61f4001S}).

It is shown that during and after the merger (through tidal interaction, collision shock, and oscillation of the remnant),
mass (of the order of $\sim 10^{-3} - 10^{-2} M_{\odot}$) is ejected dynamically at substantial velocities ($\sim 0.2c$; \citealt{2013PhRvD..87b4001H}; \citealt{2013ApJ...773...78B}; \citealt{2015MNRAS.448..541J}; \citealt{2016MNRAS.460.3255R}; \citealt{2018ApJ...869..130R}; etc.) [see panel (C) in Figure \ref{fig:key1}].
This mass, refereed to as the ``ejecta", surrounds the central engine, the birth place of the jet (see panel (D) in Figure \ref{fig:key1}).
The expanding nature of this ejecta is one key difference (in \textit{s}GRBs) compared to the collapsar model.
This ejecta is an important ingredient: 
i) it is the environment that the relativistic jet is forced to penetrate and interacts with, and is eventually shaped it (\citealt{2014ApJ...784L..28N}; \citealt{2014ApJ...788L...8M}; \citealt{2015ApJ...813...64D}),
and ii) its neutron rich composition presents a site for the r-process nucleosynthesis of heavy elements (\citealt{1989Natur.340..126E}), whose decay powers the kilonova/macronova transient (KN hereafter) [\citealt{1998ApJ...507L..59L}; \citealt{2005astro.ph.10256K}; \citealt{2010MNRAS.406.2650M}].

The \textit{s}GRB's jet-ejecta interaction has been studied with relativistic hydrodynamical simulations  (\citealt{2014ApJ...784L..28N}; \citealt{2014ApJ...788L...8M}).
Simulations have consistently shown that the jet can penetrate the ejecta and break out of it while keeping its relativistic nature (\citealt{2014ApJ...784L..28N};
\citealt{2014ApJ...788L...8M}; 
\citealt{2015ApJ...813...64D};
\citealt{2018MNRAS.473..576G};
\citealt{2020MNRAS.491.3192H};
\citealt{2020MNRAS.495.3780N};
\citealt{2021MNRAS.500..627H}; 
\citealt{2021MNRAS.502.1843N};
etc.),
in a similar way to jets in the context of active galactic nuclei (AGNs) and \textit{l}GRBs (\citealt{1974MNRAS.169..395B}; \citealt{1974MNRAS.166..513S}; \citealt{1999ApJ...524..262M}; \citealt{2011ApJ...740..100B}; \citealt{2013ApJ...777..162M}).
First, a shock structure (jet head) -- a dense and highly pressurized region where the jet outflow is slowed down and its energy is dissipated -- is formed (\citealt{1974MNRAS.169..395B}; \citealt{1974MNRAS.166..513S}).
Second, the continued influx of the jet pushes the jet head forward across the ejecta. Meanwhile, the shocked hot jet head fluid leaks sideways, creating a hot bubble that envelopes the jet in the form of a ``cocoon" (\citealt{1989ApJ...345L..21B}) [see panel (D) in Figure \ref{fig:key1}].
It is this cocoon that collimates the jet, helping it to penetrate the ejecta efficiently (for \textit{l}GRBs, see: \citealt{2002NewA....7..197R};
\citealt{2007ApJ...665..569M}; 
\citealt{2011ApJ...740..100B};
\citealt{2013ApJ...777..162M}; 
and for \textit{s}GRBs see: \citealt{2014ApJ...784L..28N}; and \citealt{2021MNRAS.500..627H} for a full analytical formulation of this collimation).
Once the shock reaches the outer edge of ejecta both the jet and cocoon can get out of the ejecta, i.e., breakout.

The cocoon is an interesting component in several aspects.
It is an intermediate component between the (highly relativistic, hot, and low rest-mass density) jet and the (non-relativistic, cold, and high rest-mass density) ejecta.
As it is expected to have decent energies (comparable to that of the jet), substantial mass ($\sim 10^{-6}-10^{-4} M_{\odot}$; see \citealt{2023MNRAS.520.1111H}), and mildly relativistic speeds ($\sim 0.4-0.9c$; 
see Figure \ref{fig:v sim}), theoretically 
the cocoon could power a unique astrophysical transient (\citealt{2017ApJ...834...28N};
\citealt{2017ApJ...848L...6L};
\citealt{2018MNRAS.473..576G}; \citealt{2018PTEP.2018d3E02I}). 
This cocoon emission, although softer and dimmer compared to the prompt emission (of \textit{s}GRB), can be visible at larger opening angles ($\sim 5^{\circ}- 30^{\circ}$) and for much longer time ($\sim 10-10^3$ s).

The first gravitational wave (GW) signal from the binary neutron star (BNS) merger event, GW170817, 
was detected by the Laser Interferometer Gravitational-Wave Observatory (LIGO) and the Virgo Consortium (LVC) observatory (\citealt{2017PhRvL.119p1101A}).
Fermi satellite detected an \textit{s}GRB, \textit{s}GRB 170817A,  just about $1.7$ s after the merger (\citealt{2017ApJ...848L..13A}).
This confirmed the scenario of NS mergers for \textit{s}GRBs (\citealt{1986ApJ...308L..43P}; \citealt{1986ApJ...308L..47G}; \citealt{1989Natur.340..126E}). 
After about $10$ hours, the merger site was localized and a large follow-up campaign across the electromagnetic (EM) spectrum followed (and is still ongoing as of the time of writing, e.g., \citealt{2022arXiv220514788B}; \citealt{2022GCN.32065....1O}). 
This marked the start of a new era of
multi-messenger astronomy (\citealt{2017ApJ...848L..13A}).

In particular, this campaign resulted in two important discoveries.
First, a KN transient was detected, and the analysis confirmed the presence of the expanding ejecta, in consistency with numerical relativity simulations, with indications of r-process nucleosynthesis (of heavy and unstable elements) [\citealt{Arcavi:2017vbi}; \citealt{Chornock:2017sdf}; \citealt{Coulter:2017wya}; \citealt{2017ApJ...848L..29D}; \citealt{2017Sci...358.1570D}; \citealt{Kilpatrick:2017mhz}; \citealt{2017Sci...358.1559K}; \citealt{Nicholl:2017ahq}; \citealt{Pian:2017gtc}; \citealt{Smartt:2017fuw}; \citealt{Shappee:2017zly}; \citealt{Soares-Santos:2017lru}; \citealt{2017PASJ...69..102T}; \citealt{Utsumi:2017cti}; \citealt{Valenti:2017ngx}), as previously predicted (\citealt{1998ApJ...507L..59L}; \citealt{2005astro.ph.10256K}; \citealt{2010MNRAS.406.2650M}).
Second, this campaign was also able to find clear evidence of a relativistic jet (\citealt{2018Natur.561..355M})
viewed from off-axis (\citealt{2018PTEP.2018d3E02I} and \citealt{2019MNRAS.487.4884I}).
These discoveries are perfectly consistent with the scenario of \textit{s}GRBs (\citealt{1986ApJ...308L..43P}; \citealt{1986ApJ...308L..47G}; \citealt{1989Natur.340..126E}; also, \citealt{2013PhRvD..87b4001H}; \citealt{2014ApJ...784L..28N}; and others).

Due to the $10$ hours gap between the GW's merger signal and GW170817's localization in the sky, GW170817 was not monitored in the first few hours,
missing an opportunity to observe early emission, 
in particular the cocoon emission (considering estimations by \citealt{2018PTEP.2018d3E02I}; \citealt{2018MNRAS.473..576G}; \citealt{2021MNRAS.500.1772N}; \citealt{2021MNRAS.502..865K}; also see \citealt{2018ApJ...855..103P} for an alternative explanation). 
However, this is expected to change in the very near future. 
With the upcoming GW campaigns by LIGO, VIRGO, and KAGRA [also, with Einstein Telescope (ET), Cosmic Explorer (CE), and Laser Interferometer Space Antenna (LISA) in the next decade], multi-messenger observations of GW170817-like events are expected to be more frequent, and sky localization is expected to be much faster, presenting more and better opportunities to study NS mergers and detect the cocoon emission.

Here, we are interested in the EM cocoon emission in the context of \textit{s}GRBs and as a counterpart to GW signal from NS mergers (NS-NS and BH-NS).
As a continuation to our companion paper (\citealt{2023MNRAS.520.1111H} dedicated to the cocoon breakout), our goal is to model the cocoon emission 
so that we can directly link the observational features
with the physical properties of central engine jets
and ejecta in NS mergers, \textit{s}GRBs, KNe, and r-process nucleosynthesis.

In this paper, we first present numerical simulations of hydrodynamical jets propagating in the dynamical ejecta of NS mergers, and use their results to understand the hydrodynamical properties of the cocoon.
It is important to focus on the ``escaped cocoon" that breaks out of (and escapes from) the ejecta because the trapped cocoon is not relevant to the cocoon emission, in contrast to the collapsar case (\citealt{2023MNRAS.520.1111H}).
We categorize the escaped cocoon into the ``relativistic cocoon" and ``non-relativistic cocoon".
We then analytically estimate the photon emission from each of these two parts, and combine them to estimate the total cocoon emission.

Despite previous work by \cite{2017ApJ...834...28N}, \cite{2017ApJ...848L...6L}, \cite{2018PTEP.2018d3E02I}, 
to our best knowledge, this is the first time that the cooling emission from the cocoon of \textit{s}GRBs has been properly modeled (analytically), 
after rigorously taking into account the critical process of the cocoon escape up to the free-expansion phase (using numerical simulations; \citealt{2023MNRAS.520.1111H}).
We present an analytic framework to evaluate observable features of the relativistic to non-relativistic cocoon from very early times (a few seconds after the merger) until the cocoon fades away (relative to the KN).

We show that the cocoon emission is within reach of current (and upcoming) observational facilities and could be detected in future GW170817-like events, 
in particular in the UV [Swift UVOT (limiting magnitude of $\sim 22$ for an exposure time of $\sim 1000$ s; \citealt{2005SSRv..120...95R}); 
and ULTRASAT (limiting magnitude of $\sim 22.4$ for an exposure time of $900$ s; \citealt{2014AJ....147...79S})]
and in soft X-rays [e.g., Swift XRT in the X-ray ($0.2-10$ keV; \citealt{2000SPIE.4140...64B}), and MAXI ($2-30$ keV; \citealt{2009PASJ...61..999M})], 
with observations in the optical/infrared being more challenging due to the spectral and temporal features of the cocoon emission (see Figure \ref{fig:NWF mag}) [e.g., Zwicky Transient Facility (ZTF; limiting magnitude of 20.8 in g-band and wide field of view; \citealt{2020ApJ...905...32B})]
although the upcoming Vera C. Rubin observatory (LSST; capable of reaching the 24th magnitude with 10 s exposure; see \citealt{2019ApJ...873..111I}; \citealt{2009arXiv0912.0201L}) is capable of detecting the early cocoon emission.
We argue that the cocoon emission can be used to identify NS mergers and \textit{s}GRBs differently,
i.e., even in cases where the jet is failed (\citealt{2018ApJ...866....3D}) or is viewed off-axis (\citealt{2017ApJ...848L...6L}),
and can bring additional information on the merger rate, on the merger process (e.g., the ejecta mass), 
on \textit{s}GRBs (e.g., their jets and the central engine), 
and on KNe (e.g., r-process nucleosynthesis and nuclear composition at early times).

It should be noted that, although the nature of \textit{s}GRB-jets is yet to be understood,
our work here is based on numerical simulations of purely hydrodynamical jets (no magnetic field).
However, as shown by \cite{2020MNRAS.498.3320G}, the propagation of weakly magnetized \textit{s}GRB-jets (and their cocoons) are quite similar to the case of pure hydrodynamical jets (in particular, see their Figures 3 and 4).
Therefore, our results can be extended to weakly magnetized jets, but not to magnetically dominated jets.

This paper is organized as follows. 
In Section \ref{sec:2}, new properties of the post-breakout cocoon as measured in numerical simulations are revealed.
In Section \ref{sec:3}, these properties are formulated analytically for later usage to calculate the cocoon emission.
In Section \ref{sec:4}, analytic modeling of the cocoon emission is presented.
In Section \ref{sec:5}, our analytic results for the cocoon emission are presented, and discussed.
Finally, a conclusion is presented in Section \ref{sec:6}.
Additionally, details related to the numerical simulations data, analytic calculation of the photospheric evolution, analytic derivation of the bolometric luminosity, the used KN model, and a full glossary of the mathematical symbols, can be found in appendices \ref{ap:cocoon}, \ref{ap:photosphere}, \ref{ap:Lbl}, \ref{ap:KN}, and \ref{ap:Glossary}, respectively.

\section{Post-breakout cocoon in numerical simulations}
\label{sec:2}
Figure \ref{fig:key1} illustrates the timeline of the simulation.
After the merger ($t\ge t_m$), the jet is launched (at $t=t_0$, until $t=t_e$; where we set $t_e-t_0=2$ s; see Table \ref{tab:1}), the jet breakout happens ($t=t_b$), and at later times ($t\sim t_1 \gg t_b$; post-breakout), the system is in the free-expansion phase (more details are given in Appendix \ref{ap:v sim}).

As a note, throughout this paper, quantities are calculated/measured in the laboratory frame (frame of the central engine), unless specified.

\begin{table}
\caption {
Our sample of jet models and their corresponding parameters (same as in \citealt{2023MNRAS.520.1111H}). 
From the left: 
The model name; 
the ejecta mass (assuming polar densities; roughly $\sim 1/5$ of the total ejecta mass when accounting for high densities in the equatorial region);
the jet initial opening angle;  
and the jet's isotropic equivalent luminosity [$L_{iso,0} = \frac{2 L_j}{1-\cos\theta_0}\simeq \frac{4 L_j}{\theta_0^2}$] where $L_j$ is the jet true luminosity (one sided);
note that the engine is active from $t_0$ to $t_e$, where $t_e-t_0=2$ s for all jet models.
Other parameters include:
the ejecta's initial density profile is taken as a power-law function with the power-law index $n=2$;
the maximum velocity of the ejecta, taken as $\beta_m=\sqrt{3/25} \approx 0.345$;
the delay time between the merger time (i.e., the launch of the ejecta) and the jet launch time is taken as $t_0-t_m = 0.160$ s.
Numerical simulations for these three models have been presented in (\citealt{2023MNRAS.520.1111H}), 
and properties of pre- and post-breakout cocoon (escaped) have been inferred using the analytic model of (\citealt{2023MNRAS.520.1111H}). 
Note that these models with their parameters are not universal; 
a diversity of ``narrow", ``wide", and ``failed" jets are expected in nature.
}
\label{tab:1}
\begin{tabular}{l|lll}
\hline
 Jet models  & $M_{e}$ [$M_\odot$] & $\theta_0$  [deg] & $L_{iso,0}$ [erg s$^{-1}$]  \\
  \hline
  Narrow & $0.002$ & $6.8$ & $5.0\times10^{50}$ \\
  Wide  & $0.002$ & $18.0$ & $5.0\times10^{50}$ \\
  Failed & $0.010$ & $18.0$ & $1.0\times10^{50}$ \\
    \hline
  \hline
 \end{tabular}
\end{table}

\subsection{Previous analysis}
The numerical results presented here are for the same sample of jet models as in \cite{2023MNRAS.520.1111H},
and as summarized in Table \ref{tab:1} (see \citealt{2017MNRAS.469.2361H} for the numerical code; also see \citealt{2020MNRAS.491.3192H} and \citealt{2021MNRAS.500..627H}).
More details can be found in Appendix \ref{ap:cocoon}.

\subsubsection{``Trapped cocoon" and ``escaped cocoon"}
\label{sec:trapped escaped ccs}
The same definition is used for separating the trapped cocoon (inside the ejecta) and the escaped cocoon (from the ejecta) as in \cite{2023MNRAS.520.1111H}:
\begin{equation}
  \beta_{inf} 
    \begin{cases}
      >\beta_m & \text{(escaped cocoon)} ,\\
      \leqslant \beta_m & \text{(trapped cocoon)} .
    \end{cases}       
    \label{eq:beta_inf cases}
\end{equation}
For the definition of $\beta_{inf}$ see equation (\ref{eq:Bernoulli Gamma}).
Here we take $\beta_m \sim 0.35$ (see Table \ref{tab:1} or Figure \ref{fig:v sim}; see \citealt{2021MNRAS.500..627H} and \citealt{2023MNRAS.520.1111H} for more details).
In the following, the trapped cocoon will not be considered because it does not contribute to the emission (see Section \ref{sec:es only}).

\subsubsection{Angular distribution of the escaped cocoon}
\label{sec:theta es 1}
In the following, we introduce the opening angle of this cone, $\theta_c^{es}$.
This angle is defined so that this cone includes $\sim 90\%$ of the escaped cocoon's mass.
Typical values are (see \citealt{2023MNRAS.520.1111H}):
\begin{equation}
  \theta_c^{es}
    \begin{cases}
      \sim 20^\circ & \text{Narrow jet model ($\theta_0=6.8^\circ$)} ,\\
      \sim 30^\circ & \text{Wide/failed jet model ($\theta_0=18.0^\circ$)} .
    \end{cases}       
    \label{eq:theta es cases}
\end{equation}

\subsubsection{``Relativistic" and ``non-relativistic" parts of the escaped cocoon}
\label{sec:escaped R NR}
The escaped cocoon is further classified as follows:
\begin{equation}
    \begin{cases}
      \beta_m<\beta_{inf}\le\beta_t & \text{(Non-relativistic)} ,\\
            \beta_t <\beta_{inf}\le\beta_{out} & \text{(Relativistic)}.
    \end{cases}       
    \label{eq:beta_t R NR cases}
\end{equation}
Here $\beta_{out}$ is the maximum velocity of the cocoon, set as $\beta_{out} =\sqrt{1-10^{-2}}\approx 0.995$
[see equation (\ref{eq:cc condition})].
$\beta_t$ is the transition velocity and is set as
\begin{eqnarray}
    \beta_t=0.8.
    \label{eq:beta_t}
\end{eqnarray}
This value of $\beta_t$ has been based on numerical simulation results showing that the profile of the escaped cocoon changes around this value (see Sections \ref{sec:den sim} and \ref{sec:Eint sim}; also see Figures \ref{fig:den sim}, \ref{fig:p sim}, and \ref{fig:key.recap}).
This transition velocity is useful as it allows one to split the cocoon into two limits, where the following approximations can be considered accordingly
\begin{equation}
    \Gamma_{inf} \beta_{inf} \approx
    \begin{cases}
\displaystyle
       \Gamma_{inf}\approx \frac{1}{\sqrt{2(1-\beta_{inf})}}  & \text{[Rela. ($\Gamma_{inf}\ge\Gamma_t$)]},      \\
      \beta_{inf} & \text{[Non-rela.  ($\beta_{inf}\le\beta_t$)]} ,
    \end{cases}       
    \label{eq:U4inf R NR cases}
\end{equation}
where $\Gamma_t=\frac{1}{\sqrt{2(1-\beta_t)}}$.
Note that this approximation of $\Gamma_t$ results in a slight discontinuity, as $\Gamma_t(\beta_t) =1$ in the non-relativistic domain, and $\Gamma_t(\beta_t)=\frac{1}{\sqrt{2(1-\beta_t)}}\approx 1.58$ in the relativistic domain. This has a very limited effect on the results.
It is worth recalling that, physically, the non-relativistic cocoon is mainly associated with the shocked ejecta part of the cocoon, and the relativistic cocoon with the shocked jet part of the cocoon (\citealt{2011ApJ...740..100B}; \citealt{2017ApJ...834...28N}).
Note that this classification is meaningful only for successful jet models, as the failed jet model lacks the relativistic part (see Figures \ref{fig:dEc sim} and \ref{fig:v sim}).

In the following the subscripts ``$r$" and ``$nr$" will be used to identify quantities
related to the ``relativistic" and ``non-relativistic" parts of the cocoon, respectively.

\subsection{Numerical profiles of the cocoon}
\label{sec:Numerical profiles of the cocoon}

\subsubsection{Density profile}
\label{sec:den sim}
\begin{figure}
    \centering
    \includegraphics[width=0.99\linewidth]{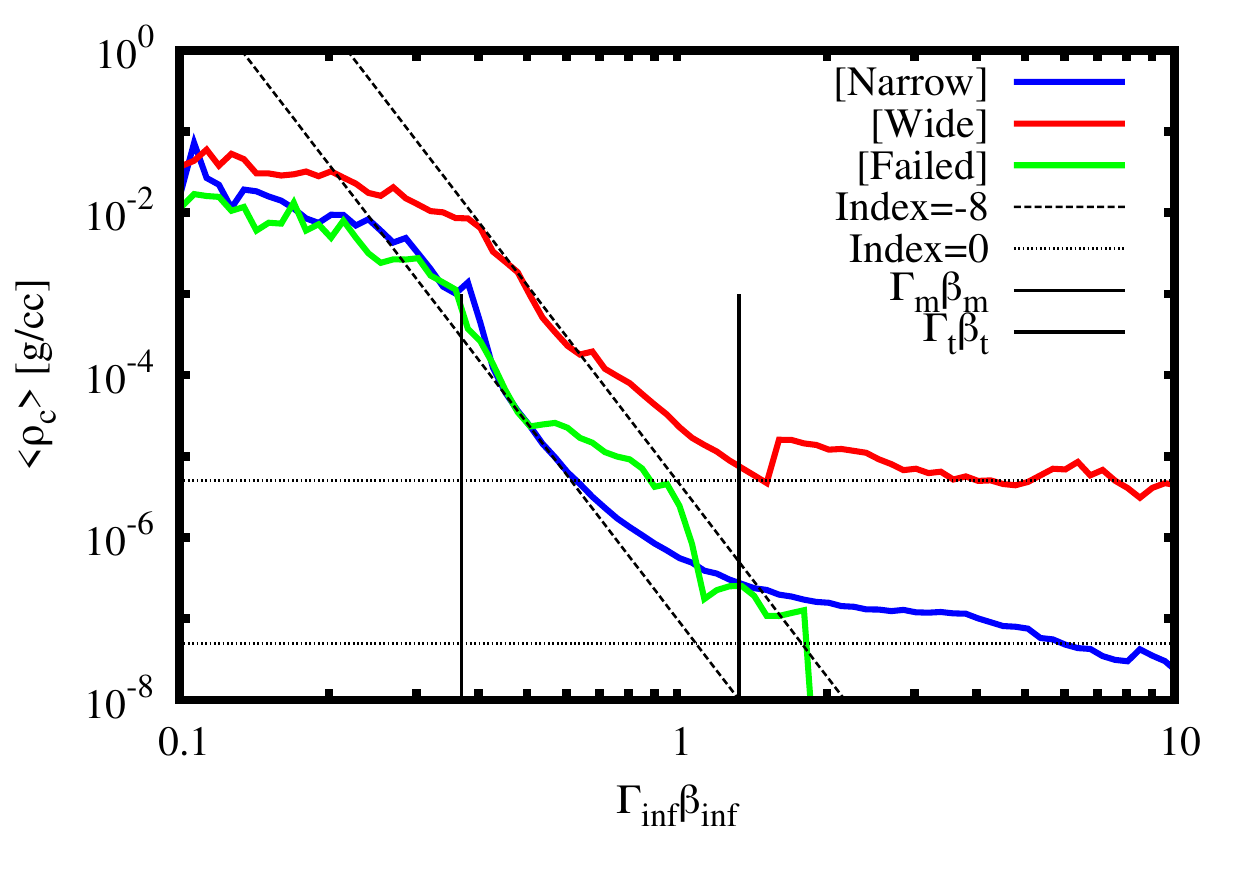} 
  
  \caption{Rest-mass density profile of the cocoon [averaged; see equation (\ref{eq:den av})], as a function of the four velocity $\Gamma_{inf}\beta_{inf}$.
  Three jet models are shown: narrow (blue), wide (red), and failed (green), at the free-expansion phase [$t_1-t_0=3$ s (narrow and wide), $=10$ s (failed)].
  The edge of the ejecta ($\Gamma_m\beta_m$), and separation between the non-relativistic and the relativistic cocoon  ($\Gamma_t\beta_t$) are shown (thin and thick solid vertical lines, respectively).
  Approximate power-law fit for the non-relativistic part of the escaped cocoon [$\Gamma_m\beta_m<\Gamma_{inf}\beta_{inf}\le\Gamma_t\beta_t$; $\propto (\Gamma_{inf}\beta_{inf})^{-8}$], and for the relativistic part of the escaped cocoon [$\Gamma_{inf}\beta_{inf}>\Gamma_t\beta_t$; $\propto (\Gamma_{inf}\beta_{inf})^{0}$] are provided (dashed and dotted lines, respectively).
  The escaped cocoon ($\Gamma_{inf}\beta_{inf}>\Gamma_m\beta_m$) has been defined by its opening angle $\theta_{c}^{es}$ (see Section \ref{sec:theta es 1}).
  Bumps in the density of the escaped cocoon just around $\Gamma_m\beta_m$ correspond to fluid elements that are about to cross the edge of the ejecta ($r \lesssim r_m$); 
  these bumps are expected to disappears at later times, as these fluid elements will be subject to lateral volume expansion soon after crossing the edge of the ejecta (similar to the failed jet model for $\Gamma_{inf}\beta_{inf}\sim 0.5-2$; see Figure \ref{fig:v sim}).
  }
  \label{fig:den sim} 
\end{figure}

In Figure \ref{fig:den sim}, the rest-mass density of the cocoon is presented as deduced from numerical simulations.
This density was extracted from simulations using a similar procedure to that in Appendix \ref{ap:v sim}.
Four-velocity bins were made [$d(\Gamma_{inf}\beta_{inf})$], and the cocoon's hydrodynamical properties were integrated in each bin.
One key difference in comparison (to Appendix \ref{ap:v sim}) is that, here, for the escaped cocoon ($\Gamma_{inf}\beta_{inf}>\Gamma_{m}\beta_m$), only fluid elements inside a certain opening angle ($\theta<\theta^{es}_c$; angle that includes $\sim 90\%$ of the escaped cocoon's mass; see \citealt{2023MNRAS.520.1111H}) were considered.
Hence, the mass-density was found using
\begin{eqnarray}
    \langle{\rho_c}\rangle =\frac{\int_{\Gamma_{inf}\beta_{inf}}^{\Gamma_{inf}\beta_{inf}+d(\Gamma_{inf}\beta_{inf})} \delta M_{c}}{\int_{\Gamma_{inf}\beta_{inf}}^{\Gamma_{inf}\beta_{inf}+d(\Gamma_{inf}\beta_{inf})} \delta V_c},
    \label{eq:den av}
\end{eqnarray}
where $\delta V_c$ is the differential volume of the cocoon.

Considering the homologous expansion
[$r\propto \beta\propto \beta_{inf}$; see equation (\ref{eq:av homologous})] and the scaling of the four-vector velocity in the two extremes [see equation (\ref{eq:U4inf R NR cases})],
Figure \ref{fig:den sim} shows that, first on average, the trapped cocoon ($\beta_{inf}\leqslant \beta_m$) follows the same density profile of the ejecta ($\propto \beta^{-n}\propto \beta^{-2}$).
This is understandable, considering that the escaped cocoon's mass is very small in comparison (\citealt{2023MNRAS.520.1111H}).
The non-relativistic part of the escaped cocoon shows a steep density profile, with an index $m\sim 8$; this density structure is quite similar to that of the outer ejecta in supernova explosions (see \citealt{1989ApJ...341..867C}; and \citealt{1999ApJ...510..379M}).
It is worth mentioning that, similarly to \cite{2021ApJ...909..209P}, our analytic model here also indicates that the exact value of $m$
does not significantly affect the results as long as $m\gg 1$.
The density follows a different trend in the relativistic part of the escaped cocoon (absent in the failed jet model), roughly as $\propto \Gamma_{}^l\propto \Gamma_{}^0\propto \rm{Const}$. 
To our best knowledge, this is the first time that density structure of the cocoon has been revealed from the non-relativistic limit to the relativistic limit.
In summary
\begin{equation}
\langle{\rho_c}\rangle \equiv \frac{dM_{c}}{dV_c} \propto 
    \begin{cases}
      \beta_{}^{-n} \:[\beta_{m}\geqslant\beta_{}] & \text{(Trapped)} ,\\
      \beta_{}^{-m} \:[\beta_t \geqslant\beta_{} > \beta_m] & \text{(Escaped; non-rela.)},\\
      \Gamma^{l} \:[\Gamma_{out} \ge \Gamma_{} > \Gamma_{t}] & \text{(Escaped; rela.)},    
    \end{cases}       
    \label{eq:dm/dv cases}
\end{equation}
where from our simulations, we have $n\approx 2$, $m\approx 8$, and $l\approx 0$. Figure \ref{fig:key.recap} (top) gives an analytic overview of the density profile throughout the cocoon.

As a note, in our simulations we used a relatively dense circumstellar density (CSM) $\rho_{CSM} =10^{-10}$ g cm$^{-3}$ for numerical reasons. This could have affected the relativistic cocoon (the narrow jet model in particular) mass and internal energy. However, this effect has been limited by: i) only considering the cocoon with $\theta<\theta_c^{es}$, and ii) taking $t_1\gg t_b$ but not too large ($3$ s for successful jet models, and $10$ s for the failed jet model). Also, it should be noted that, in reality as the ejecta could be surrounded by a fast tail component (see Figure 8 in \citealt{2020MNRAS.491.3192H}), there would be such dense component around the ejecta (even denser than what has been considered here), and it would naturally affect the escaped cocoon (and its emission) in a similar way.

\subsubsection{Internal energy distribution}
\label{sec:Eint sim}

\begin{figure}
    \centering
    \includegraphics[width=0.99\linewidth]{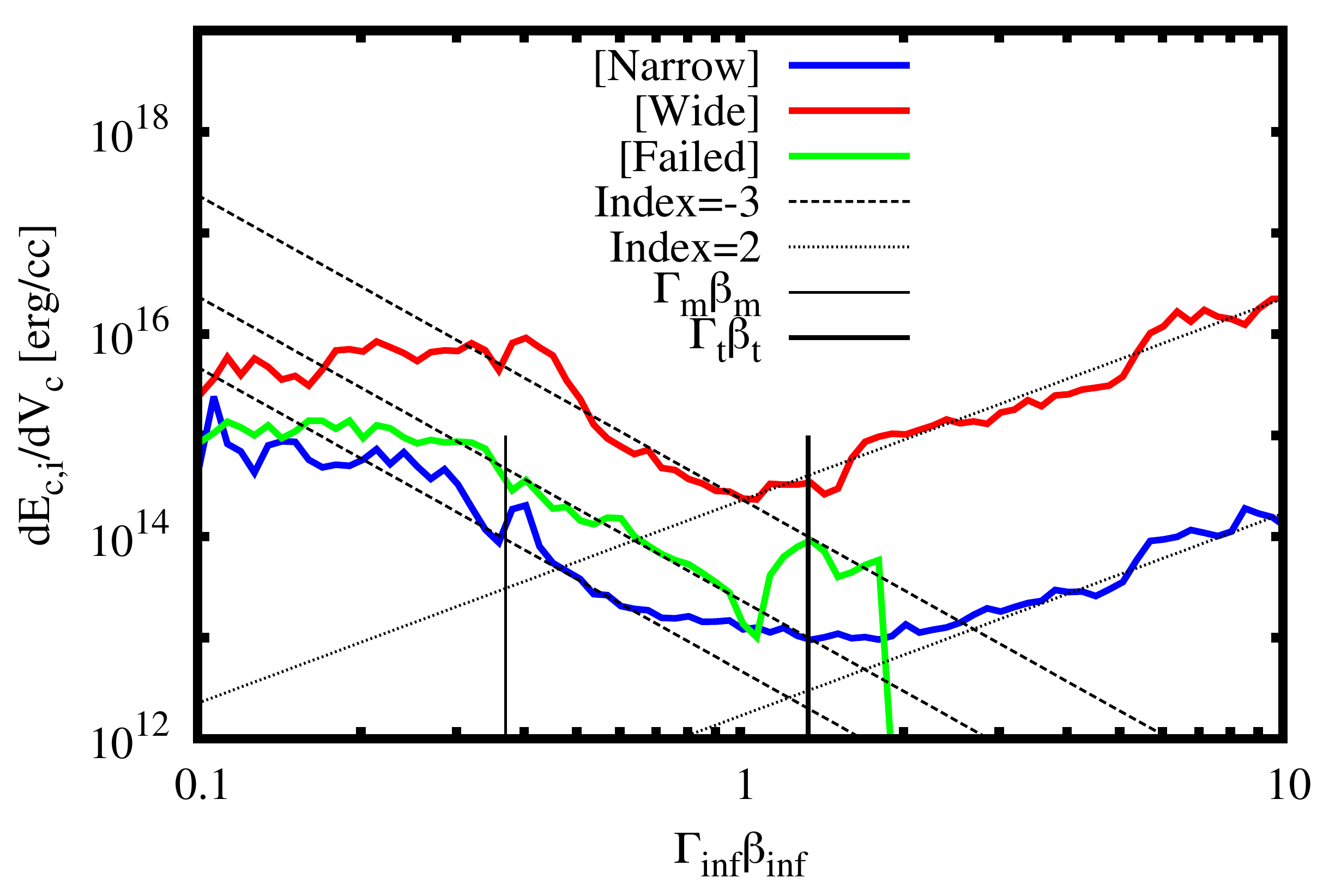} 
  \caption{Same as Figure \ref{fig:den sim} for the cocoon's internal energy density. 
  Approximate power-law fits for the non-relativistic part of the escaped cocoon [$\Gamma_m\beta_m<\Gamma_{inf}\beta_{inf}\le\Gamma_t\beta_t$; $\propto (\Gamma_{inf}\beta_{inf})^{-3}$], and for the relativistic part of the escaped cocoon [$\Gamma_{inf}\beta_{inf}>\Gamma_t\beta_t$; $\propto (\Gamma_{inf}\beta_{inf})^{2}$] are provided (dashed and dotted lines, respectively).
  Also, same as in Figure \ref{fig:den sim} concerning the bumps around $\beta_m$.
  }
  \label{fig:p sim} 
\end{figure}

In Figure \ref{fig:p sim}, the internal energy density of the cocoon as a function of $\Gamma_{inf}\beta_{inf}$ at the free-expansion phase is presented.
Similarly to the density in Section \ref{sec:den sim}, the profile of the internal energy density was extracted from numerical simulations using
\begin{eqnarray}
    \frac{dE_{c,i}}{dV_c} =\frac{\int_{\Gamma_{inf}\beta_{inf}}^{\Gamma_{inf}\beta_{inf}+d(\Gamma_{inf}\beta_{inf})} \delta E_{c,i}}{\int_{\Gamma_{inf}\beta_{inf}}^{\Gamma_{inf}\beta_{inf}+d(\Gamma_{inf}\beta_{inf})} \delta V_c},
    \label{eq:Eint average}
\end{eqnarray}
where $\delta E_{c,i}$ is the internal energy of a given fluid element.

Here too, considering $r\propto \beta\propto \beta_{inf}$ [see equation (\ref{eq:av homologous})] and the scaling of the four-vector velocity in the two limits [see equation (\ref{eq:U4inf R NR cases})], Figure \ref{fig:p sim} shows that the internal energy density of the trapped cocoon is $\propto \beta_{}^0$ (i.e., the comoving pressure $P=\rm{Const.}$). 
For the non-relativistic part of the escaped cocoon ($\Gamma\beta<\Gamma_t \beta_t$), the internal energy density varies as $\propto (\Gamma\beta_{})^{-3}\propto \beta_{}^{-3}$ (follows the volume increase).
For the relativistic part ($\Gamma\beta>\Gamma_t \beta_t$), it can be seen that the internal energy density varies as $\propto (\Gamma\beta_{})^{2}\propto \Gamma_{}^{2}$.
This reflects a constant pressure because the internal energy density in the relativistic limit can be written as
\begin{eqnarray}
    \frac{dE_{c,i}}{dV_c}=P\left[\left(\frac{\Gamma_a}{\Gamma_a-1}\right)\Gamma^2-1\right],
    \label{eq:Gamma_a and P}
\end{eqnarray}
and for $\Gamma \gg 1$ [see equation (\ref{eq:U4inf R NR cases})], $dE_{c,i}/dV_c =(4\Gamma^2-1)P\approx 4\Gamma^2 P\propto \Gamma^2$ (for an adiabatic index $\Gamma_a=4/3$).

Hence, at later times ($\beta\approx \beta_{inf}$, see Appendix \ref{ap:v sim}), and using the approximation in equation (\ref{eq:U4inf R NR cases}), the internal energy densities (according to Figure \ref{fig:p sim}) can be summarized as:
\begin{equation}
\frac{dE_{c,i}}{dV_c} \propto 
    \begin{cases}
      \beta_{}^0 \:[\beta_{m}\geqslant\beta_{}] & \text{(Trapped)} ,\\
      \beta_{}^{-3} \:[\beta_t \geqslant\beta_{} > \beta_m] & \text{(Escaped; non-rela.)},\\
      \Gamma^2 \:[\Gamma_{out} \ge \Gamma_{} > \Gamma_{t}] & \text{(Escaped; rela.)}.    \end{cases}       
    \label{eq:Eint cases}
\end{equation}
Consequently, the escaped cocoon has an equal amount of internal energy
per each logarithmic four-velocity $\Gamma\beta$
[see equation (\ref{eq:Eijcum R+NR})].
For an analytic summary, see Figure \ref{fig:key.recap} (bottom).

This distribution can be interpreted as follow.
First, in the (inner) trapped cocoon the pressure is (roughly) constant and internal energy is homogeneously distributed (already discussed in the literature: \citealt{2011ApJ...740..100B}; Figure 2 in \citealt{2017ApJ...834...28N}; Figure 5 in \citealt{2018MNRAS.477.2128H}; also \citealt{2021MNRAS.500..627H} and Figure 1 in \citealt{2023MNRAS.520.1111H} for the case of expanding medium).
Second, in the non-relativistic escaped cocoon expanding homologously ($r\propto \beta$), the volume (as the only fast varying term) $\propto \int \beta^2d\beta\propto \beta^3$ gives the $\propto \beta^{-3}$ dependency.
Third, at the relativistic limit of the escaped cocoon, 
the internal energy density is $\propto \Gamma^2$ [see equation (\ref{eq:Gamma_a and P})]. 
This is the first time that this internal energy structure of the escaped cocoon has been revealed.

Here too, it is worth noting that the relatively dense CSM ($\rho_{CSM} =10^{-10}$ g cm$^{-3}$) might have affected the internal energy of the relativistic cocoon (i.e., shock heating). However, as explained in Section \ref{sec:den sim}, such effect has been minimized, and is expected to eventually arise from the fast tail in reality. Another possible numerical issue here is the baryon diffusion near the jet (\citealt{2013ApJ...777..162M}), enhancing the cocoon with high-enthalpy jet outflow that contributes to the relativistic cocoon after the breakout (see Appendix \ref{ap:dEc sim}). As shown in \cite{2013ApJ...777..162M}, baryon loading is almost inevitable in numerical simulations. However, as show by \cite{2019MNRAS.490.4271M} and \cite{2021MNRAS.500.3511G}, 3D jets are inherently unstable, and dissipation of highly relativistic jet outflow into the cocoon is a natural physical phenomenon. Hence, although numerical artifacts may have contributed to the relativistic cocoon's internal energy here, we expect that features found here would still be similar to physically realistic ones.

\section{Analytic modeling of the escaped cocoon}
\label{sec:3}
\subsection{The parameters}
In the following the cocoon's hydrodynamical properties will be modeled, in order to estimate its emission. 
The key parameters are classified as i) parameters of the ejecta: mass $M_e$, density profile's power-law index $n$ (here $n=2$), and maximum velocity $\beta_m$,
and ii) parameters of the jet: isotropic equivalent luminosity $L_{iso,0}$, opening angle $\theta_0$, and delay time between the merger and the jet launch $t_0-t_m$ (see Table \ref{tab:1}).

Therefore, these quantities will be considered as known, and will be used in the following analytic formulation.
As shown in \cite{2023MNRAS.520.1111H}, these parameters allows one to analytically find the properties of the cocoon, relevant to the cocoon emission (i.e., the escaped part; see Section \ref{sec:es only}), in terms of rest-mass ($M_c^{es}$), internal energy ($E_{c,i}^{es}\equiv E_{c,i}^{j}$ here), and opening angle ($\theta_c^{es}$).

\subsection{Escaped cocoon and the cocoon emission}
\label{sec:es cc}

\subsubsection{Typical viewing angle to detect the cocoon emission}
\label{sec:theta_v}
The jet after its breakout has a bulk Lorentz factor $\Gamma_j\sim 10-100$.
The jet opening angle is approximately found as 
$\theta_j\sim \Gamma_j^{-1}\lesssim 6^\circ$.
As explained in Section \ref{sec:theta es 1}, the escaped cocoon approximately takes a conical geometry and is characterized by the opening angle $\theta_c^{es} \sim 20$--$30^{\circ}$ [see equation (\ref{eq:theta es cases})].

Let's consider an observer with a line of sight (LOS) with an angle $\theta_v$ relative to the jet (polar) axis.
For an on-axis observer (within the jet cone; $\theta_v<\theta_j$), after the prompt emission, afterglow emission comes from the decelerating jet, outshining the cocoon emission.
Therefore, in order for a clean detection of the cocoon emission to be achieved, in the following we consider the ideal scenario where the observer's viewing is in the range $\theta_j<\theta_v<\theta_{c}^{es}$ (so that there is no contamination from the early jet afterglow),
roughly $\theta_v\sim 6^\circ -20^\circ$.

\subsubsection{The escaped cocoon to power the cocoon emission}
\label{sec:es only}
The trapped cocoon is surrounded by the ejecta,
in addition to being covered by the escaped cocoon (in the radial direction).
Since the opening angle of the escaped cocoon $\theta_c^{es}$ is larger than that of the trapped cocoon (ellipsoidal shape; \citealt{2021MNRAS.500..627H}),
the emission from the trapped cocoon is hidden by the ejecta for a typical viewing angle (see Section \ref{sec:theta_v}).
Even if the viewing angle is luckily within the opening angle of the trapped cocoon, 
emission from the trapped cocoon is not observable until the escaped cocoon becomes optically thin.

As r-process heating, and the KN emission, dominates at late times, the cocoon emission is dominant only at early times.
Consequently, at later times, by the time the escaped cocoon (and the outer ejecta) becomes optically thin and the trapped cocoon is visible, most of its internal energy provided by jet-shock heating would have been cooled adiabatically, and internal energy from r-process heating dominates.
Therefore, at much later times after the breakout, the expected emission from the trapped cocoon is indistinguishable from that of the ejecta (i.e., the KN)\footnote{\label{foot:Klion+1}There is one exception; in the case where jet propagation through the ejecta forms a wide low density region around the polar (jet) axis.
Such structure would allow photons from the inner hot regions to escaped earlier on, before their internal energy cools adiabatically (\citealt{2021MNRAS.502..865K}; see their Figure 5).
If so, the trapped cocoon (its cooling emission) would be relevant. 
Still, this is in the extreme limit where the jet is launched almost immediately after the merger with intense luminosities, dramatically deforming the ejecta (see Table 1 in \citealt{2021MNRAS.502..865K}).}.

Hence, from an observational point of view, the escaped cocoon [see its definition in equation (\ref{eq:beta_inf cases})] is the relevant part for the cocoon emission.
In the following, we will refer to the escaped part of the cocoon simply as ``cocoon". 

\subsubsection{Angular dependency and 1D approximation}
\label{sec:Angular dep}
Simulations indicate that the mass and energy inside the escaped cocoon's cone (with $\theta_c^{es}$ as its opening angle) do have angular dependencies.
In particular, immediately after the breakout, internal energy of the escaped cocoon is more concentrated in the region near the jet axis (shocked jet part in particular).
However, in the following, 
we make a conservative estimate
by considering that 
the smeared-out energy distribution 
is relevant for a typical viewing angle. 
Therefore, we average out the angular distribution, and only consider the radial [or $\beta_r \sim \beta \sim \beta_{inf}$ dependencies; see equation (\ref{eq:Bernoulli Gamma}); also see Figure \ref{fig:v sim}] dependencies. 
This is reasonable considering our goal of analytically estimating the cocoon emission for an observer with a viewing angle $\theta_v$ (i.e., a line of sight; LOS) within the cocoon's cone, but outside (i.e., off-axis) of the jet's cone ($\theta_c^{es} > \theta_v\gtrsim \theta_j$; see Section \ref{sec:theta_v}).

\subsection{Velocity and Lorentz factor}
\label{sec:v profile}
At $t > t_1\gg t_b$, the expansion of the cocoon is considered as homologous (see Appendix \ref{ap:v sim} and Figure \ref{fig:v sim}).
In other words, $\beta_{inf}\equiv \beta$ and
\begin{eqnarray}
    r \approx c\beta t,
    \label{eq:r=cbt}
\end{eqnarray}
where $t$ is the time since the merger (setting $t_m=0$) [see equation (\ref{eq:av homologous})].
Also, it is worth recalling that in the following, the cocoon is divided in terms of four-velocity as follows [see equations (\ref{eq:beta_t}) and (\ref{eq:U4inf R NR cases}), respectively; also see Section \ref{sec:escaped R NR}]:
\begin{equation}
    \Gamma_{} \beta_{} \approx
    \begin{cases}
      \Gamma_{}\approx
      \displaystyle
      \frac{1}{\sqrt{2(1-\beta_{})}}  & \text{[Relativistic cocoon ($\Gamma\ge\Gamma_t$)]},\\
      \beta_{} & \text{[Non-relativistic cocoon ($\beta\le\beta_t$)]}.
    \end{cases}       
    \label{eq:U4 R NR cases}
\end{equation}
In other words, we take $\beta_t\lesssim\beta\sim 1$ (i.e., $\Gamma_t\lesssim \Gamma \sim [2(1-\beta)]^{-1/2}$), and $\Gamma_t\gtrsim \Gamma\sim 1$,
respectively.

It is important to note that, for the non-relativistic cocoon, although we approximate $\Gamma\sim 1$, the term $1-\beta$ (and only this term) is not approximated as $1-\beta\sim 1$, considering the still decent velocities of the non-relativistic cocoon [$\beta\sim 0.8-0.4$; see equation (\ref{eq:beta_t R NR cases})]. The term $1-\beta$ is particularly important for finding the observed time $t_{obs}$ appropriately [see equation (\ref{eq:t obs lab})], even at later times, at the limit when $\beta\sim\beta_m$ (see Sections \ref{sec:Lbl} and \ref{sec:Tobs}). This same $1-\beta$ term appears in the expression of the optical depth [equation (\ref{eq:tau def})], where it is also retained.

In the following, the rest-mass and internal energy densities of the (escaped) cocoon will be expressed as a functions of $\Gamma$ for the relativistic part, and as a function of $\beta$ for the non-relativistic part. 

\subsection{Analytic density profile}
\label{sec:Analytic density profile}
\begin{figure}
    \centering
    \begin{subfigure}
    \centering
    \includegraphics[width=0.99\linewidth]{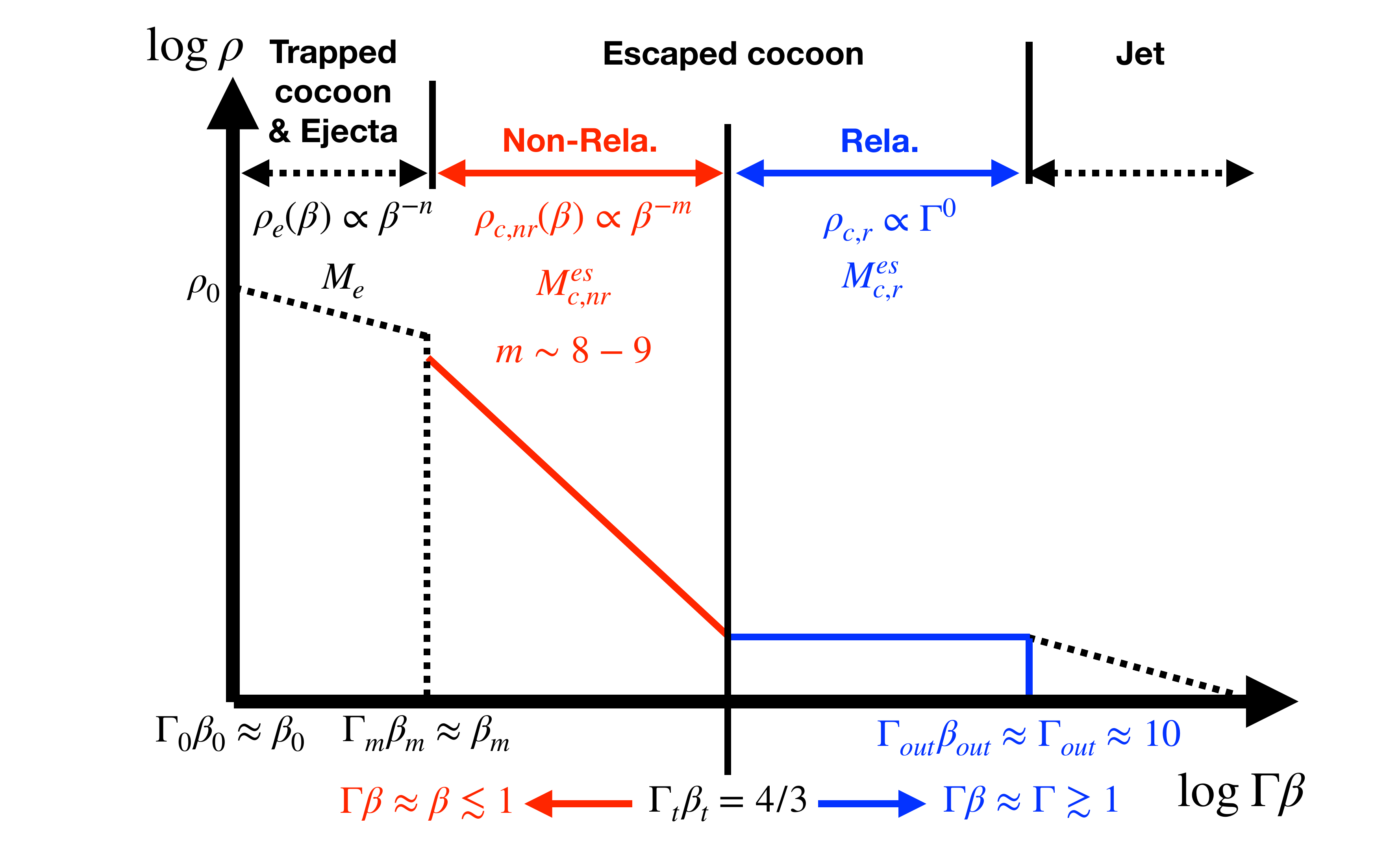}
  \end{subfigure}
  \begin{subfigure}
    \centering
    \includegraphics[width=0.99\linewidth]{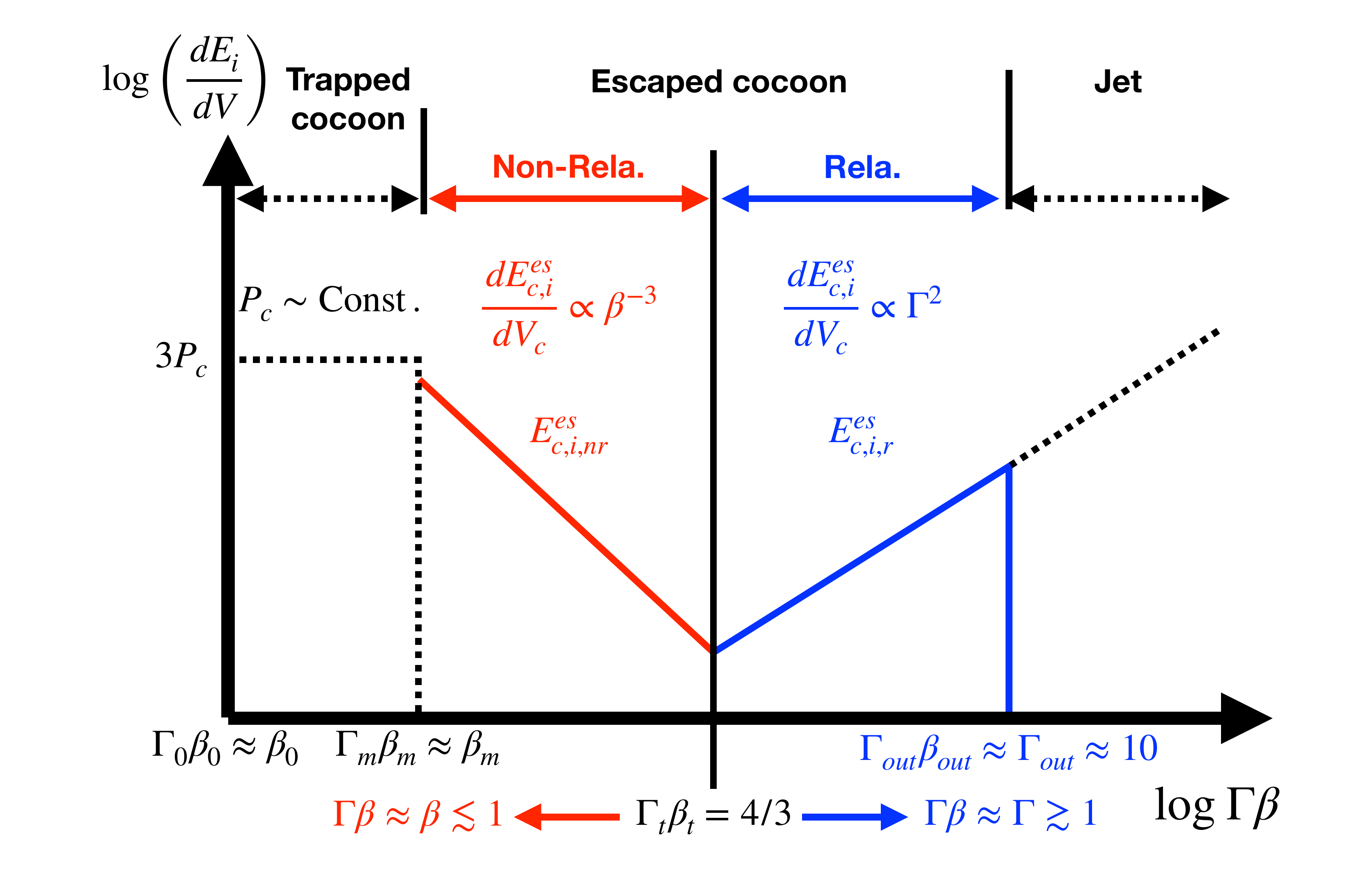}
  \end{subfigure}
  \caption{Top: mass-density (laboratory) profile of the post-breakout cocoon ($t\gg t_b$); 
  non-relativistic escaped cocoon (red) and relativistic escaped cocoon (blue) [for the original data see Figure \ref{fig:den sim}].
  Bottom: the same for the internal energy density (laboratory) [see Figure \ref{fig:p sim}].
  In the failed jet model, the escaped cocoon only composed of the red non-relativistic part (see Section \ref{sec:2} and Appendix \ref{ap:cocoon}).
  In both panels, the jet and the trapped cocoon (with the ejecta) are indicated in black dotted lines.
  }
  \label{fig:key.recap} 
\end{figure}

Figure \ref{fig:key.recap} illustrates the rest-mass density profile (top panel) for the post-breakout cocoon, with a focus on the escaped cocoon [see Figure \ref{fig:den sim}; and equation (\ref{eq:dm/dv cases}) in Section \ref{sec:den sim}].
By simplifying the expression of velocity [see equation (\ref{eq:U4 R NR cases})], the equations describing the spatial and temporal evolution of the relativistic cocoon's density $\rho_{c,r}(\Gamma\beta,t)\equiv  \rho_{c,r}(\Gamma,t)$ and the non-relativistic cocoon's density $\rho_{c,nr}(\Gamma\beta,t) \equiv  \rho_{c,nr}(\beta,t)$ can be simplified (respectively) as
\begin{equation}
    \begin{cases}
\rho_{c,r}(\Gamma,t) &=
\displaystyle \rho_t(t)\left(\frac{\Gamma}{\Gamma_t}\right)^{-l}, \\
\\
\rho_{c,nr}(\beta,t) &=
\displaystyle \rho_t(t)\left(\frac{\beta}{\beta_t}\right)^{-m} ,
\end{cases}       
    \label{eq:den R+NR cases}
\end{equation}
with the total escaped cocoon mass ${M^{es}_{c}}={M^{es}_{c,r}}+{M^{es}_{c,nr}}$, $M_{c,r}^{es}= \int_{\Gamma_t}^{\Gamma_{out}} \rho_{c,r}(\Gamma,t) dV_c$ [with $dV_c=\Omega(ct)^3\Gamma^{-3} d\Gamma $; see equations (\ref{eq:r=cbt}) and (\ref{eq:U4 R NR cases})], and
$M_{c,nr}^{es}= \int_{\beta_m}^{\beta_{t}} \rho_{c,nr}(\beta,t) dV_c$ [with $dV_c=\Omega(ct)^3\beta^2 d\beta$].
We can find the normalization as
\begin{equation}
\begin{split}
    &\rho_t(t)=\frac{M_c^{es}}{\Omega (ct)^3}\times\\
    &\left[\frac{\Gamma_t^l}{2+l}\left(\frac{1}{\Gamma_t^{2+l}}-\frac{1}{\Gamma_{out}^{2+l}}\right)+\frac{\beta_t^m}{m-3}\left(\frac{1}{\beta_m^{m-3}}-\frac{1}{\beta_{t}^{m-3}}\right)\right]^{-1}
    \end{split}
    \label{eq:rho_t}
\end{equation}
where $t\ge t_1$, $\Gamma\approx [2(1-\beta)]^{-1/2}$  (see Sections \ref{sec:trapped escaped ccs} and \ref{sec:escaped R NR} for the values of $\beta_m$, $\beta_t$, $\Gamma_{t}$, and $\Gamma_{out}$),
$m\approx 8\gg 1$ here, $l= 0$ here, and the solid angle of the escaped cocoon can be found as [in both hemispheres; see equation (\ref{eq:theta es cases}) for the values of $\theta_c^{es}$]
\begin{equation}
    \Omega=4\pi(1-\cos{\theta_c^{es}}).
    \label{eq:Omega}
\end{equation}
Note that $M_c^{es}$ is determined using the analytic model of \cite{2023MNRAS.520.1111H}.

Also, one can find the cumulative mass in the relativistic cocoon $M^{es}_{c,r}(>\Gamma)=\int_{\Gamma\ge\Gamma_t}^{\Gamma_{out}}  \rho_{c,r}(\Gamma,t) dV_c$, and in the non-relativistic cocoon $M^{es}_{c,nr}(>\beta)=\int_{\beta\ge\beta_m}^{\beta_{t}}  \rho_{c,nr}(\beta,t) dV_c$, respectively as
\begin{equation}
\begin{split}
M^{es}_{c,r}(>\Gamma) =& {M^{es}_{c,r}} \left[\frac{\Gamma^{-(2+l)}-\Gamma_{out}^{-(2+l)}}{\Gamma_t^{-(2+l)}-\Gamma_{out}^{-(2+l)}}\right] ,
\\
M^{es}_{c,nr}(>\beta)=& M^{es}_{c,nr} \left[\frac{\beta^{-(m-3)}-\beta_t^{-(m-3)}}{\beta_m^{-(m-3)}-\beta_t^{-(m-3)}}\right],
\end{split}
\label{eq:Mcum R+NR}
\end{equation}
where, ${M^{es}_{c,r}}\equiv {M^{es}_{c,r}}(>\Gamma_{t}) =\left[ \frac{\rho_t(t)\Omega (ct)^3\Gamma_t^{l}}{2+l}\right]\left\{\Gamma_{t}^{-(2+l)}-\Gamma_{out}^{-(2+l)}\right\}$ is the total mass of the relativistic cocoon part, and ${M^{es}_{c,nr}}\equiv {M^{es}_{c,nr}}(>\beta_{m}) =\left[ \frac{\rho_t(t)\Omega (ct)^3\beta_t^{m}}{m-3}\right]\left\{\beta_{m}^{-(m-3)}-\beta_{t}^{-(m-3)}\right\}$ is the total mass of the non-relativistic cocoon part.
This also gives alternative ways to write $\rho_t(t)$ as follows:
\begin{equation}
    \begin{split}
        \rho_t(t) =&  
        \left[
        \frac{M^{es}_{c,r}(2+l)}{\Omega (ct)^3\Gamma_t^{l}}
        \right]
        \left\{\Gamma_{t}^{-(2+l)}-\Gamma_{out}^{-(2+l)}\right\}^{-1},\\
         =&  \left[ \frac{M^{es}_{c,nr}(m-3)}{\Omega (ct)^3\beta_t^{m}}\right]\left\{\beta_{m}^{-(m-3)}-\beta_{t}^{-(m-3)}\right\}^{-1}.
    \end{split}
    \label{eq:rho_t v2}
\end{equation}

\subsection{Internal energy}
\label{sec:Eint}
The cocoon emission has two sources of internal energy.

\subsubsection{Internal energy from jet-shock heating}
\label{sec:Eint jet}
Figure \ref{fig:key.recap} illustrates the internal energy density from the jet-shock heating (bottom panel) for the post-breakout cocoon, with a focus on the escaped cocoon [see Figure \ref{fig:p sim}; and equation (\ref{eq:Eint cases}) in Section \ref{sec:Eint sim}].
Here too (similarly to the rest-mass density in Section \ref{sec:Analytic density profile}), by simplifying the expression of velocity [see equation (\ref{eq:U4 R NR cases})], equations describing the spatial and temporal evolution of this internal energy density for the relativistic cocoon part $e_{c,i,r}^j(\Gamma\beta,t)\equiv  e_{c,i,r}^j(\Gamma,t)$ and for the non-relativistic cocoon part $e_{c,i,nr}^j(\Gamma\beta,t) \equiv  e_{c,i,nr}^j(\beta,t)$ can be simplified (respectively) as
\begin{equation}
    \begin{cases}
e_{c,i,r}^j(\Gamma,t) &=
\displaystyle e_t^j(t)\left(\frac{\Gamma}{\Gamma_t}\right)^{2}, \\
\\
e_{c,i,nr}^j(\beta,t) &=
\displaystyle e_t^j(t)\left(\frac{\beta}{\beta_t}\right)^{-3} ,
\end{cases}       
    \label{eq:Eint/v R+NR cases}
\end{equation}
with the total internal energy (from the jet-shock heating) ${E^{j}_{c,i}}(t)={E^{j}_{c,i,r}}(t)+E^{j}_{c,i,nr}(t)$, $E^{j}_{c,i,r}(t)= \int_{\Gamma_t}^{\Gamma_{out}} e_{c,i,r}(\Gamma,t) dV_c$ [with $dV_c=\Omega(ct)^3\Gamma^{-3} d\Gamma $; see equations (\ref{eq:r=cbt}) and (\ref{eq:U4 R NR cases})], and $E^{j}_{c,i,nr}(t)= \int_{\beta_m}^{\beta_t} e_{c,i,nr}(\beta,t) dV_c$ [with $dV_c=\Omega(ct)^3\beta^2 d\beta$]. One can find
\begin{equation}
    e_t^j(t)=\frac{E_{c,i}^{j}(t)}{\Omega (ct)^3}
    \left[\Gamma_t^{-2}\ln\left(\frac{\Gamma_{out}}{\Gamma_t}\right)+\beta_t^{3}\ln\left(\frac{\beta_{t}}{\beta_m}\right)\right]^{-1}
    \label{eq:Eint_t}
\end{equation}
where $t\ge t_1$ and $\Gamma\approx [2(1-\beta)]^{-1/2}$ (see Sections \ref{sec:trapped escaped ccs} and \ref{sec:escaped R NR} for the values of $\beta_m$, $\beta_t$, $\Gamma_{t}$, and $\Gamma_{out}$).
The analytic modeling presented in \cite{2023MNRAS.520.1111H} allows us to find $E_{c,i}^{j}(t_1)\equiv E_{c,i}^{es}(t_1)$ reasonably well.
By taking into account the adiabatic expansion [in the free-expansion phase, see equation (\ref{eq:r=cbt})], we can write [see equation (\ref{eq:E conserved rp+adiabatic}) for the general case]
\begin{eqnarray}
    E_{c,i}^{j}(t) = E_{c,i}^{j}(t_1)\left(\frac{t_1}{t}\right).
    \label{eq:Adiabatic 1/t}
\end{eqnarray}
Hence, equations (\ref{eq:Eint_t}) and (\ref{eq:Eint/v R+NR cases}) are closed, allowing us to find the internal energy (from the jet-shock heating) of the escaped cocoon at later times as a function of $\Gamma$ (for the relativistic part) and $\beta$ (for the non-relativistic part).

We also calculate the cumulative internal energy, in the relativistic cocoon as $E_{c,i,r}^j(>\Gamma,t)=\int_{\Gamma\ge \Gamma_t}^{\Gamma_{out}}  e_{c,i,r}^j(\Gamma,t) dV_c$, and in the non-relativistic cocoon as $E_{c,i,nr}^j(>\beta,t)=\int_{\beta\ge\beta_m}^{\beta_{t}}  e_{c,i,nr}^j(\beta,t) dV_c$; they can be found (respectively) as [using equations (\ref{eq:Eint/v R+NR cases}) and (\ref{eq:Eint_t})]
\begin{equation}
\begin{split}
{E^{j}_{c,i,r}(>\Gamma,t)} =& {E^{j}_{c,i,r}(t)} \left[\frac{\ln{\left({\Gamma_{out}}/{\Gamma}\right)}}{\ln{\left({\Gamma_{out}}/{\Gamma_t}\right)}}\right], \\
    {E^{j}_{c,i,nr}(>\beta,t)}=& {E^{j}_{c,i,nr}(t)}  \left[\frac{\ln(\beta_{t}/\beta)}{\ln(\beta_{t}/\beta_m)}\right],
\end{split}
\label{eq:Eijcum R+NR}
\end{equation}
where, at a given time $t$,
$E^{j}_{c,i,r}(t)\equiv E^{j}_{c,i,r}(>\Gamma_{t}, t) =\left[ \frac{e_t^j(t)\Omega (ct)^3}{\Gamma_t^2}\right]\ln\left(\frac{\Gamma_{out}}{\Gamma_t}\right)$ is the total internal energy in the relativistic cocoon part and $E^{j}_{c,i,nr}(t)\equiv E^{j}_{c,i,nr}(>\beta_m, t)=\left[\frac{e_t^j(t)\Omega (ct)^3}{\beta_t^{-3}}\right]\ln\left(\frac{\beta_t}{\beta_m}\right)$
is the total internal energy in the non-relativistic cocoon part [with $E^{j}_{c,i}(t) = E^{j}_{c,i,r}(t) + E^{j}_{c,i,nr}(t)$].
Consequently, the internal energy can be understood as equally distributed per logarithmic four-velocity $\Gamma\beta$.

\subsubsection{Internal energy from radioactive heating}
\label{sec:Edep rp}
The dynamical ejecta in NS mergers is neutron rich (low $Y_e$). 
It presents a favorable environment for the nucleosynthesis of heavy elements through rapid neutron capture (i.e., r-process).
These elements are unstable; they decay (e.g., $\alpha$,  $\beta$ decay, and nuclear fission) heating up the ejecta, and eventually powering the KN emission (\citealt{1989Natur.340..126E}).
This process is also expected to heat up the cocoon (\citealt{2017ApJ...834...28N}).

This radioactive heating rate has been studied numerically (\citealt{2014ApJ...789L..39W}), and it has been shown that its time evolution can be approximated to two phases, a flat phase, and a power-law phase, both of which take place in the comoving frame of the expanding fluid, as follow (see Figure 2 in \citealt{2021ApJ...922..185I}):
\begin{equation}
    \dot{\varepsilon}(t')= \begin{cases} \dot{\varepsilon}_0\left(\frac{t'}{t_h'}\right)^{0} & \left(t'<t_{h}'\right) ,\\  
    \dot{\varepsilon}_0\left(\frac{t'}{t_h'}\right)^{-k} & \left(t' \geqslant t_{h}'\right),
    \end{cases} 
    \label{eq:Epsilon dot}
\end{equation}
where,
$t'=t/\Gamma$ is the comoving time,
$\dot{\varepsilon}_0$ is the heating rate at $t_h'$, 
$t_h'\sim 0.1$ s (considering that the cocoon originates from the polar region of the ejecta with $Y_e \sim 0.3$),
and $k\sim 1.3$ (\citealt{2014ApJ...789L..39W} \citealt{2017MNRAS.468...91H}; \citealt{2021ApJ...922..185I}; etc.).
The corresponding initial heating rate is $\dot{\varepsilon}_0\sim 1\times 10^{18}$ erg g$^{-1}$ s$^{-1}$ (or $\sim 2\times 10^{10}$ erg g$^{-1}$ s$^{-1}$ at 1 day; see Figure 2 in \citealt{2021ApJ...922..185I}).

The ``rest-mass" of a given shell of the radioactive cocoon in the laboratory frame, $dM_c^{es}$, is the same as that in the comoving frame $dM^{es}_c = {dM^{es}_c}'$.
Also, the power (i.e., energy deposition rate here) in the laboratory frame is the same as in the comoving frame\footnote{\label{foot:L=L'}
With $dt=\Gamma dt'$, and $dE= \Gamma dE'$ (where we are considering internal energy $E \equiv E_i$), 
the power $\frac{dE}{dt} \equiv \dot{E}$ in the laboratory frame is the same as that in the comoving frame $\frac{dE}{dt} = \frac{dE'}{dt'}$ [see equation (4.87) in \citealt{1979rpa..book.....R}].
Note that this is a simplification that could results in a factor $4/3$ in the relativistic limit [see \citealt{2013MNRAS.433.2107N}; in particular their equation (4)].}.
With the energy deposition rate being the product of the rest-mass and the heating rate $\dot \varepsilon(t')$, one can deduce that $\dot \varepsilon_0$ is also the same for both frames.
Hence, the internal energy deposition rate into the cocoon beyond a certain velocity (i.e., $>\Gamma\beta$) at a given time in the laboratory frame can be calculated as:
\begin{equation}
    \dot{E}_{c,i}^{rp}(>\Gamma\beta,t)= 
    \begin{cases} 
    \int_{>\Gamma\beta}\rho_c(\Gamma\beta,t)dV_c \dot{\varepsilon}_0\left(\frac{t}{\Gamma t_h'}\right)^{0} & \left(\frac{t}{\Gamma}<t_{h}'\right), \\
    \\
    \int_{>\Gamma\beta}\rho_c(\Gamma\beta,t)dV_c
    \dot{\varepsilon}_0\left(\frac{t}{\Gamma t_h'}\right)^{-k} & \left(\frac{t}{\Gamma}\geqslant t_{h}'\right).
    \end{cases}
    \label{eq:rp dep cases}
\end{equation}

It should be noted that, 
first, for the escaped cocoon, 
internal energy from the jet is overwhelmingly dominant until late times when the inner part of the escaped cocoon ($\gtrsim \beta_m$) becomes optically thin, at $t_{obs}\sim t \sim 100 -1000$ s $\gg t_h'$ [see equation (\ref{eq:t obs lab}) and Figure \ref{fig:NWF L v T}].
Second, as the system expands homologously, internal energy decays as $\propto 1/t$ [adiabatic cooling; see equation (\ref{eq:Adiabatic 1/t})].
At times when the cocoon emission becomes relevant $t \gg t_h'$, contribution (in terms of internal energy) from the early flat phase is insignificant, and can be ignored.
Third, taking into account radioactive heating
and the adiabatic cooling (but not the emission of the internal energy yet), energy conservation gives (regardless of the frame; see footnote \ref{foot:L=L'})
\begin{eqnarray}
\frac{\partial E_{c,i}^{rp}(>\Gamma\beta,t)}{\partial t}=-\frac{E_{c,i}^{rp}(>\Gamma\beta,t)}{t}+\dot{E}^{rp}_{c,i}(>\Gamma\beta,t).
\label{eq:E conserved rp+adiabatic}
\end{eqnarray}
Hence, at later times (for $t\gg \Gamma t_h'$ and $k\sim 1.3$), ${E}_{c,i}^{rp}(>\Gamma\beta,t)=\frac{1}{t}\int_{>\Gamma t_h'}^t\dot{E}_{c,i}^{rp}(>\Gamma\beta,t)tdt \sim \dot{E}_{c,i}^{rp}(>\Gamma\beta,t)t \propto t^{1-k}$;
just as in the case of the KN emission, 
the remaining total internal energy is dominated by r-process internal energy deposition and can be approximated as the product of energy deposition rate ($\dot{E}_{c,i}^{rp}$) and time ($t$) [as shown in equation (\ref{eq:Erp cum R+NR})]. 

The internal energy deposition rate beyond a certain shell (defined by $\beta$ or $\Gamma$) can be found at a given time, for the relativistic and for the non-relativistic cocoon, respectively, as [using equations (\ref{eq:U4 R NR cases}), deriving equation (\ref{eq:Mcum R+NR}) and replacing in equation (\ref{eq:rp dep cases}); while taking $\Gamma^k\sim 1$ for the non-relativistic cocoon; and 
$t=\rm{Const.}$]
\begin{equation}
\begin{split}
\dot{E}_{c,i,r}^{rp}(>\Gamma,t) =& \dot{\varepsilon}_0 {M^{es}_{c,r}}\left(\frac{t}{ t_h'}\right)^{-k}\left(\frac{l+2}{l+2-k}\right)\\
&\times \left[\frac{\Gamma^{-(2+l-k)}-\Gamma_{out}^{-(2+l-k)}}{\Gamma_t^{-(2+l)}-\Gamma_{out}^{-(2+l)}}\right], \\
\dot{E}_{c,i,nr}^{rp}(>\beta,t) =& \dot{\varepsilon}_0 {M^{es}_{c,nr}}\left(\frac{t}{t_h'}\right)^{-k}\left[\frac{\beta^{-(m-3)}-\beta_t^{-(m-3)}}{\beta_m^{-(m-3)}-\beta_t^{-(m-3)}}\right] ,
\end{split}
\label{eq:Edep cum R+NR}
\end{equation}
with $\dot{E}_{c,i,r}^{rp}(t) =\dot{E}_{c,i,r}^{rp}(>\Gamma,t) + \dot{E}_{c,i,r}^{rp}(<\Gamma,t)$ and $\dot{E}_{c,i,nr}^{rp}(t) =\dot{E}_{c,i,nr}^{rp}(>\beta,t) + \dot{E}_{c,i,nr}^{rp}(<\beta,t)$.
The stored r-process internal energy beyond a certain shell at a time (for $t\gg \Gamma t_h'$) can then be found, for the relativistic and for the non-relativistic cocoon, respectively,
as [using equation (\ref{eq:E conserved rp+adiabatic}); for $l=0$]
\begin{equation}
\begin{split}
E_{c,i,r}^{rp}(>\Gamma,t) \approx& \frac{t\dot{E}^{rp}_{c,i,r}(>\Gamma,t)}{2-k} , \\
E_{c,i,nr}^{rp}(>\beta,t) \approx& \frac{t\dot{E}^{rp}_{c,i,nr}(>\beta,t)}{2-k} .
\end{split}
\label{eq:Erp cum R+NR}
\end{equation}

\section{Cocoon emission: Analytic modeling}
\label{sec:4}

\subsection{Main approximations}
\label{sec:cc emission basics}
We follow the same hydrodynamical treatment as in \cite{2023MNRAS.520.1111H}, e.g.,
the internal energy is dominated by radiation (adiabatic index $\Gamma_a =4/3$; see Section \ref{sec:Eint sim}).

As explained in Sections \ref{sec:2} and \ref{sec:v profile}, 
at $t>t_1\gg t_b$, acceleration by radiation is negligible.
Therefore, the system is considered as in the coasting phase and the velocity is almost constant $\beta\sim \beta_{inf}$
(see Figure \ref{fig:v sim}).

In regard to radiation, we use the luminosity shell approximation;
at the diffusion shell where the diffusion time is comparable to the dynamical time, internal energy is instantly (and completely) transformed into photons (as in \citealt{2012ApJ...747...88N}).
Hence, with the inner regions being more opaque than the outer region, the luminosity shell tend to move inward (in a Lagrangian coordinate), from high velocity shells to low velocity shells.

Note that, in the following we do not take into account the shock breakout emission, nor attempt to solve the emission in the very early time soon after the cocoon breakout.
Therefore, our model is applicable for laboratory times that are much later than the breakout time $t\gtrsim t_1 \gg t_b$.

\subsection{Optical depth}
\label{sec:Optical depth}
In the relativistic limit, the optical depth for an observer with a LOS with a viewing angle $\theta_v$ (relative to the polar axis; a direction $z$), and located at a distance $z=Z_{obs}$ (away from the central engine) can be found as a function of the observed time $t_{obs}$ as (\citealt{1991ApJ...369..175A})
\begin{equation}
\tau( t_{obs})=
\int_{z}^{Z_{obs}} \kappa \rho\left(z,t_{\mathrm{obs}}-\frac{Z_{\mathrm{obs}}-z}{c}\right) \left[1-\beta(z) \cos \Theta\right] dz,
\label{eq:tau def}
\end{equation}
where $\kappa$ is the opacity, 
$\rho =\Gamma \rho'$ is the density in the laboratory frame (with $\rho'$ as the density in the comoving frame),
$\theta$ is the angle of the fluid to the polar axis and $\Theta$ is the angle between the LOS and the velocity vector $\vec{\beta}$.

In order to estimate the values taken by $\cos\Theta$, we consider the relativistic ($\Gamma\gg 1$) and the non-relativistic ($\beta\sim \beta_m$) limits, and the fact that the cocoon distribution is axial symmetric.
At early times ($\Gamma\gg 1$), as only a small region of the relativistic cocoon is visible to the observer, 
centered around the fluid element with $\theta=\theta_v$, 
we have $\Theta\le 1/\Gamma$, 
and $\cos\Theta \sim 1$ (velocity vectors in the emitting region can be approximated as well alighted with the LOS). 
At late times, (as $\Gamma\sim 1$), and at some point [when $\Gamma\sim 1/\sin(\theta_v+\theta_c^{es})$], the outermost edge of the non-relativistic cocoon (with its angle $\theta_c^{es}$) becomes visible to the observer and $|\Theta|\le\theta_c^{es}+\theta_v$.
Considering that velocities at this limit, $\beta\sim \beta_m$, are no longer at the vicinity of 1, the term $1-\beta\cos\Theta$ is slowly varying with $\beta$.
Also, considering $\theta_v$ and that the cocoon fluid is distributed at overall small angles [$\theta_c^{es}\sim 20^\circ-30^\circ$; see equation (\ref{eq:theta es cases}); also see Section \ref{sec:theta_v}], the minimum values taken by $\cos\Theta$ (at the non-relativistic limit) are still close to 1.
Therefore, it is reasonable (and very convenient) to approximate $\cos\Theta\sim 1$
inside the emitting region.

This approximation is combined with other simplifications. In particular, the observed region (e.g., photosphere, or inner boundary of the diffusion region) is approximated to a shell with the same radius and velocity (same $r$ and $\beta$; regardless of $\theta$). Also, the arrival time of photons is simplified as it is considered as the same for all fluid elements of the same shell (in reality the arrival time is between $t_{obs}$ and $\sim 2t_{obs}$ for one shell).
This allows us to substitute $z\equiv r$ ($z\approx r\approx c\beta t$,  and $Z_{obs}\approx R_{obs}$) giving
\begin{equation}
\tau( t_{obs})\approx \int_{r}^{r_{out}} \kappa \rho\left(r,t_{\mathrm{obs}}-\frac{R_{\mathrm{obs}}-r}{c}\right) \left[1-\beta(r) \right] d r ,
\end{equation}
where $r_{out} \approx c\beta_{out}t$ is the outer edge of the cocoon [see equation (\ref{eq:beta_t R NR cases})].

For an observer at a distance $R_{obs}$, as photons and GWs travel with the speed of light, one can relate the observed time and laboratory time as $t_{obs} = t +R_{obs}/c-r/c$.
For simplicity, the merger time has been set as the reference time at the laboratory frame ($t_m=0$).
The merger time at the observer's frame $t_{obs,m}\equiv t_m+R_{obs}/c =R_{obs}/c$ is also set as the reference time for the observer, i.e., $t_{obs}-t_{obs,m}\equiv t_{obs}$, giving $t_{obs} = t -r/c$.
Hence, using equation (\ref{eq:r=cbt}), the observed time and the laboratory time can be related as 
\begin{eqnarray}
t_{obs} =(1-\beta)t,
\label{eq:t obs lab}
\end{eqnarray}
where both times here are measured since the merger\footnote{It is worth stressing that, as explained in Section \ref{sec:v profile},
even in the non-relativistic cocoon's case, 
we retain $1-\beta$ (and only this term) avoiding the approximation $1-\beta \sim 1$; this is precise yet a concise approach.}.
Additionally, one can find that $r=ct_{obs}\frac{\beta}{1-\beta}$, and   $\frac{dr}{d\beta}\mathrel{\Big|}_{t_{obs}=\rm{Const.}}=\frac{ct_{obs}}{(1-\beta)^2}$.
This allows one to 
replace $r$ with the more useful $\beta$ in the above integration
\begin{equation}
\tau( t_{obs})=\kappa\int_{\beta}^{\beta_{out}}  \rho\left(\beta,\frac{t_{obs}}{1-\beta}\right) \left(\frac{ct_{obs}}{1-\beta}\right) d \beta .
\label{eq:tau final}
\end{equation}
Here, we take $\kappa \sim 1$ cm$^2$ g$^{-1}$, although, in reality, $\kappa$ depends on the comoving temperature $T'$, and $T'$ varies with $t$ and $\beta$. This is a reasonable fiducial value (corresponding to $T'\sim 10^5$ K; see Figure 8 in \citealt{2020ApJ...901...29B}; also see Figure 3 in \citealt{2022ApJ...934..117B}).

\subsubsection{The relativistic cocoon's observed time $t_{obs}(\Gamma)$}
\label{sec:tobs R}
For the relativistic cocoon part, the optical depth is refereed to as $\tau_r(t_{obs})$.
It can be written as a function of $t_{obs}$ and $\Gamma$ as follows [using equations (\ref{eq:U4 R NR cases}), (\ref{eq:den R+NR cases}), (\ref{eq:rho_t v2}), and (\ref{eq:tau final})]:
\begin{equation}
\tau_r(t_{obs})= \frac{1}{t_{obs}^2}\left[\frac{\kappa M_{c,r}^{es}}{c^2\Omega}\frac{2+l}{4(6+l)}\right]\left[\frac{\Gamma^{-(6+l)}-\Gamma_{out}^{-(6+l)}}{\Gamma_t^{-(2+l)}-\Gamma_{out}^{-(2+l)}}\right] .
\label{eq:tau R}
\end{equation}
Consequently, the observer time can then be deduced as
\begin{equation}
t_{obs}(\Gamma)= \left\{\frac{1}{\tau_{r}}\left[\frac{\kappa M_{c,r}^{es}}{c^2\Omega}\frac{2+l}{4(6+l)}\right]\frac{\Gamma^{-(6+l)}-\Gamma_{out}^{-(6+l)}}{\Gamma_t^{-(2+l)}-\Gamma_{out}^{-(2+l)}}\right\}^{1/2} ,
\label{eq:tobs R}
\end{equation}
where $t_{obs}(\Gamma)$ is the time when photons emitted from the shell moving with a Lorentz factor $\Gamma$ are observed [the condition for photon diffusion defines $\tau_r$ and is introduced in equation (\ref{eq:tau_d R})].

It should be noted that this formulation underestimates the observed times for photons emitted from $\Gamma \sim \Gamma_{out}$ (at extremely early times) because the homologous approximation is not ideal at times of the order of $t_{obs}(\Gamma_{out})$, which [with equation (\ref{eq:t obs lab})] would correspond to laboratory times earlier than the start of the free-expansion phase ($t_1$).

\subsubsection{The non-relativistic cocoon's observed time $t_{obs}(\beta)$}
\label{sec:tobs NR}
For the non-relativistic cocoon part, the optical depth is refereed to as $\tau_{nr}(t_{obs})$.
It can be written as a function of $t_{obs}$ and $\beta$ as follows [using equations (\ref{eq:U4 R NR cases}), (\ref{eq:den R+NR cases}), (\ref{eq:rho_t v2}), and (\ref{eq:tau final})]
\begin{equation}
\tau_{nr}(t_{obs}) = \frac{1}{t_{obs}^2}\frac{\kappa M_{c,nr}^{es}(m-3)}{\Omega c^2 \left(\beta_{m}^{3-m}-{\beta_t}^{3-m}\right)}\int_{\beta}^{\beta_t}\frac{(1-\beta)^2}{\beta^{m}}d\beta.
\label{eq:tau NR exact}
\end{equation}
This integration can be solved analytically.
Simplifications can be made for a cleaner analytic form, especially at the non-relativistic limit when $\beta\sim \beta_m$.
Considering that $m\gg1$, the term $\beta^{1-m}$ contributes more significantly to the change in $\tau_{nr}$ than the term $(1-\beta)^2$, and the integration can be simplified as
\begin{equation}
\begin{split}
\tau_{nr}(t_{obs}) &\approx \frac{1}{t_{obs}^2}\frac{\kappa M_{c,nr}^{es}(m-3)\left[\frac{(1-\beta)^2}{\beta^{m-1}}-\frac{(1-\beta_{t})^2}{\beta_{t}^{m-1}}\right]}{\Omega c^2(m-1) \left(\beta_{m}^{3-m}-{\beta_t}^{3-m}\right)}\\
&\approx \frac{1}{t_{obs}^2}\frac{\kappa M_{c,nr}^{es}\beta_{m}^{m-3}}{\Omega c^2}\left[\frac{(1-\beta)^2}{\beta^{m-1}}-\frac{(1-\beta_{t})^2}{\beta_{t}^{m-1}}\right],
\end{split}
\label{eq:tau NR}
\end{equation}
where, intuitively, we only integrate the rapidly varying part $\beta^{1-m}$; mathematically, we used $[f(\beta)g(\beta)]_{a}^{b} = \int_{a}^{b}{ f(\beta) g'(\beta) d\beta} +\int_{a}^{b}{ f'(\beta)g(\beta)d\beta}$, where $f(\beta)=(1-\beta)^2$ and $g(\beta)=\beta^{1-m}$, and the last term $\int_a^b f'(\beta)g(\beta)d\beta=\int_{a}^{b}\frac{2\Gamma^2\beta(1+\beta)}{m-2}{ f(\beta) g'(\beta) d\beta}$ has been neglected in the non-relativistic limit [considering $m\approx 8$ and $\Gamma\approx 1$ for $\beta\sim \beta_m$; see equation (\ref{eq:U4 R NR cases})]. This is reasonable considering $\beta\lesssim \beta_t$, and results in an acceptable error ($\sim2$ at very early times when $\beta=\beta_t$, but $\sim \frac{7}{6}\sim 1$ at later times when $\beta\sim \beta_m$).

Similarly to Section \ref{sec:tobs R}, the observer time can (approximately)
be found as
\begin{equation}
t_{obs}(\beta) \approx \left\{
\frac{1}{\tau_{nr}}\frac{\kappa M_{c,nr}^{es}\beta_{m}^{m-3}}{\Omega c^2 }\left[\frac{(1-\beta)^2}{\beta^{m-1}}-\frac{(1-\beta_{t})^2}{\beta_{t}^{m-1}}\right]
\right\}^{1/2},    
\label{eq:tobs NR}
\end{equation}
where $t_{obs}(\beta)$ is the time when photons emitted from the shell moving with a velocity $\beta$ are observed [the condition for photon diffusion defines $\tau_{nr}$ in equation (\ref{eq:tau_d late NR})].
Note that, the optical depth of the outer relativistic part (i.e., for successful jet models) has been ignored with this formulation. This is reasonable considering the steep density profile, and its limited effect (except at very early times).

\subsection{Photon diffusion condition}
\label{sec:photon diff}
\subsubsection{Critical differences compared to the SN scenario}
The observed photons have been released from a given radius moving with a velocity $\beta_d$ or $\Gamma_d =[2(1-\beta_d)]^{-1/2}$
via diffusion.
In order to find the observed time $t_{obs}$ of these photons
[to solve equations (\ref{eq:tobs R}) and (\ref{eq:tobs NR})], the corresponding $\tau(t_{obs})$ has to be obtained as a function of $\beta_d$ (or $\Gamma_d$).

The classical condition for photon diffusion (e.g., for SN emission etc.) is $\tau\sim c/v_d \sim 1/\beta_d$ (\citealt{1980ApJ...237..541A} and others).
However, the situation here is different.
First, photons start to diffuse early on, and hence the emitting region is very ``thin" ($\Delta r\ll r$, where $\Delta r$ is the size of the photon diffusing region; see \citealt{2015ApJ...802..119K}; also see \citealt{2015MNRAS.451.2656K}).
Second, the outer part of the cocoon ($\Gamma>\Gamma_t$) is mildly relativistic $\beta \sim 1$ ($\Gamma\gg 1$).
Third, the density profile of the non-relativistic cocoon is very steep $\rho_{c,nr} \propto \beta^{-8}$, and $\tau \propto \beta^{-7}$.
Hence, the optical depth is mostly dominated by the region very near to the diffusion radius with $\beta_d$ (see Figure \ref{fig:key.tau}), and the common one-zone approximation of mean density ($\rho\sim M/V\sim \rm{Const.}$) is not ideal.

\subsubsection{Random walk approximation}
\label{sec:random walk}
In the following photon scattering is approximated to a pure elastic scattering process [in reality, photons are constantly being absorbed and released by electrons (bound-bound transitions), and their energy (and frequency) is subject to change].

Early on, the medium is optically thick $\tau\gg 1$, so that photons are continually scattered, 
making their net forward motion significantly slower than $c$.
We consider the random walk approximation for these photons (\citealt{1979rpa..book.....R}; also see \citealt{2015ApJ...802..119K} for a similar discussion).

First, let's consider photon propagation in the comoving frame of the fluid (moving with $\beta_d$ in the laboratory frame), so that the fluid is static in this frame ($\beta_d'=0$).
Photons propagate across a certain length $\Delta r'$, over which the density is constant $\rho'\sim \text{Const}$.
The random walk approximation gives the net motion of photons forward as $\Delta r'\sim  {l_{mfp}}' \sqrt{N} $, where ${l_{mfp}}'=\Delta r'/\tau$ is the mean free path of the photon, and $N$ is the number of scatterings experienced by photons ($N=N'$ and $\tau=\tau'$).
$N$ is related to the optical depth as (\citealt{1979rpa..book.....R})
\begin{equation}
  N
    \begin{cases}
      \sim \tau^2 & \text{($\tau\gg 1$ : optically thick medium)} ,\\
      \sim \tau & \text{($\tau \ll 1$ : optically thin medium)} .
    \end{cases}       
    \label{eq:N cases}
\end{equation}
In the region $\Delta r'$, we consider $\tau \gg 1$, hence $N\sim \tau^2$.
As photons travel with the speed of light $c$ between two scatterings, $\Delta t_{}' =\frac{{l_{mfp}}' N}{c}$ is the time for photons to propagate through $\Delta r'$.
Then, one can find $\Delta t' =\frac{\Delta r'}{c}\tau$, and the effective velocity of photons across the fluid is $\frac{\Delta r'}{\Delta t'}\sim \frac{c}{\tau}$.
Photons are able to diffuse (across $\Delta r'$ during $\Delta t'$) if they are faster than the fluid velocity $c\Delta \beta'$ at $\Delta r'$ in the comoving frame.
Hence, the diffusion condition can be found (by equalling the two velocities) as
\begin{eqnarray}
\tau \sim \frac{1}{\Delta\beta'} \:,
\label{eq:tau_d classic}
\end{eqnarray}
where $\Delta \beta'$ is the relative velocity between the diffusion radius (at $\beta_d'$) and the outer edge of the medium.

\subsubsection{Relativistic cocoon}
\label{sec:diff R}
Let's consider the relativistic cocoon part, where the outer velocity is $\beta_{out}$, and the diffusion radius is at $\beta_d$ with $\beta_t\le\beta_d\le\beta_{out}$.
The relative velocity in the comoving frame $\Delta \beta'$ can be found using the velocity-addition formula as
\begin{eqnarray}
\Delta \beta ' = \frac{\beta_{out} -\beta_d}{1-\beta_{out}\beta_d} .
\label{eq:U+V}
\end{eqnarray}
In the relativistic limit, taking $\beta_{out}\sim 1$, the diffusion condition can be found using equation (\ref{eq:tau_d classic}) as
\begin{eqnarray}
\tau_r(t_{obs},\beta_d)\equiv \tau_r(t_{obs}) \sim 1 .
\label{eq:tau_d R}
\end{eqnarray}
In other words, the relative velocity across the causally connected region
is $\sim c$ in the comoving frame, i.e., $\Delta \beta'\sim 1$ in equation (\ref{eq:tau_d classic}).

Hence, in the relativistic limit the diffusion radius converges to the photospheric radius (see Section \ref{sec:Photospheric velocity}) [also see \citealt{1991ApJ...369..175A}; \citealt{2009ApJ...700L..47L}; \citealt{2011ApJ...732...26M}].

\subsubsection{Non-Relativistic cocoon}
\label{sec:diff NR}
\begin{figure}
    \centering
    \includegraphics[width=0.999\linewidth]{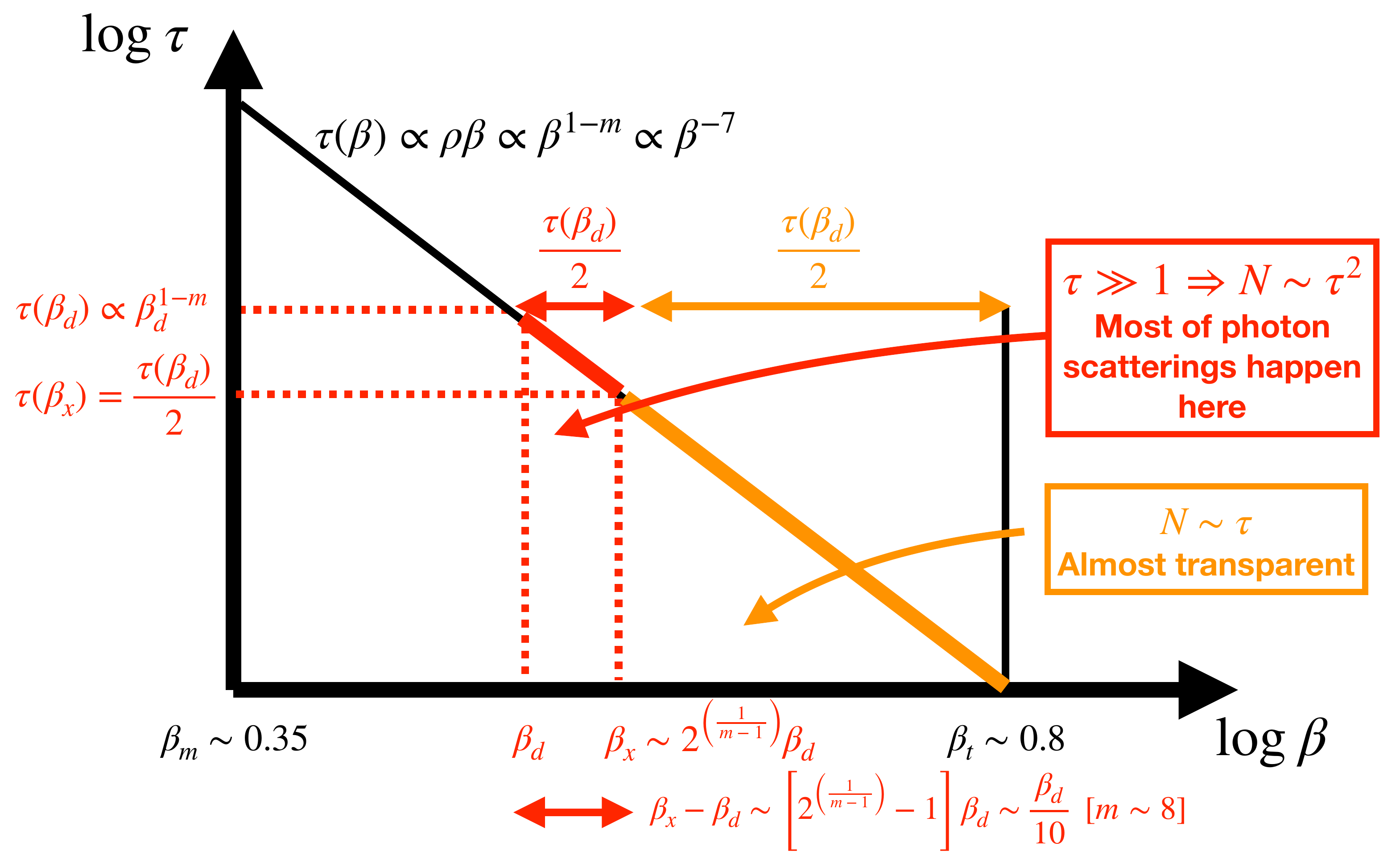} 
  \caption{Illustration of the optical depth as a function of the velocity $\beta$ at later times, from near the outer edge of the non-relativistic cocoon ($\lesssim \beta_t$) to the diffusion velocity ($\beta_d$).
  The approximation that most photon scattering events $N$ happen in the inner half of the optical depth [$\tau_{nr}(\beta_d)/2$, from $\beta_d$ to $\beta_x\sim 2^{\frac{1}{m-1}} \beta_d$] is shown (red) [see equation (\ref{eq:tau_d late NR})]. 
  As the optical depth drops rapidly ($m\approx 8 \gg 1$) through the outer half of the optical depth (from $\beta_x$ to $\beta_t$), it can be considered that the outer part is almost transparent for photons as $N\sim\tau$ (orange) [see Section \ref{sec:diff NR}].
  }
  \label{fig:key.tau} 
\end{figure}

For the non-relativistic cocoon ($\beta_m\le\beta_d\le\beta_t$), the density profile is quite steep $m\approx 8\gg 1$ (see Figure \ref{fig:den sim}).
As a result, the optical depth is dominated by the inner region, near $\beta_d\ll \beta_t$.
In the following, we introduce the velocity $\beta_x$, which is the velocity of the shell where the inner half of the total optical depth $\tau_{nr}(\beta_x) =\int_{\beta_d}^{\beta_x}d\tau_{nr}(\beta)$ equals the outer half $\tau_{nr}(\beta_d) - \tau_{nr}(\beta_x) =\int_{\beta_x}^{\beta_t}d\tau_{nr}(\beta)$ [see Figure \ref{fig:key.tau}]. 
Hence,
\begin{eqnarray}
\tau_{nr}(\beta_x) = \tau_{nr}(\beta_d)/2.
\label{eq:betax}
\end{eqnarray}
One can find that, with $\tau_{nr} \propto \rho\beta \propto \beta^{1-m}$, roughly $\beta_x \sim 2^{\frac{1}{m-1}}\beta_d \sim 1.1 \beta_d$ (for $m\approx  8$).
In other words, the optical depth from $\sim 1.1\beta_d$ to $\beta_d$ is the same as that from $\beta_x$ to $\beta_d$, despite the large difference in distance (see Figure \ref{fig:key.tau}).
As $\tau_{nr} \propto \beta^{1-m}$, one can find that $\tau_{nr}(\beta_d<\beta<\beta_x) \gg 1$ [opacity is highly concentrated in the inner part].
In contrast, throughout the majority (of the length scale) of the outer part $\tau_{nr}(\beta_x \lesssim \beta<\beta_t) \ll 1$.
Also, this definition of $\beta_x$ ensures that $\rho(\beta_d<\beta<\beta_x)\sim \text{Const.}$ (within a factor $\sim 2$), allowing us to use the random walk approximation throughout this region (as defined in Section \ref{sec:random walk}).
Hence, as shown in Figure \ref{fig:key.tau}, one can approximate that in the inner part $N\sim \tau^2\gg 1$, and in the outer part $N\sim \tau \ll 1$ [see equation (\ref{eq:N cases})]. 
Therefore, for a given photon, most scattering events happen in the inner region $\beta<\beta_x$. Consequently, once photons diffuse out of this inner region ($\beta > \beta_x$), they can be approximated as decoupled from matter, traveling with $\sim c$\footnote{This argument is true except at the very early times when $\beta_d\sim\beta_x\sim \beta_t$.}.

Hence, taking into account that most photon scattering takes place in the inner part ($\sim 1/2$) of optical depth (from $\beta_x$ to $\beta_d$),
equations (\ref{eq:tau_d classic}) and (\ref{eq:U+V}) give:

\begin{eqnarray}
\begin{split}
\tau_{nr}(\beta_x)=\int_{\beta_d}^{\beta_x}d\tau_{nr}(\beta)&\sim \frac{1-\beta_x\beta_d}{\beta_x-\beta_d}.
\end{split}
\label{eq:tau_d true NR}
\end{eqnarray}
Considering that, at later times (as $\beta_d \lesssim \beta_t$), $\tau_{nr}\propto \beta^{1-m}_d$, $\beta_x \sim 2^{\frac{1}{m-1}}\beta_d \sim 1.1 \beta_d$,
and $1-\beta_x\beta_d\sim \Gamma_d^{-2}$, 
and taking the non-relativistic limit ($\beta_d\ll 1$, and $\Gamma_d\sim 1$;
see Section \ref{sec:v profile}),
we find that
\begin{eqnarray}
\begin{split}
\tau_{nr}(\beta_d)&\sim \frac{2}{\Gamma_d^2(\beta_x-\beta_d)}
&\sim \frac{20}{\beta_d} ,
\end{split}
\label{eq:tau_d late NR}
\end{eqnarray}
for a steep density profile with an index $m\approx 8$ [and using the definition of $\beta_x$ in equation (\ref{eq:betax})].
This is significantly different from the classical $\tau \sim 1/\beta_d$ condition\footnote{The factor 20 in equation (\ref{eq:tau_d late NR}) is explained by: i) the consideration of the inner optically thick region $\beta_x-\beta_d$ with half of the optical depth, giving a factor of $2$; and ii) the term $[2^{\frac{1}{m-1}}-1]^{-1}\sim 10$ as a result of the steep density profile (for $m=8$).}.

This approach, using $\beta_x$, is important as it allows us to separate the fluid in two limits: optically thick region ($\tau\gtrsim 1$ for $\beta<\beta_x$), and optically thin region ($\tau\lesssim 1$ for $\beta>\beta_x$) [see Figure \ref{fig:key.tau}]. 
Therefore, this allows us to apply the random-walk approximation, only in the sufficiently optically thick limit (inner region), where most of the photos scatterings happen (as $\tau\gg1$ and $N\propto \tau^2$); and allows us to find the diffusion condition in such a strongly variable density (and optical depth) environment.
Hence, this is a more realistic approach than just taking $\tau \sim 1/\beta_d$.

However, it should be stressed that this criteria [$\tau_{nr}(\beta_x) = \tau_{nr}(\beta_d)/2$; as in equation (\ref{eq:betax})] to define $\beta_x$ is quite simplistic.
In reality, the situation is more complex, especially considering the large uncertainties in the opacity ($\kappa$).
For instance, $\kappa$ should also depend on temperature, and the level of ionization.  
As the diffusion shell moves inward, the outer regions (in particular $\beta>\beta_x$) cool down, and temperature is expected to drop sharply.
This is expected to drastically affect the opacity (see Figure 7 in \citealt{2020ApJ...901...29B}; also see \citealt{2020MNRAS.496.1369T}) reducing the optical depth in the cooler regions, and reducing $\tau_{nr}(\beta_x)$ in particular.
Here, our choice of $\tau_{nr}(\beta_x) = \tau_{nr}(\beta_d)/2$ was motivated by these considerations, and was chosen so that it guarantees that the inner optical depth (for $\beta_d<\beta<\beta_x$) is, at least, dominant over the outer optical depth.

Therefore, although this definition [and the expression in equation (\ref{eq:tau_d late NR})] should not be regarded as universal, considering the goal of our study (order of magnitude estimation of the cocoon emission), and the complexity of the problem, this is a reasonable simplification.
Future works should put this criteria to the test, through comparison with numerical radiative transfer methods.

Hence, it is possible to find the time when an observer detects photons released from the diffusion radius (tagged with $\Gamma_d$ or $\beta_d$), with equations (\ref{eq:tobs R}) and (\ref{eq:tau_d R}), and (\ref{eq:tobs NR}) and (\ref{eq:tau_d late NR})
for the relativistic and non-relativistic cocoon, respectively.

\subsection{The observed photospheric velocity $\Gamma_{ph}\beta_{ph}$}
\label{sec:Photospheric velocity}
The photospheric velocity $\beta_{ph}$ (or $\Gamma_{ph}$) is necessary for evaluating the observed temperature (see Section \ref{sec:Tobs}). 
The velocity $\beta_{ph}$ (and the radius $r_{ph}\approx c\beta_{ph}t$) is found using (\citealt{1991ApJ...369..175A}):
\begin{eqnarray}
\tau_{ph} = 1,
\label{eq:tau ph}
\end{eqnarray}
with equations (\ref{eq:tobs R}) and (\ref{eq:tobs NR}) for the relativistic and the non-relativistic cocoon, respectively.

\subsubsection{The overall time evolution of $\Gamma_{ph}\beta_{ph}$}
\label{sec:beta ph R+NR}
The time evolution of the observed photospheric four-velocity is summarized in the successful jet case and the failed jet case (respectively) as follows [using equations (\ref{eq:U4 R NR cases}), (\ref{eq:Gamma ph R simple}), and (\ref{eq:beta ph NR simple}); for more details see Appendix \ref{ap:photosphere}]
\begin{equation}
\begin{split}
&\Gamma_{ph}\beta_{ph}
\begin{cases}
\text{Successful jet case:}\\
\approx\Gamma_t\left\{\frac{t_{obs}}{t_{obs}(\Gamma_{t})}\right\}^{-\frac{1}{3}} &[ t_{obs}<t_{obs}(\Gamma_t)]\\
\text{(Rela. cocoon)} ,\\
\\
\approx \beta_m\left\{\frac{t_{obs}}{t_{obs}(\beta_m)}\right\}^{-\frac{2}{7}} &[ t_{obs}>t_{obs}(\Gamma_t)]\\
\text{(N. Rela. cocoon)},\\
\end{cases}
\\
\\
&\Gamma_{ph}\beta_{ph}
\begin{cases}
\text{Failed jet case:}\\
\approx \Gamma_t\beta_t &\left[t_{obs}<{\frac{t_{obs}(\beta_m)}{(\beta_t/\beta_m)^{\frac{7}{2}}} }\right] \\
\text{(N. Rela. cocoon)} ,\\
\\
\approx \beta_m\left\{\frac{t_{obs}}{t_{obs}(\beta_m)}\right\}^{-\frac{2}{7}} &\left[t_{obs}>{\frac{t_{obs}(\beta_m)}{(\beta_t/\beta_m)^{\frac{7}{2}}} }\right] \\
\text{(N. Rela. cocoon)}.\\
\end{cases} 
\end{split}
    \label{eq:U4 ph R+NR cases}
\end{equation}
The timescales $t_{obs}(\Gamma_{t})$ and $t_{obs}(\beta_{m})$ can be found using equations (\ref{eq:tobs R}) and (\ref{eq:tobs NR}), for $\Gamma=\Gamma_t$ and $\beta=\beta_m$, respectively, with equation (\ref{eq:tau ph}).

In Section \ref{sec:U4 ph NWF}, these solutions are presented and discussed (see Figure \ref{fig:NWF L v T}).

\subsection{Bolometric isotropic luminosities}
\label{sec:Lbl}
The total isotropic-equivalent bolometric luminosity $L_{bl}^{}$ of the cocoon is the sum of the two contributions,
jet-shock heating and r-process heating (see Sections \ref{sec:Eint jet} and \ref{sec:Edep rp}, respectively), as
\begin{eqnarray}
L_{bl}^{}(t_{obs})\approx L_{bl}^{j}(t_{obs}) + L_{bl}^{rp}(t_{obs}) .
\label{eq:Lbl tot}
\end{eqnarray}
In the following we avoided the one-zone approximation $L\sim E_i/t$ (\citealt{1980ApJ...237..541A}) because this could give inaccurate results, 
in particular in the case of the relativistic cocoon [see equation (\ref{eq:Lbl R j S})].

The detailed derivative procedure is presented in Appendix \ref{ap:Lbl}, and bellow are the final results.

\subsubsection{Luminosity in a successful jet case}
\label{sec:Lbl S}
There are two phases for a successful jet case,
emission from the relativistic cocoon (early times), and emission from the non-relativistic cocoon (late times).

\paragraph{Luminosity from the relativistic cocoon [$\Gamma_d>\Gamma_t$]:}
\label{sec:Lbl R S}
Using equation (\ref{eq:U4 R NR cases}), inserting equation (\ref{eq:Eijcum R+NR}) in equation (\ref{eq:Lbl j}), and using equations (\ref{eq:dt dtobs cases}) and (\ref{eq:partial Gamma_d beta_d cases}), the isotropic luminosity from the jet-shock heated cocoon (relativistic part) can be found as
\begin{equation}
L_{bl,r}^j(t_{obs})\approx \frac{E_{c,i,r}^j(>\Gamma_d,t)\left[\frac{1/3}{\ln\left({\Gamma_{out}}/{\Gamma_d}\right)}\right]}{t_{obs}} \times\frac{4\pi}{\Omega_{EM}} .   
\label{eq:Lbl R j S} 
\end{equation}
The total luminosity from r-process heating is the sum of two contributions, cooling emission from the stored internal energy and the direct emission from the diffusively thin part.
For the relativistic cocoon, 
the first term can be found using equations (\ref{eq:Edep cum R+NR}), (\ref{eq:Erp cum R+NR}), (\ref{eq:dt dtobs cases}), (\ref{eq:partial Gamma_d beta_d cases}), and (\ref{eq:Lbl rp in}), and the second term using (\ref{eq:Erp cum R+NR}), (\ref{eq:dt dtobs cases}), and (\ref{eq:Lbl rp in}).
Then the summed isotropic luminosity can be found as
\begin{equation}
L_{bl,r}^{rp}(t_{obs})\approx \frac{{E}_{c,i,r}^{rp}(>\Gamma_d,t_{})\left[\frac{2(2-k)}{3}\right]}{t_{obs}}\times\frac{4\pi}{\Omega_{EM}}  .   
\label{eq:Lbl R rp S} 
\end{equation}
The total luminosity from the relativistic cocoon $L_{bl,r}$ can be found by plugging equations (\ref{eq:Lbl R j S}) and (\ref{eq:Lbl R rp S}) in equation (\ref{eq:Lbl tot}).
It is worth mentioning that the above r-process powered luminously is quite dim in this phase (relative to $L_{bl,r}^j$).

\paragraph{Luminosity from the non-relativistic cocoon [$\beta_d\gtrsim\beta_m$]:}
\label{sec:Lbl NR S}
Once the relativistic cocoon becomes transparent [$t_{obs}(\Gamma_d)=t_{obs}(\Gamma_t)$; see equations (\ref{eq:tobs R}) and (\ref{eq:tau_d R})], this phase is set to start.
Using equations (\ref{eq:Eijcum R+NR}), (\ref{eq:dt dtobs cases}), (\ref{eq:partial Gamma_d beta_d cases}), and (\ref{eq:Lbl j}),
the isotropic luminosity of the jet-shock heated non-relativistic cocoon can be found as
\begin{equation}
L_{bl,nr}^j(t_{obs})\approx\frac{E_{c,i,nr}^j(>\beta_d,t)}{t_{obs}}\left[\frac{\frac{4\pi}{\Omega_{EM}}}{\ln\left({\beta_t}/{\beta_d}\right) \left(\frac{m-2}{2}+\frac{\beta_d}{1-\beta_d}\right)}\right] .
\label{eq:Lbl NR j S} 
\end{equation}
Similarly to Section \ref{sec:Lbl R S},
the total luminosity from r-process heating is the sum of two contributions.
For the non-relativistic cocoon, the first term can be found using equations (\ref{eq:Edep cum R+NR}), (\ref{eq:Erp cum R+NR}), (\ref{eq:Lbl rp in}), (\ref{eq:dt dtobs cases}), and (\ref{eq:partial Gamma_d beta_d cases}),
and the second term can be found using (\ref{eq:Erp cum R+NR}), (\ref{eq:dt dtobs cases}), and (\ref{eq:Lbl rp in}).
The summed isotropic luminosity can be found as (at late times, for $\beta_d \gtrsim \beta_m$)\footnote{This luminosity is very dim at early times, especially compared to $L_{bl,nr}^{j}$; hence, it is more relevant at late times.}
\begin{equation}
\begin{split}
L_{bl,nr}^{rp}(t_{obs})\approx  \frac{{E}^{rp}_{c,i,nr}(>\beta,t)}{t_{obs}}\frac{4\pi}{\Omega_{EM}}
    \times\left[\frac{4-k}{1+\frac{\beta_d}{1-\beta_d}\left(\frac{2}{m-2}\right)^{}}\right].
\end{split}
\label{eq:Lbl NR rp S} 
\end{equation}
The total isotropic luminosity from the non-relativistic cocoon $L_{bl,nr}$ is given by plugging equations (\ref{eq:Lbl NR j S}) and (\ref{eq:Lbl NR rp S}) in equation (\ref{eq:Lbl tot}).

\subsubsection{Luminosity in a failed jet case}
\label{sec:Lbl F}
In the failed jet case, the only relevant component is the non-relativistic cocoon, and there are two phases to consider.
\paragraph{Luminosity from the non-relativistic cocoon [$\beta_t>\beta_d>\beta_m$]}
\label{sec:Lbl NR F}
First, at early times $\beta_d\sim \beta_t\sim \text{Const.}$, the isotropic luminosity of the jet-shock heated cocoon can be found using equations (\ref{eq:Eijcum R+NR}), (\ref{eq:Lbl j}), (\ref{eq:dt dtobs cases}), and (\ref{eq:partial Gamma_d beta_d cases}).
Second, at later times, for $\beta_d \gtrsim \beta_m$
and similarly to Section \ref{sec:Lbl NR S} [again, using equations (\ref{eq:Eijcum R+NR}), (\ref{eq:dt dtobs cases}), (\ref{eq:partial Gamma_d beta_d cases}), and (\ref{eq:Lbl j})], the isotropic luminosity of the jet-shock heated cocoon can also be found.
This second phase is set to start once the approximation $\beta_d\sim \beta_t\sim \text{Const.}$ is no longer valid. Here, we artificially set it to start at $\beta_d\sim 0.8\beta_t$. This approximation gives a broken power-law function for the luminosity of the cocoon, which has been confirmed to be reasonably consistent with numerical solutions.
These are summarized as
\begin{equation}
\begin{split}
L_{bl,nr}^j(t_{obs})\approx
\begin{cases}
\text{Failed jet case:}\\
\approx \frac{E_{c,i,nr}^j(>\beta_d,t)}{t_{obs}}\frac{4\pi}{\Omega_{EM}} &\\
\left[\beta_d\lesssim \beta_t \right]\\
\text{(N. Rela. cocoon)},\\
\\
\approx \frac{E_{c,i,nr}^j(>\beta_d,t)}{t_{obs}}\left[\frac{\frac{4\pi}{\Omega_{EM}}}{\ln\left({\beta_t}/{\beta_d}\right) \left(\frac{m-2}{2}+\frac{\beta_d}{1-\beta_d}\right)}\right] &\\
[\text{$\beta_d\gtrsim \beta_m$}]\\
\text{(N. Rela.  cocoon)} .\\
\end{cases} 
\end{split}
\label{eq:Lbl NR j F} 
\end{equation}
The isotropic luminosity from r-process heating is the same as in equation (\ref{eq:Lbl NR rp S}).
The total isotropic luminosity $L_{bl,nr}$ is given by plugging equations (\ref{eq:Lbl NR j F}) and (\ref{eq:Lbl NR rp S}) in equation (\ref{eq:Lbl tot}).

\subsubsection{Summary of the temporal evolution of $L_{bl}^j$}
\label{sec:Lbl^j R+NR}
Summarizing Sections \ref{sec:Lbl S} and \ref{sec:Lbl F} results, 
the time evolution of the bolometric isotropic luminosity from the jet-shock heated cocoon (which dominates over r-process at early times) can be found as [using equations (\ref{eq:Lbl R j S}) and (\ref{eq:Lbl NR j S}); also see equation (\ref{eq:theta_EM})]
\begin{equation}
\begin{split}
&L_{bl}^j
\begin{cases}
\text{Successful jet:}\\
\propto t_{obs}^{-4/3},  & [\Gamma_d>1/\theta_c^{es}] \\
\text{(Rela.  cocoon)},\\
\propto t_{obs}^{-2}  & [\Gamma_d<1/\theta_c^{es}]   ,\\
\text{(Rela.  cocoon)},\\
\propto t_{obs}^{-2}   & [\beta_d\gtrsim \beta_m]   ,\\
\text{(N. Rela.  cocoon)},
\end{cases}
\\
\\
&L_{bl}^j
\begin{cases}
\text{Failed jet:}\\
\propto t_{obs}^{-1}   & [\beta_d\lesssim \beta_t]   ,\\
\text{(N. Rela.  cocoon)},\\
\propto t_{obs}^{-2}   & [\beta_d\gtrsim \beta_m]   ,\\
\text{(N. Rela.  cocoon)}.
\end{cases}
\end{split}
\label{eq:Lbl^j R+NR}
\end{equation}
Note that $l=0$ and $m=8$ are used here.

\subsection{Temperature}
\label{sec:Tobs}
The isotropic equivalent luminosity in the relativistic limit can roughly be written as $\sim \Omega r^2c\epsilon' \Gamma^2\times (\frac{4\pi}{\Omega_{EM}})$ [i.e., $\Omega r^2c\epsilon' \Gamma^2$ is the true power and $\frac{4\pi}{\Omega_{EM}}$ is the beaming factor;
see equations (\ref{eq:Omega}) and (\ref{eq:Omega_EM})].
One can find that the $\Gamma^2$ term arises from the combination of Lorentz contraction (volume) and the blueshift (energy).
Note that, this above estimation should include an averaging geometrical factor, but for simplicity, we take the non-relativistic limit (resulting in a $\sim 1/4$ factor).
Hence, the observed temperature ($T_{obs} = \Gamma T'$) can be found as
\begin{equation}
T_{obs}(t_{obs})  \approx 
      \left\{\frac{L_{bl}^{}(t_{obs})}{4\pi\sigma [r_{ph}(t_{obs})/\Gamma_{ph}(t_{obs})]^2 }\left(\frac{\Omega_{EM}}{\Omega}\right)\right\}^{\frac{1}{4}},
    \label{eq:Teff}
\end{equation}
where $\sigma$ is the Stefan-Boltzmann constant, $L_{bl}(t_{obs})$ refers to the isotropic equivalent luminosity,
$t_{obs}(\beta_{ph})=t_{obs}(\beta_{d})=t_{obs}$\footnote{\label{foot:tobs ph 2 variables Teff} For a given $t_{obs}$,
$\beta_d=\beta_{ph}$ for the relativistic cocoon, while $\beta_{d}\neq \beta_{ph}$ 
(but almost the same)
for the non-relativistic cocoon,
corresponding to different optical depth
($\tau_{nr} \sim 20/\beta_{d}$ or $\tau_{nr}\sim 1$, respectively)
in equation (\ref{eq:tobs NR}).
Therefore, while luminosity depends on $\beta_{d}$,
photospheric quantities depend on both $\beta_{ph}$ and $\beta_{d}$.},
and the photospheric radius can be written as [see equation (\ref{eq:t obs lab})]\footnote{\label{foot:rph}
Remember that in general we are keeping the term $1-\beta$ without approximating $1-\beta \sim 1$ (see Section \ref{sec:v profile}).
Therefore, we use $r_{ph}(t_{obs}) \approx c\beta_{ph}t \neq c\beta_{ph}t_{obs}$.}
\begin{equation}
r_{ph}(t_{obs}) \approx c t_{obs}\frac{\beta_{ph}(t_{obs}) }{1-\beta_{ph}(t_{obs})}.
\label{eq:rph}
\end{equation}

\subsubsection{The temporal evolution of the observed temperature from the jet-shock heated cocoon}
\label{sec:Tobs^j R+NR}
For the jet-shock heated cocoon (without r-process heating),
the observed temperature evolves as [using equations (\ref{eq:U4 ph R+NR cases}), (\ref{eq:Lbl^j R+NR}), and (\ref{eq:Teff})]
\begin{equation}
\begin{split}
&T_{obs}(t_{obs})  
\begin{cases}
\text{Successful jet:}\\
\propto t_{obs}^{-2/3} & [\Gamma_d>1/\theta_c^{es}]\\ 
\text{(Rela.  cocoon)},\\
\propto t_{obs}^{-2/3} & [\Gamma_d<1/\theta_c^{es}]\\ 
\text{(Rela.  cocoon)},\\
\propto t_{obs}^{-6/7} &[\beta_d\gtrsim \beta_m]\\ 
\text{(N. Rela.  cocoon)},\\
\end{cases}
\\
\\
&T_{obs}(t_{obs})  
\begin{cases}
\text{Failed jet:}\\
\propto t_{obs}^{-3/4} &[\beta_d\lesssim \beta_t]\\ 
\text{(N. Rela.  cocoon)},\\
\propto t_{obs}^{-6/7} &[\beta_d\gtrsim \beta_m]\\ 
\text{(N. Rela.  cocoon)}.\\
\end{cases}
\end{split}
\label{eq:Tobs R+NR}
\end{equation}
It should be noted that, in reality, at late times, r-process heating makes the time dependence much shallower than $\propto t_{obs}^{-6/7}$ (see Figure \ref{fig:NWF L v T}).

\subsubsection{Inferring the photospheric velocity from observations}
\label{sec:Tobs interpretations}

Combining equations (\ref{eq:Teff}) for the observed temperature $T_{obs}$ with (\ref{eq:rph}) for $r_{ph}$, one can find that by measuring $L_{bl}$ and $T_{obs}$ as a function of $t_{obs}$, the following photospheric quantity is measurable,
\begin{eqnarray}
\begin{split}
\frac{\beta_{ph}}{(1-\beta_{ph})\Gamma_{ph}}=&\left[\frac{L_{bl}}{4\pi \sigma  T_{obs}^4 (c t_{obs})^2}\right]^{1/2}
\left(\frac{\Omega}{\Omega_{EM}}\right)^{-1/2},
\end{split}
    \label{eq:ph mesure}
\end{eqnarray}
where $\frac{\Omega}{\Omega_{EM}}=1$ at the relativistic limit (for $\Gamma_{ph}>1/\theta_c^{es}$) and $\frac{\Omega}{\Omega_{EM}}\sim 1-\cos{\theta_c^{es}}$ at the non-relativistic limit (for $\Gamma_{ph}\sim 1$) [see equations (\ref{eq:Omega}) and (\ref{eq:Omega_EM})].
This is a generalized description of the (measurable) photospheric velocity, including the relativistic limit, so that:
\begin{itemize}
    \item In the relativistic limit ($\Gamma_{ph}\gg 1$, i.e., relativistic cocoon here), $\left[\frac{L_{bl}}{4\pi \sigma  T_{obs}^4 (c t_{obs})^2}\right]^{1/2}$ measures a combination of the photospheric velocity and Lorentz factor, more precisely, the photospheric velocity times the Doppler factor $\delta_{ph} =1/[(1-\beta_{ph})\Gamma_{ph}]$.
    Note that the effect of the viewing angle, which could have an appreciable effect on the Doppler factor, has not been considered here for simplicity (i.e., $\cos \Theta_{}\sim 1$; see Section \ref{sec:Optical depth}).
    \item In the non-relativistic case where $\Gamma_{ph}\sim 1$ but $\beta_{ph}$ is not negligible compared to unity (roughly in the range $\sim 0.4-0.8$),
    $\left[\frac{L_{bl}}{4\pi \sigma  T_{obs}^4 (c t_{obs})^2}\right]^{1/2}$
    gives a measurement of $\beta_{ph}/(1-\beta_{ph})$, times $\left({\frac{\Omega}{\Omega_{EM}}}\right)^{1/2}$ which is roughly constant ($\sim \frac{1}{3}-\frac{1}{2}$).
    \item In the non-relativistic limit where $\beta_{ph}\ll 1$ (e.g., KN or SN case), $\left[\frac{L_{bl}}{4\pi \sigma  T_{obs}^4 (c t_{obs})^2}\right]^{1/2}$ simply gives the photospheric velocity $\beta_{ph}$, times a geometrical constant $1-\cos{\theta_c^{es}}$ ($\sim 1$ in the KN or SN case).
\end{itemize}

Therefore, by observing $L_{bl}$ (isotropic equivalent luminosity) and $T_{obs}$ (observed temperature), and using the combination $\left[\frac{L_{bl}}{4\pi \sigma  T_{obs}^4 (c t_{obs})^2}\right]^{1/2}$,
one can deduce the relativistic nature of the emitting region; velocity and Lorentz factor of the cocoon.

\subsection{Magnitudes}
\label{sec:mag}
AB magnitude can be found using (\citealt{1974ApJS...27...21O}):
\begin{eqnarray}
m_{A B}=-2.5 \log _{10}\left(\frac{F_{\nu}}{\mathrm{erg\:s}^{-1} \mathrm{cm}^{-2} \mathrm{\:Hz}^{-1}}\right)-48.6 .
\label{eq:mag ap}
\end{eqnarray}
For a blackbody radiation, $F_\nu$ is the observed blackbody flux density at a frequency $\nu$, which is given by
\begin{eqnarray}
F_\nu (T_{obs},\nu)= \frac{\pi B_\nu}{\sigma T_{obs}^4}\frac{L_{bl}}{4\pi D^2} ,
\label{eq:Fnu}
\end{eqnarray}
where $B_{\nu}(T_{obs})=\frac{2 h\nu^3}{c^2}\frac{1}{e^\frac{h\nu}{k_b T_{obs}}-1}$,
$T_{obs}$
is the observed temperature, $L_{bl}$ is the bolometric isotropic luminosity,
and $D$ is the luminosity distance, taken as $D=40$ Mpc here for GW170817 (\citealt{2017ApJ...848L..13A}).

The absolute magnitude is 
\begin{eqnarray}
M_{{AB}}=m_{{AB}}-5 \log _{10}\left(\frac{D}{10 \mathrm{pc}}\right) \approx m_{{AB}}-33,
\label{eq:mag ab}
\end{eqnarray}
where $D = 40$ Mpc is used in the last equality.

\section{Results \& Discussion}
\label{sec:5}
In the following the analytic model presented in Section \ref{sec:4} is applied to different jet models:
narrow, wide and failed in Tables \ref{tab:1}.

\subsection{Bolometric light curves}
\label{sec:Lbl results}

\subsubsection{jet-shock heating vs. r-process heating}
\label{sec:Lbl wide}
Figure \ref{fig:Lbl} shows the different contributions to the bolometric luminosity of the cocoon, for the wide jet model as an example.
It is worth recalling that there are two (escaped) cocoon components, relativistic and non-relativistic,
and that each component is the sum of contributions from jet-shock heating and r-process heating [see equation (\ref{eq:Lbl tot})].

As explained in Section \ref{sec:4}, early emission originates form the outer relativistic cocoon component.
The timescale of this early emission (here $\sim 200$ s for our wide jet parameters) depends on the mass of the relativistic cocoon (here, for our successful jets, this is about $\sim 2 \%$ of the mass of the escaped cocoon).
As in Figure \ref{fig:Lbl}, 
we found that this early emission is overwhelmingly dominated by the jet-shock heating component (dashed line).
At $t_{obs}(\Gamma_d\sim \Gamma_t)\sim 200$ s, there is a transition (in terms of the emitting region) from the relativistic cocoon to the non-relativistic cocoon.

At late times (here $> 200$ s for our wide jet model), as the relativistic cocoon becomes transparent, the non-relativistic cocoon starts to contribute to the luminosity.
Until about $\sim 500$ s, the jet-shock heating dominates in terms of luminosity. 
The r-process heating takes over
at the very late times
due to the steep density profile ($m\approx 8$)
with most of the cocoon mass located at the slower inner shells.
This late times dominance of r-process heating (over jet-shock heating) has already been pointed out (\citealt{2018MNRAS.473..576G}; \citealt{2021MNRAS.502..865K} in the context of the blue KN).  
However, this r-process contribution could be dimmer than the KN emission.
For clear detection of the cocoon,
the early time emission, mostly dominated by jet-shock heating $L_{bl}^j$ [see equation (\ref{eq:Lbl^j R+NR})], is the most relevant part (the first $\sim 500$ s after the merger for our wide jet model; see Section \ref{sec:Lbl NWF} for more details).

It should be noted that the timescale of this time window is highly dependent on the parameters of the cocoon luminosity [see equation (\ref{eq:Lbl^j S estiamte})] and the parameters of the early KN (here, roughly assumed as similar to GW170817, which is not always the case; see Appendix \ref{ap:KN}). 
Therefore, for brighter cocoons or dimmer KNe, this time window can be much longer (as in \citealt{2018MNRAS.473..576G}; and \citealt{2020MNRAS.498.3320G}), and vice versa.
In particular, the opacity (in the early cocoon, and in the KN) highly dependent on the chemical composition, and on temperature (i.e., ionization level), is a major source of incertitude (see \citealt{2020MNRAS.496.1369T}).
Early observations of NS mergers could give an idea about such parameters (e.g., opacity), by measuring this timescale (during which the cocoon, as a KN-excess, lasts).

As a remark, due to our approximation in equation (\ref{eq:U4 R NR cases}) [useful for treating the whole cocoon], our luminosity (also $T_{obs}$) is slightly discontinuous at this very transition time. This is an artifact, and it is dealt with by extrapolating $L_{bl,nr}(t_{obs})$ [same procedure is also followed for $T_{obs}$] at the vicinity of $t_{obs}(\Gamma_t)$. This is a reasonable treatment in consistency with numerical results (see Figures \ref{fig:NWF L v T} and \ref{fig:NWF mag}).

For our failed jet case, the relativistic cocoon part is absent.
Therefore, the late time emission is identical, but the early time emission comes from the non-relativistic cocoon.
Similarly to the successful jet case, jet-shock heating dominates at early times, although the time dependence of the luminosity is different [$\propto t_{obs}^{-1}$; see equation (\ref{eq:Lbl^j R+NR})].

\subsubsection{Parameter dependence for the cocoon emission}
\label{sec:Lbl para}
Let's make an order-of-magnitude calculation and clarify the parameter dependence for the cocoon luminosity from jet-shock heating (analytically calculated in Section \ref{sec:Lbl}).

First, the jet energy deposited into the cocoon, from the jet launch time $t_0$ up to the breakout time $t_b$, can roughly be 
estimated (taking $t_m=0$ for convenience) as $\sim 2L_j(t_b-t_0)(1-\beta_m\frac{t_b}{t_b-t_0})\sim L_j(t_b-t_0)$ [$2L_j$ for two jets in both hemispheres, and the term $\beta_m\frac{t_b}{t_b-t_0}\sim 0.4-0.6$ (i.e., the average jet head velocity until the breakout) has been roughly approximated to $\sim 0.5$].
As explained in \cite{2021MNRAS.500..627H}, 
at the breakout time of \textit{s}GRB jets,
the energy is approximately equipartitioned with the half into thermal energy and the other half into kinetic energy (irrelevant to the cocoon emission).
Hence, at the breakout time, the total internal energy of the cocoon is roughly $\sim L_j(t_b-t_0)/2$.

Second, as explained in \cite{2023MNRAS.520.1111H}, only a fraction of the cocoon mass and internal energy escapes the ejecta, 
where $M_{c}^{es}/M_{c}^{}\sim 0.005 - 0.05$ 
and $E_{c,i}^{es}/E_{c,i}^{}\sim 0.1 -0.6$; which is highly dependent on mixing in the cocoon, represented analytically by the mixing parameter $f_{mix}$ (with $f_{mix}\sim 2/3$; see \citealt{2023MNRAS.520.1111H}).
Also, as explained in \cite{2023MNRAS.520.1111H}, after the jet breakout, the cocoon and the escaped cocoon continue to grow by a factor $f_g\sim 2$ until the free-expansion phase is reached ($t_1$).
Hence, for $t>t_1$, the escaped cocoon's internal energy is $\sim L_j(t_b-t_0)\frac{E_{c,i}^{es}}{E_{c,i}^{}}\frac{t_b}{t}$ (after taking into account the adiabatic cooling), roughly half of which is contained in the relativistic cocoon $E_{c,i,r}^j(t)$ [for the case of a successful jet], and the other half is contained in the non-relativistic cocoon $E_{c,i,nr}^j(t)$ [see Section \ref{sec:Eint jet}].
Note that, $t=t_{obs}/(1-\beta_d)$ in equation (\ref{eq:t obs lab}) should be used here to substitute $t$ with $t_{obs}$.

The order of the luminosity can be calculated using the above internal energy in the relativistic cocoon $E_{c,i,r}^{j}(t)$ and dividing it by the observed time. 
However, as shown in equation (\ref{eq:Lbl R j S}), this is quite inaccurate.
Instead, we recommend using equation (\ref{eq:Lbl R j S}), which [with equation (\ref{eq:Eijcum R+NR})] gives $L_{bl}^j \sim \frac{E_{c,i,r}^j(t)}{t_{obs}} [3\ln(\Gamma_{out}/\Gamma_t)(1-\cos\theta_c^{es})]^{-1}$ in the first phase [$\Gamma_d>1/\theta_c^{es}$ and $\Omega_{EM}=\Omega$; see equation (\ref{eq:Omega_EM})].
Consequently, using equation (\ref{eq:Lbl^j R+NR}) [with equation (\ref{eq:t obs lab})], the isotropic luminosity from jet-shock heating in the different phases [calibrating to the parameters of the wide (successful) jet model; see Table \ref{tab:1}] can be estimated reasonably well as
\begin{equation}
\begin{split}
&L_{bl}^j \sim 3.3\times 10^{43} \text{ erg s$^{-1}$}\\ 
&\left(\frac{L_{iso,0}}{5\times10^{50}\text{ erg s$^{-1}$}}\right)
\left(\frac{t_b-t_0}{0.46 \text{ s}}\right)
\left(\frac{t_b}{0.62 \text{ s}}\right)
\left(\frac{E_{c,i,r}^{es}/E_{c,i}^{es}}{0.52}\right)\\
&\left(\frac{E_{c,i}^{es}/E_{c,i}}{0.63}\right)
\left(\frac{\theta_0}{18^{\circ}}\right)^{2} 
\left(\frac{\kappa}{\text{$1$ cm$^{2}$ g$^{-1}$}}\right)^{\frac{p-2}{2}}
{\left(\frac{t_{obs}}{75 \text{ s}}\right)^{-p}}, \\
\end{split}
\label{eq:Lbl^j S estiamte}
\end{equation}
where the the time index $p$ has been introduced to reproduce the temporal properties of the luminosity [$p=4/3$ for $t_{obs}<t_{obs}(\Gamma_d=1/\theta_c^{es})$, and $p=2$ for $t_{obs}>t_{obs}(\Gamma_d=1/\theta_c^{es})$; see equation (\ref{eq:Lbl^j R+NR})];
and from equation (\ref{eq:tobs R}) one can find the timescale $t_{obs}(\Gamma_d=1/\theta_c^{es})\sim \left[\frac{\kappa M_{c,r}^{es}(\theta_c^{es})^6}{12c^2\Omega\Gamma_t^{-2}}\right]^{1/2}$ ($\sim 75$ s here for the wide jet case; and $\sim 6$ s for the narrow jet case) [using $\kappa\sim 1$ cm$^2$ g$^{-1}$; see \citealt{2023arXiv230405810B}].
Note that the $t_{obs}(\Gamma_d=1/\theta_c^{es})$ dependency on quantities other than $\kappa$ [e.g., $M_{c,r}^{es}$ (i.e., $M_{ej}$), and $\theta_c^{es}$] has been abbreviated in this simplistic formulation [refer to equations (\ref{eq:Lbl R j S}) and (\ref{eq:Lbl NR j S}) for a more accurate formulation].
Also, note that the isotropic jet luminosity is
$L_{iso,0}\approx (4/\theta_0^2) L_j$ (see Table \ref{tab:1}),
and we used $3\ln(\Gamma_{out}/\Gamma_t)(1-\cos\theta_c^{es}) \sim 0.74$
in the above equation.

For the failed jet case the situation is similar,
although the relativistic cocoon is absent,
and the non-relativistic cocoon powers the cocoon emission from early times.
Also, the total jet energy injected into the cocoon is, more precisely, $2L_j(t_e-t_0)$ [where $t_e-t_0=2$ s here; see Table \ref{tab:1}].
The same logic can be applied to find $E_{c,i,nr}^j(t)$, 
and at early times as $\beta_{d}\sim \beta_t$, the parameter dependency of the luminosity can be found using equations (\ref{eq:Eijcum R+NR}) and (\ref{eq:Lbl NR j F}).
This early phase, with $L_{bl,nr}^j\propto t_{obs}^{-1}$, 
can roughly be approximated to continue until $\beta_d \sim 0.8\beta_t$, i.e., until $t_{obs}(\beta_d=0.8\beta_t)$ [$\sim 134$ s for the failed jet model considered here; using equations (\ref{eq:tobs NR}) and (\ref{eq:tau_d late NR})],
and the luminosity at this time can be found using equation (\ref{eq:Lbl NR j F}) [as $L_{bl,nr}^j(\beta_d=0.8\beta_t)\sim 4.4\times 10^{43}$ erg$^{-1}$ s; for the failed jet model considered here].
Hence, for our failed jet model (see Table \ref{tab:1}),
the result is as follows:
\begin{equation}
\begin{split}
&L_{bl}^j\equiv L_{bl,nr}^j \sim 4.4\times 10^{43} \text{ erg s$^{-1}$}\\ 
&\left(\frac{L_{iso,0}}{1\times10^{50}\text{ erg s$^{-1}$}}\right)
\left(\frac{t_e-t_0}{2 \text{ s}}\right)
\left(\frac{t_b}{3.13 \text{ s}}\right)
\left(\frac{E_{c,i,nr}^{es}/E_{c,i}^{es}}{1}\right)\\
&\left(\frac{E_{c,i}^{es}/E_{c,i}}{0.11}\right)
\left(\frac{\theta_0}{18^{\circ}}\right)^{2} 
\left(\frac{\kappa}{\text{$1$ cm$^{2}$ g$^{-1}$}}\right)^{\frac{p-2}{2}}
{\left(\frac{t_{obs}}{134 \text{ s}}\right)^{-p}}, \\
\end{split}
\label{eq:Lbl^j F estiamte}
\end{equation}
where, roughly, for $t_{obs}< t_{obs}(\beta_d=0.8\beta_t)$ we have $p=1$, 
and for $t_{obs}> t_{obs}(\beta_d=0.8\beta_t)$ we have $p=2$ [with $t_{obs}(\beta_d=0.8\beta_t)\sim 134$ s here; using equations (\ref{eq:tobs NR}) and (\ref{eq:tau_d late NR})].

The cocoon luminosity is proportional to the square of the jet opening angle $\theta_0$ and to the square of the breakout time [$\propto (t_b-t_0)t_b$ or $\propto (t_e-t_0)t_b$ (failed);
combined with the delays from the merger time $t_m$ and the jet launch time $t_0$].
Hence, strong jets with a large opening angle and/or late breakout time (e.g., large ejecta mass, fast tail, and late jet launch) produce significantly bright cocoon emissions.

It should be stressed that our estimates are very conservative.
First,
we did not take into account the fast tail component of the ejecta,
which if abundant, 
would boost the escaped cocoon's internal energy at later times (after the jet breakout time from the ejecta $t_b$) through interactions of the jet and escaped cocoon with the fast tail. 
Considering the time dependency of the adiabatic cooling process [see equation (\ref{eq:Adiabatic 1/t})], such late time heating processes can be impactful at increasing the cocoon luminosity.
Second, the delay time between the merger and the jet launch ($t_0-t_m$),   taken as $0.16$ s here, could be much longer (considering the $\sim1.7$ s delay in GW170817; see \citealt{2017ApJ...848L..13A}, and Figure 9 in \citealt{2020MNRAS.491.3192H}), which would too significantly increase the cocoon luminosity.
Finally, as we used 2D jet simulations, and considering 3D hydrodynamical jet's instabilities (\citealt{2019MNRAS.490.4271M}; although this is much less severe for \textit{s}GRBs' jets, see \citealt{2021MNRAS.500.3511G}), the energy deposited into the cocoon by the jet can be a bit higher than considered here.

\subsubsection{Luminosity for different jet models}
\label{sec:Lbl NWF}
In Figure \ref{fig:NWF L v T} (top panel), the total isotropic luminosity [using equation (\ref{eq:Lbl tot}); see Appendices \ref{ap:Lbl j} and \ref{ap:Lbl rp}] is presented for the three jet models [narrow (in blue), wide (in red), and failed (in green);
see Table \ref{tab:1}].
In addition, the expected luminosity of the KN  (dotted; using \citealt{2015ApJ...802..119K}, see Appendix \ref{ap:KN} for the used parameters of the ejecta) as well as observations recorded for GW170817 (grey circles; \citealt{2017Sci...358.1559K}; \citealt{2017Sci...358.1570D}) are shown.

First, the cocoon emission in the narrow jet model is much dimmer (by more than one order of magnitude than the wide jet model) and its timescale (i.e., timescale during which the cocoon emission dominates over the KN) is significantly shorter (by about 2 orders of magnitude than the wide jet model).
This dim luminosity is due to i) the much less internal energy delivered by the central engine (than the wide jet case; as $L_{j}\approx L_{iso,0}\theta_0^2/4 \propto \theta_0^2$) and ii) the short breakout time of the jet [see equation (\ref{eq:Lbl^j S estiamte})].
This makes it challenging to detect the cocoon emission for narrow jets.

Second, the wide and the failed jet models 
(with $10^{45}-10^{43}$ erg s$^{-1}$ in the first $\sim 10^3$ s)
likely surpass the early KN emission.
These decent luminosities are, as mentioned above, due to the large $\theta_0$ and $t_b-t_0$ [see equations (\ref{eq:Lbl^j S estiamte}) and (\ref{eq:Lbl^j F estiamte})].
Their similarity is because 
the 
total internal energy of the escaped cocoon happens to be comparable for these two models. 
Although early emission follows different power-law indices [$-4/3$ (wide) and $-1$ (failed); see equation (\ref{eq:Lbl j})], it might be difficult to discriminate between these two models only based on the bolometric luminosities.

\begin{figure}
    \centering
    \includegraphics[width=0.99\linewidth]{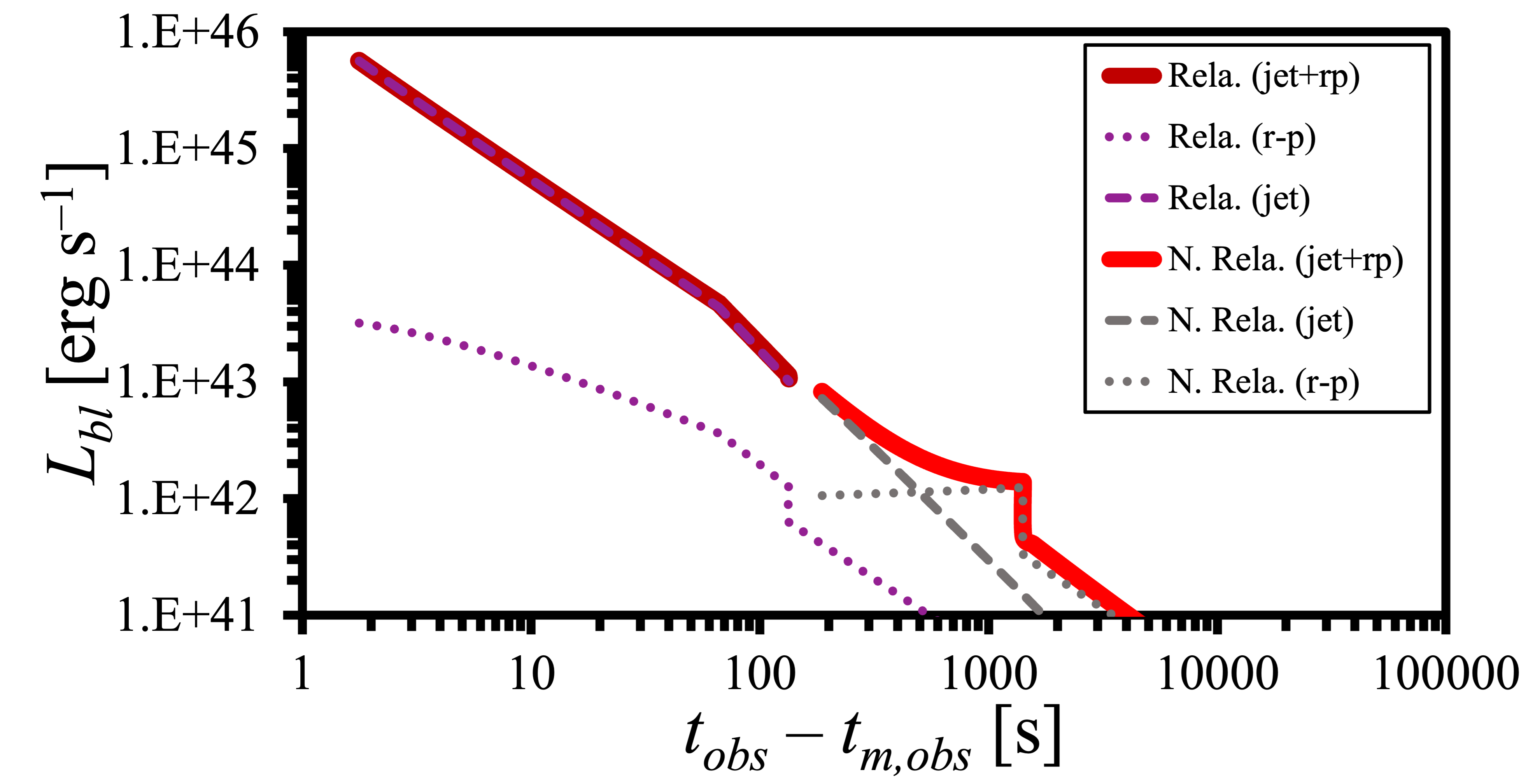} 
  \caption{Components contributing to the bolometric isotropic luminosity of the cocoon for the wide jet model
  in Table \ref{tab:1}.
  At early times, the relativistic cocoon contributes [solid dark red; equations (\ref{eq:Lbl tot})],
  which is the sum of components from jet-shock heating [dashed purple; equation (\ref{eq:Lbl R j S})] and from r-process heating [dotted purple; equation (\ref{eq:Lbl R rp S})].
  At later times ($t_{obs} > 200$ s), the non-relativistic cocoon contributes [solid red; equations (\ref{eq:Lbl tot})], which is the sum of components from jet-shock heating [dashed grey; equation (\ref{eq:Lbl NR j S})] and from r-process heating [dotted grey; equation (\ref{eq:Lbl NR rp S})].
  }
  \label{fig:Lbl} 
\end{figure}
\begin{figure*}
    \centering
    \begin{subfigure}
    \centering
    \includegraphics[width=0.79\linewidth]{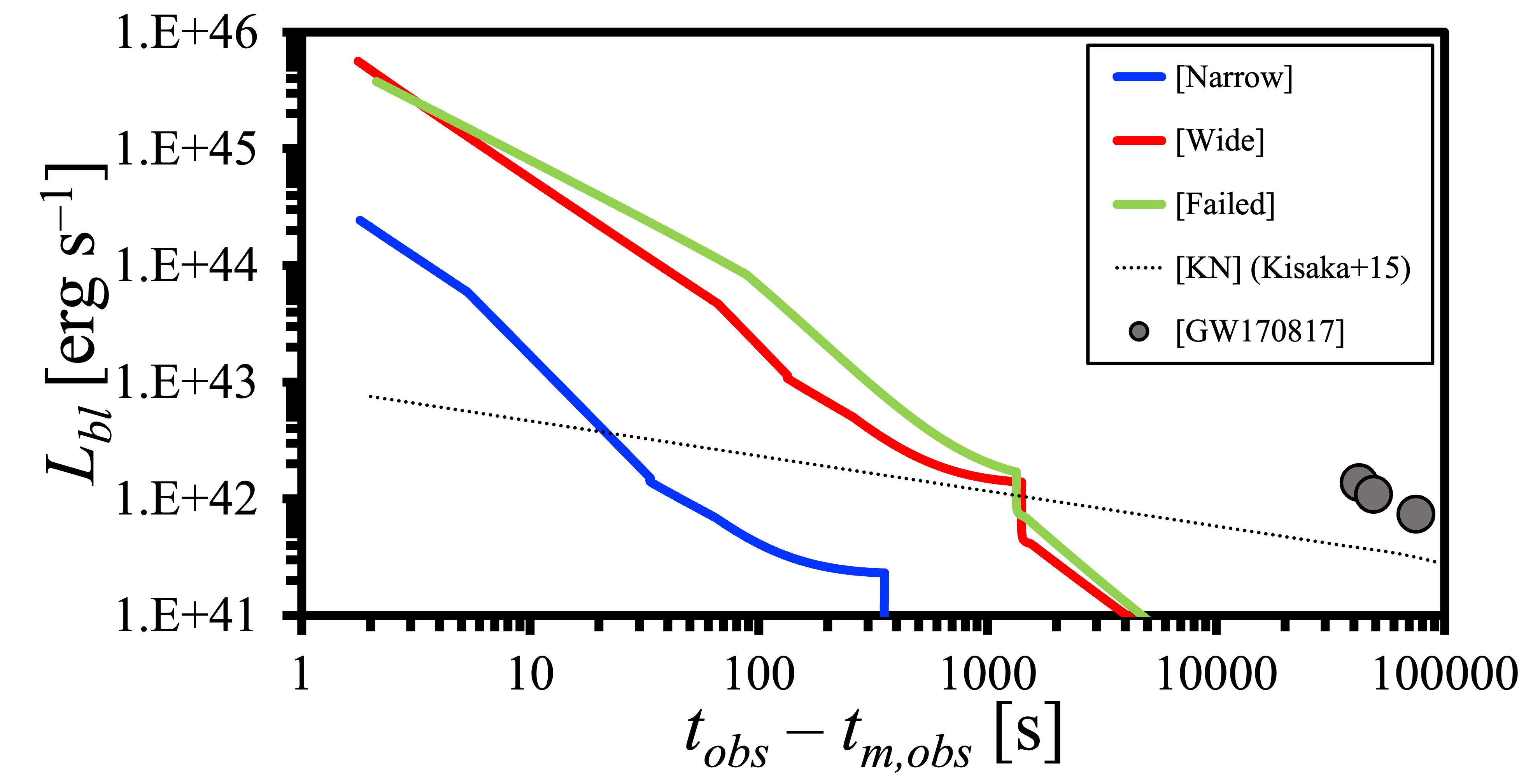}
  \end{subfigure}
  \begin{subfigure}
    \centering
    \includegraphics[width=0.79\linewidth]{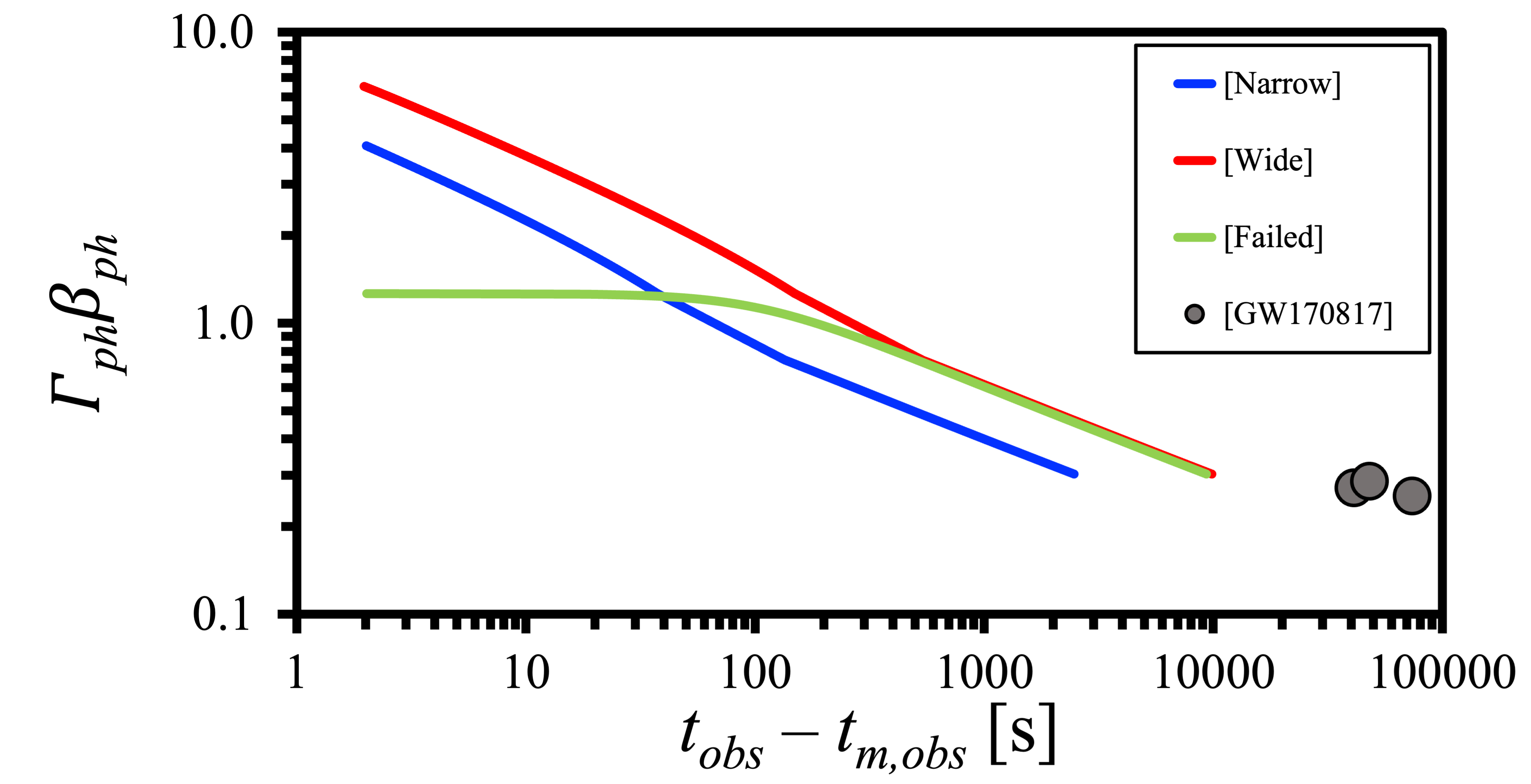}
  \end{subfigure}
    \begin{subfigure}
    \centering
    \includegraphics[width=0.79\linewidth]{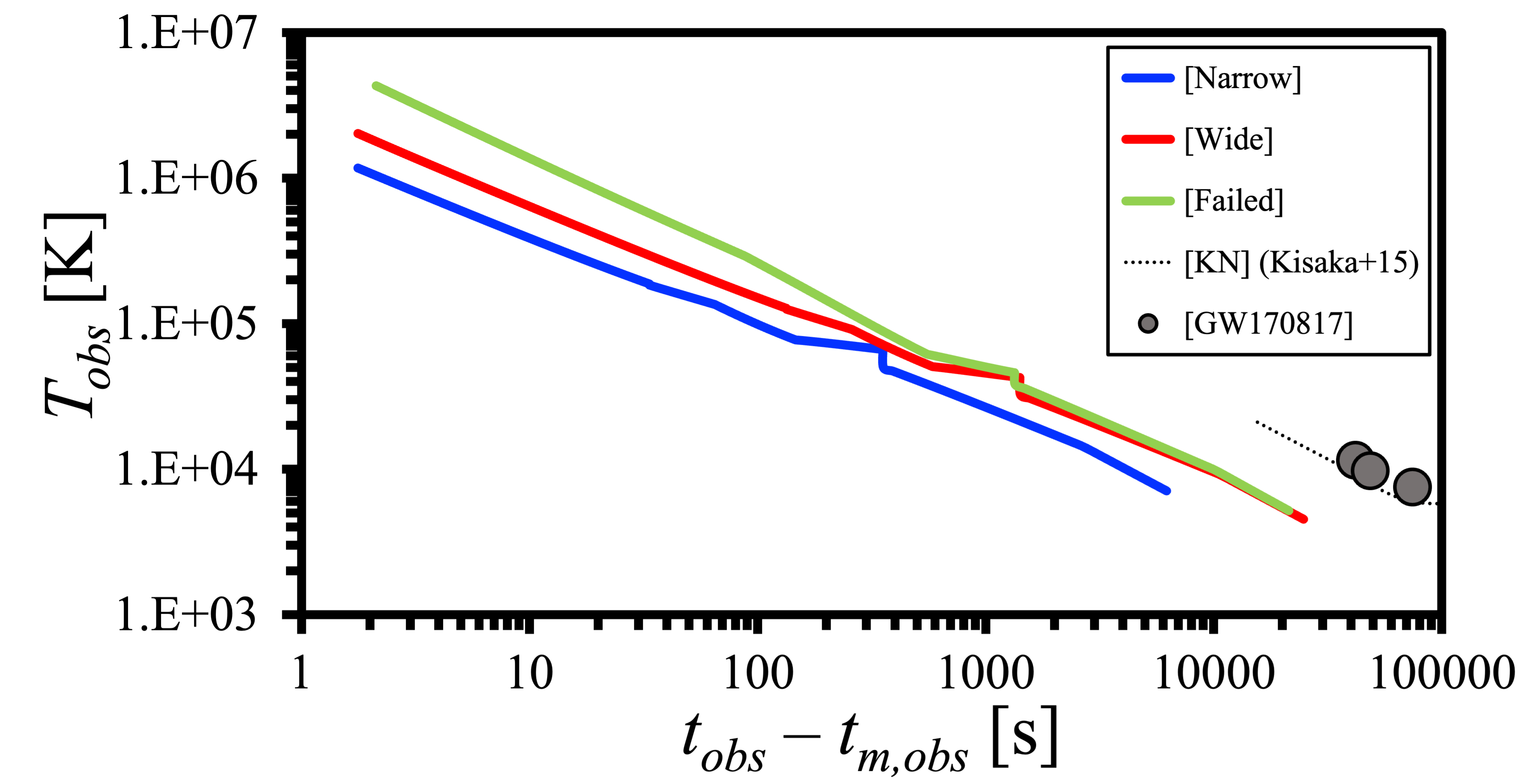}
  \end{subfigure}
  \caption{
  Bolometric isotropic luminosity (top), photospheric four-velocity (middle), and observed temperature (bottom), for three representative jet models [narrow (blue), wide (red), and failed (green) in Table \ref{tab:1}], as a function of the observed time since the merger.
  The predicted early KN is shown (dotted black; following the analytic model by \citealt{2015ApJ...802..119K}; see Appendix \ref{ap:KN}), as well as the recorded measurements on GW170817 (grey circles; \citealt{2017Sci...358.1559K}; \citealt{2017Sci...358.1570D}).
  This illustrates the expected imprint of the cocoon depending on the jet model in future GW170817-like events (NS mergers with or without a \textit{s}GRB).
  }
  \label{fig:NWF L v T} 
\end{figure*}

\subsection{Photospheric four-velocity}
\label{sec:U4 ph NWF}
Figure \ref{fig:NWF L v T} (middle panel) shows the evolution of $\log\, \Gamma_{ph}\beta_{ph}$ in an observed time $\log\, t_{obs}$ [using equations (\ref{eq:tobs R}) and (\ref{eq:tau NR exact}), with equation (\ref{eq:tau ph}); also see Section \ref{sec:Photospheric velocity}]
for the three jet models in Table \ref{tab:1} with data of GW170817.
Note that the photospheric four-velocity $\Gamma_{ph} \beta_{ph}$ is measurable from the observed quantities
[see Section \ref{sec:Tobs interpretations} and equation (\ref{eq:ph mesure})].

First, in successful jet models,  
$\Gamma_{ph}\beta_{ph}$ follows a power-law function with an index in the range $\sim \frac{1}{3} - \frac{1}{3.5}$ [see equation (\ref{eq:U4 ph R+NR cases})].
Because the observed time $t_{obs}$ for a given velocity $\Gamma_{ph}\beta_{ph}$ is $\propto \sqrt{\kappa M_c^{es}}\propto \sqrt{M_c^{es}}$ [see equation (\ref{eq:tobs R})],
the velocity at a given observed time is roughly $\propto {M_c^{es}}^{\left({1}/{7}\right)}$.
For instance, in the wide jet model ${M_c^{es}}$ is $\sim 30$ times larger than that in the narrow jet model, 
resulting in $\sim 1.7$ times larger $\Gamma_{ph}\beta_{ph}$ at the same $t_{obs}$.

Second in the failed jet model (no relativistic cocoon component), $\Gamma_{ph}\beta_{ph}$ is constant early on (the first $\sim 100$ s, although this timescale is subject to change depending on the parameters) before following a power-law function with an index of $\sim \frac{2}{7}$ (similar to successful jet models).

Therefore, 
by deducing $\Gamma_{ph}\beta_{ph}$ 
from observations and monitoring it,
it can provide the evidence of a cocoon component and insights on the nature of the jet: 
failure or success, and physical quantities such as the mass of the escaped cocoon ${M_c^{es}}$.

\subsection{Temperature}
\label{sec:Tobs NWF}
In Figure \ref{fig:NWF L v T} (bottom panel), the observed temperature [as calculated using equation (\ref{eq:Teff})] is shown for the three different jet models.
First, the failed jet model displays the highest temperature.
This is due to the significantly lower value of $\Gamma_{ph}\beta_{ph}$ at early times (see the middle panel of Figure \ref{fig:NWF L v T})
and also to the slightly high $L_{bl}$ (see the top panel).

\begin{figure*}
    \centering
    \begin{subfigure}
    \centering
    \includegraphics[width=0.79\linewidth]{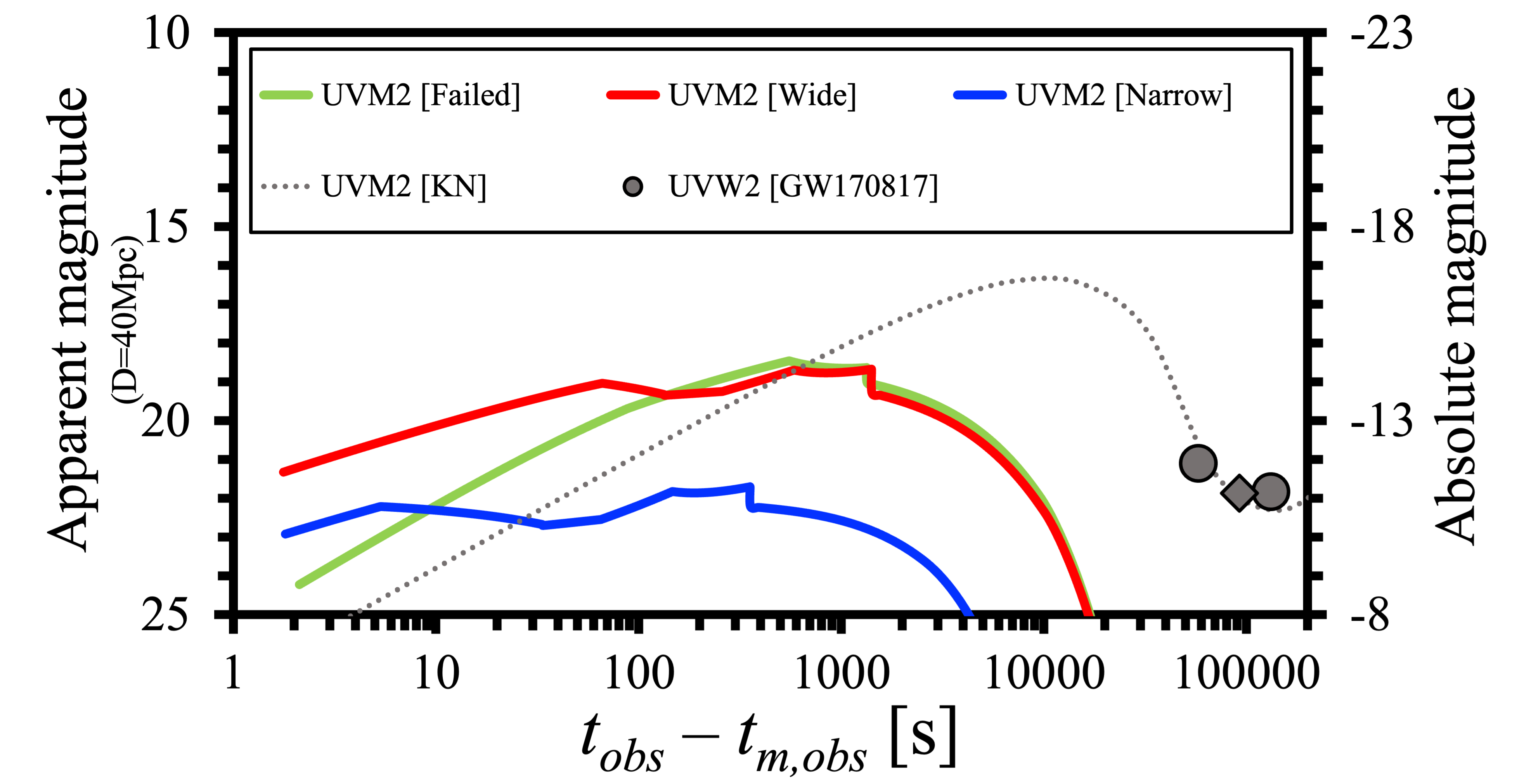}
  \end{subfigure}
  \begin{subfigure}
    \centering
    \includegraphics[width=0.79\linewidth]{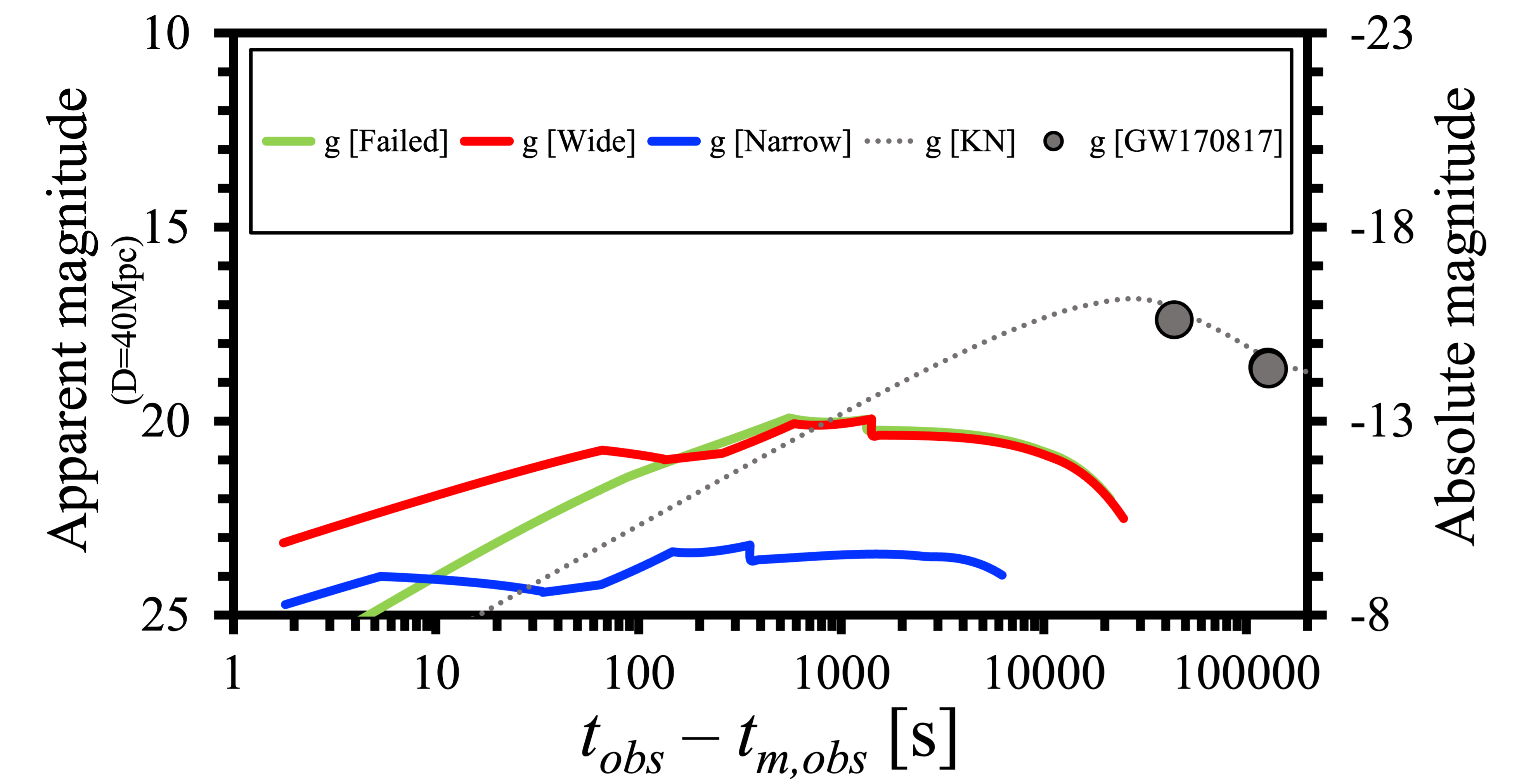} 
  \end{subfigure}
    \begin{subfigure}
    \centering
    \includegraphics[width=0.79\linewidth]{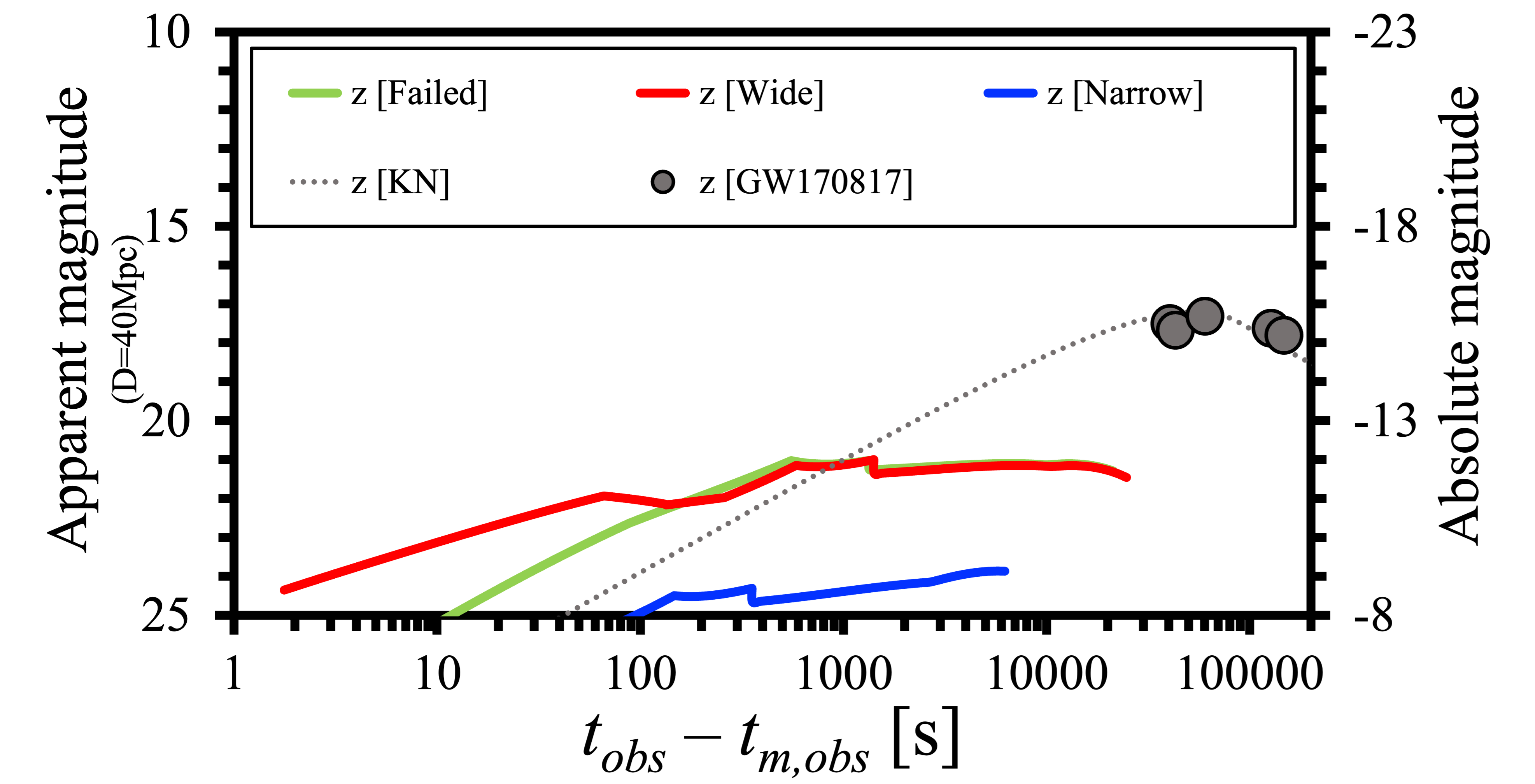}
  \end{subfigure}
  \caption{Apparent (left axis) and absolute (right axis) magnitudes of the cocoon [see equations (\ref{eq:mag ap}) and (\ref{eq:mag ab}); respectively, with $D=40$ Mpc] for three jet models [narrow (blue), wide (red), and failed (green) in Table \ref{tab:1}] and for three astronomical bands [UVM2 (top), g (middle), and z (bottom)].
  The magnitude of the early KN (dotted black line; following \citealt{2015ApJ...802..119K}; see Appendix \ref{ap:KN}) and the data on GW170817 (grey circles, and diamonds for upper limits; \citealt{2017Sci...358.1559K}; \citealt{2017Sci...358.1570D}) are also shown.
 This illustrates the imprint of the cocoon excepted at early times in future GW170817-like events (NS mergers associated with or without a \textit{s}GRB).
  }
  \label{fig:NWF mag} 
\end{figure*}

Second, the temperature in the wide and narrow jet models is lower than that in the failed jet model (by a factor of a few) at early times.
Nevertheless, these are very decent temperature (e.g., compared to the KN), and the cocoon spectrum is expected to peak at short wavelengths during its observable timescale (e.g., UV band; see Figure \ref{fig:NWF mag}).
This is consistent with numerical results in \cite{2018MNRAS.473..576G}.

Finally, the temperatures at early times ($t_{obs}\lesssim 10$ s), especially in the failed jet case, are higher than the UV range, 
reaching the soft X-ray range,
$h \nu_{\text {peak }}\approx 2.4\, \mathrm{keV}  \left({T_{obs}}/{10^7 \text{K}}\right)$
with Wien's displacement law.
Hence, theoretically, we predict that the cocoon in NS mergers is a bright X-ray source if observed early (with the right viewing angle; see Section \ref{sec:theta_v}); especially in cases where the jet is failed (Hamidani \& Ioka in preparation).

\subsection{Magnitudes}
\label{sec:Mab NWF}
In Figure \ref{fig:NWF mag}, magnitudes [apparent (for $D=40$Mpc, similar to GW170817; \citealt{2017ApJ...848L..13A}) in the left axis, and absolute in the right axis; see Section \ref{sec:mag}] for three bands [UVM2 (top); g (middle) and; z (bottom)] are shown
in the three jet models,
together with the expected KN magnitudes (see Appendix \ref{ap:KN}) and recorded observations on GW170817.

First, our cocoon magnitudes are the lowest (i.e., brightest) at shorter wavelengths (UV).
This is expected from
the high blackbody temperatures of the cocoon (see the bottom panel in Figure \ref{fig:NWF L v T}).

Second, our models suggest than the cocoon is brighter than the KN at early times, and can only be identified in this time window.
The timescale of this time window differs for the different jet types, as it strongly depends on the parameters of the jet and the ejecta [as it depends on $L_{bl}$ and $T_{obs}$; see equation (\ref{eq:Fnu})].
For our parameter-space, this timescale is the longest for the wide and failed jet models ($\sim 500$ s in UV), and much shorter for the narrow jet model ($\sim 30$ s in UV).

Third, at early times, our successful jet models diverge from the failed jet model.
While successful jets give a small cocoon magnitude (i.e., bright) from the start ($ m_{AB}\sim 21-19$ in the UV band, in the first $\sim 100$ s, for the wide jet model), for our failed jet model the magnitude at early times is much higher ($\sim 24 - 20$).
This is due to the contribution of the relativistic cocoon, which is absent in the failed jet case.
At later times, the evolution of the magnitude is determined by the non-relativistic cocoon, and the behaviors converge.

Therefore, considering our jet models, early measurements of magnitudes can be very useful for identifying and possibly discriminating between the different jet types.
Among our models, 
the cocoon of wide jets is the brightest and easiest to detect, closely followed by the cocoon of the failed jets.

It should be noted that the face values of the above timescales are not universal; depending on the cocoon and the KN parameters (set here to reproduce GW170817) this timescale can be either extended or shortened (with either brighter cocoon parameters or dimmer KN, and vice versa; see Section \ref{sec:Lbl wide}).
For example, in \cite{2018MNRAS.473..576G} and \cite{2020MNRAS.498.3320G}, relatively brighter cocoons and dimmer KN parameters resulted in much longer timescales where the cocoon emission is detectable.

\subsection{Impact of the escaped cocoon mass}
In \cite{2023MNRAS.520.1111H} we demonstrated that only a small fraction of the cocoon escapes the ejecta, with the overwhelming majority of the cocoon being trapped (in particular in terms of mass, as $M_{c}^{es}/M_{c}^{}\sim 0.005 - 0.05$) and unable to contribute to the cocoon emission as it is concealed by the optically-thick escaped cocoon and merger ejecta (until much later times, and by then the adiabatic cooling would have made emission from the trapped cocoon very faint; see Section \ref{sec:es only}).
Additionally, as shown in Section \ref{sec:3} (in particular see Figure \ref{fig:key.recap}), the density profile of the escaped cocoon is quite steep (with the power-law index $m\sim 8$).
This means that there is a substantial evolution of the density (and optical thickness) throughout the escaped cocoon, and results in
a very fast photospheric evolution (across the escaped cocoon).

Although the general trend of the bolometric luminosity would not be significantly affected (as the light curve would just be shifted to later timescales while still scaling as $L_{bl} \propto t_{obs}^{-2}$; i.e., same amount of energy is released at $t_{obs}$ after adiabatic cooling) and would almost be parallel to the current light curve [see Figures \ref{fig:Lbl} and \ref{fig:NWF L v T} (top)];
the observed cocoon temperature is very sensitive to such properties of the escaped cocoon (e.g., steepness of its density profile).
If a rough one-zone approximation is used (as $\rho \sim M/V$, i.e., $m=0$; or $v\sim\sqrt{E/M}$), the expected observed temperature at a given observe time (compared to our realistic results here) would be much lower (down to optical range; rather than the X-ray/UV range) and the peak (in terms of magnitudes) would be at much later times (given the same cocoon mass).

Therefore, our work here is a substantial improvement to previous estimates (e.g., \citealt{2017ApJ...834...28N}; \citealt{2018ApJ...855..103P}; \citealt{2018PTEP.2018d3E02I}), and we expect these previous analytic models to give more accurate results if the mass and energy of the escaped cocoon are properly implemented.

\subsection{Observational prospects}
\label{sec:Discussion}
Our results (Figures \ref{fig:NWF L v T} and \ref{fig:NWF mag}) demonstrate that the cocoon emission is bright enough to be detected by current facilities, especially in the UV band [Swift's UVOT with a limiting magnitude of $\sim 22$ for an exposure time of $\sim 1000$ s (\citealt{2005SSRv..120...95R}); or with the upcoming UV satellite ULTRASAT, with a limiting magnitude of $\sim 22.4$ for an exposure time of $900$ s, \citealt{2014AJ....147...79S}],
up to a distance $D\lesssim 100$ Mpc, for the wide and failed jet cases ($\sim 40$ Mpc for the narrow jet case).
Note that our estimation has still been very conservative (e.g., did not include the fast tail component of the ejecta).
Also, note that the follow-up happens after the burst is localized by X-ray.

Observations in the optical (and infrared) are more challenging due to the high temperature of the cocoon at early times, making the cocoon emission much fainter in the optical (and infrared) band, and by the time the cocoon temperature decreases, the KN emission is expected to dominate over the cocoon emission by several orders of magnitude [see Figure \ref{fig:NWF mag} (middle)].
Hence, the cocoon emission is difficult to detect by optical instruments such as ZTF [with its wide field of view and a limiting magnitude of 20.8 in the g-band (\citealt{2020ApJ...905...32B}); unless $D\lesssim 40$ Mpc which is very rare considering the rate of NS mergers (see \citealt{2020ApJ...892L...3A}; \citealt{2022arXiv221005695R})].
Nevertheless, with the upcoming Vera C. Rubin Observatory (i.e., LSST; with its wide field of view and a limiting magnitude of 24 from a 10 s exposure; see \citealt{2019ApJ...873..111I}; \citealt{2009arXiv0912.0201L}), the early (optical) cocoon emission can be detected (at $\sim 40$ Mpc for our narrow jet model; and at $\sim 100 - 200$ Mpc for our wide and failed jet models).

The cocoon as a prime UV target has already been pointed out by \cite{2018MNRAS.473..576G} numerically and for three individual jet models.
Here, we presented a new analytic framework that explains features of the cocoon emission, as a function of the physical parameters of the jet and the ejecta.
Hence, our analytic model can be used in future GW170817-like NS mergers events, with early multi-messenger observations (e.g., with LIGO O4; and especially in the next decade with ET, CE, and LISA).

We showed for the first time that the cocoon, in particular in terms of its relativistic nature, varies dramatically between successful jets and the failed jet.
We also showed that the relativistic nature of the cocoon can be deduced directly from observations (see Section \ref{sec:Tobs interpretations}; also see middle panel in Figure \ref{fig:NWF L v T}), hence providing an opportunity to infer the type of the jet (narrow, wide, or failed). 
However, constraining the type of jet would require, at least, knowing the ejecta mass [ideally, from the combination of: the GW signal (e.g., \citealt{2017ApJ...850L..39A}); numerical relativity calculations (e.g., \citealt{2013PhRvD..87b4001H}); and the KN emission (e.g., \citealt{2017Sci...358.1559K})] and its early opacity [with atomic data (e.g., \citealt{2020MNRAS.496.1369T}); and with the EM monitoring of the early KN (e.g., \citealt{2022ApJ...934..117B})].
This would still be challenging considering the systematic uncertainties.

In ideal circumstances where observations (from early times, before the KN emission starts to contaminate the cocoon emission, up to the KN 
nebular phase) have been achieved, and especially when observations of the different EM counterparts have also been achieved (e.g., afterglow, early and late KN spectra; independently constraining the jet and the ejecta),
our model can be used in coordination to provide additional constraints on important quantities such as 
the mass of the escaped cocoon [and hence the mass of the dynamical ejecta (see Section \ref{sec:U4 ph NWF})],
the jet power and/or opening angle,
the delay of jet launch
(see Section \ref{sec:Mab NWF}), etc.
This might be quite challenging for the next a few GW170817-like events. 
However, we expect our model to be implemented at its best with the next generation of GW detectors (ET and CE in particular) and the with upcoming EM facilities (LSST and ULTRASAT), as the quality of the data and the statistics (of GW170817-like events) are expected to drastically improve (expected to reach $\sim 10^3$ GW170817-like event per year; e.g., see Figure 11 in \citealt{2023ApJ...942...88Z}).

It should be noted that there are different possible scenarios for jet failure (e.g., dense ejecta, low jet luminosity, brief engine activity, large jet opening angle).
Also, failed jets can be collimated (as for our model here) or uncollimated (\citealt{2011ApJ...740..100B}).
Therefore, our failed jet model here does not entirely reflect the large possibilities for failed (or chocked) jets.
Nevertheless, we expect that as long as the jet is chocked inside the ejecta with similar energy budget, the overall relativistic nature of the cocoon should be similar to the failed jet model described here, in contrast with successful jets (lacking the $\Gamma\beta\gg 1$ relativistic component; see Figure \ref{fig:dEc sim}).

Furthermore, our results show for the first time that the \textit{s}GRB-cocoon (cooling emission) is potentially bright in soft X-rays in the first few seconds after the GW signal of the merger\footnote{As a note, in a different context, \cite{2022MNRAS.512.3627D} found that, similarly, the cooling emission of \textit{l}GRB-cocoon can be bright in soft X-ray at early times (see their Figure 4).
Also, \citealt{2017ApJ...834...28N} previously estimated that the cocoon's ``afterglow emission" in \textit{l}GRBs can be bright in the X-ray (see their Table 1). 
}.
This is especially the case for failed jets, with their decent luminosities and relatively compact photospheric radii at early times [see equation (\ref{eq:Teff}) and Section \ref{sec:Tobs NWF}; also see Figure \ref{fig:NWF L v T} (bottom)].
We estimate that with reasonable parameters, early cocoon emission could even explain some of the X-Ray Flashes (XRFs; see \citealt{2002ApJ...571L..31Y}), 
and in particular MAXI Unidentified Short Soft Transient (MUSST) detected by MAXI (in the range $2-10$ keV; see \citealt{2014PASJ...66...87S}), but missed by Swift/BAT (in the range $15-50$ keV) and also missed by Swift/XRT follow-up at later times indicating their short timescale (e.g., \citealt{2016GCN.20206....1D}; \citealt{2015ATel.7960....1M}; \citealt{2015GCN.17772....1H}; \citealt{2014GCN.16686....1U}; etc.); 
all in consistency with the properties of the cocoon emission found here (in particular for the failed jet cocoon).

We estimate that future wide field X-ray satellite missions will be able to better observe this soft X-ray cocoon transient. 
The Einstein Probe (EP; planned for launch in 2023),  
and the HiZ-GUNDAM (High-z Gamma-ray bursts for Unraveling the Dark Ages Mission; planned for launch in 2030) with their wide-field ($\sim 3600$ deg$^2$, and  $\sim 1$ steradian; respectively) soft X-ray (in $0.3-10$, and $0.5-4$ keV; respectively) detectors, are two particularly relevant space missions (\citealt{2022arXiv220909763Y}; and \citealt{2020SPIE11444E..2ZY}; respectively).

Note that this soft-X-ray emission is different from the emission that is expected from shock breakouts (\citealt{2010ApJ...725..904N}; \citealt{2012ApJ...747...88N}) or from the jet-cocoon breakout (although similar in photon energy).
Here, the cocoon cooling emission is thermal (blackbody) while in shock breakouts the emission is expected to be non-thermal (or at least a combination of blackbody and power-law; see \citealt{2019MNRAS.484.3502I}; \citealt{2020MNRAS.499.4961I}). 
In terms of the overall evolution of bolometric luminosity, for the cocoon emission ($L_{bl}^j\propto t_{obs}^{-2}$) it is different from that in shock breakouts (see Figure 1 in \citealt{2012ApJ...747...88N}).
Furthermore, relativistic shock breakouts are characterized by the closure relation (e.g., linking luminosity to temperature) which could also be used to identify (and differentiate them) from the cocoon emission (see \citealt{2012ApJ...747...88N}; \citealt{2019MNRAS.484.3502I}).

In the context of multi-messenger astronomy,
the cocoon emission is another electromagnetic counterparts
to NS mergers,
in addition to the other well studied sources: 
\textit{s}GRBs (prompt emission) from a relativistic jet,
its afterglow,
and KN.
The cocoon emission has advantages such as, larger opening angles 
and longer timescales, compared to \textit{s}GRBs, 
and it can be detected even if the jet is failed; without \textit{s}GRBs and afterglows.


\section{Summary \& Conclusion}
\label{sec:6}
We present the first analytic model of the cocoon emission in NS mergers (and NS-BH mergers; see Figure \ref{fig:key1}) that is directly based on numerical simulations of \textit{s}GRB-jets (see Table \ref{tab:1}), and that allows one to analytically determine the cocoon emission (in terms of luminosity, temperature, and photospheric velocity) at very early times (<1000 s) directly as a function of the parameters of the jet and the ejecta.
We made substantial improvements to previous analytical models (see \citealt{2017ApJ...834...28N}; \citealt{2018PTEP.2018d3E02I}; \citealt{2018ApJ...855..103P}),
and calculated the cocoon emission by
taking into account two heating processes:
r-process heating and jet-shock heating; and the latter process was found to largely dominate the early emission.
We only considered prompt emission's jets, and did not consider late time engine activity (e.g., \citealt{2015ApJ...802..119K}; \citealt{2017ApJ...846..142K}; \citealt{2019ApJ...887L..16K}; etc.).
In Section \ref{sec:2}, we used numerical simulations to model the cocoon after the breakout in the free-expansion phase (homologous expansion; see Figure \ref{fig:v sim}).
We showed that the cocoon, as an intermediate component between the jet and the ejecta, can be split into a non-relativistic part ($\Gamma\beta\sim \beta$) and a relativistic part ($\Gamma\beta\sim \Gamma$).
In Section \ref{sec:3}, using numerical simulation results (see Figures \ref{fig:den sim} and \ref{fig:p sim}), we modeled the cocoon's density and internal energy along these two parts (see Figure \ref{fig:key.recap}).

Then, in Section \ref{sec:4}, we calculated the optical depth for an observer with a viewing angle $\theta_j<\theta_v\sim 6^{\circ}-20^{\circ}$ (outside of the jet cone; where $\theta_j\sim 1/\Gamma_j$ and $\Gamma_j\sim 10-100$).
We used the approximation of a sharp diffusion shell (see \citealt{2012ApJ...747...88N}), and showed that this shell can be found using $\tau\sim 1$ in the relativistic cocoon ($\beta\sim 1$), and $\tau \sim 20/\beta_d$ in the non-relativistic cocoon (due to its steep density profile; see Figure \ref{fig:key.tau}).

This allowed us to estimated the photospheric velocity ($\Gamma_{ph}\beta_{ph}$) analytically, as a function of the observer's time [see equation (\ref{eq:U4 ph R+NR cases})].

Also, we carefully estimated the isotropic equivalent luminosity of the cocoon using differential formulation (see Section \ref{sec:Lbl}), and taking into account contributions from jet-shock heating and r-process heating, we presented it in the form of simple analytical equations (see Section \ref{sec:Lbl S} and \ref{sec:Lbl F}).
We analysed the contribution of jet-shock heating and r-process heating in each of the relativistic (early times) and non-relativistic (later times) cocoon parts, 
and confirmed that jet-shock heating is the relevant component, as it dominates at early times, before the KN emission becomes too bright making observations of the cocoon challenging (see Figure \ref{fig:Lbl}).
Furthermore, analyzing the temporal evolution of the cocoon luminosity from jet-shock heating [see equation (\ref{eq:Lbl^j R+NR})],
we later presented a much simpler, yet reasonably accurate, equation that (in theory) gives this cocoon luminosity directly as a function of the parameters of the jet and the ejecta, for successful and failed jets [see equations (\ref{eq:Lbl^j S estiamte}) and (\ref{eq:Lbl^j F estiamte}), respectively].
It should be noted that these equations were based on a limited number of simulations with a simplified setup (e.g., for the jet and the ejecta).
Therefore, dependency on specific values used here [notably, the density profile of the ejecta ($n=2$); also, the density profile of the escaped relativistic cocoon ($l=0$), the engine luminosity function and its timescale ($L_j(t) \propto t^0$ and $t_e-t_0=2$ s)] should be taken into account for accurate probing of the parameters of the jet and the ejecta (in cases where such values can not be constrained independently).

We also calculated the observed temperature of the cocoon [see equation (\ref{eq:Tobs R+NR})], 
and explained that measuring the combination of the bolometric isotropic equivalent luminosity and temperature can be used to effectively measure the relativistic velocity of the cocoon's photosphere (see Section \ref{sec:Tobs interpretations}).

In Section \ref{sec:5} we presented our results for three conservative jet models: narrow, wide, and failed (conservative in terms of jet luminosities, short delay time, and absence of the fast tail component of the ejecta; see Table \ref{tab:1}).
We explained that our wide and failed jet models are the brightest in terms of luminosity [see Figure \ref{fig:NWF L v T} (top)], in particular due to their large jet opening angle (and consequently their high true jet luminosities, as $L_j \propto L_{iso,0}\theta_0^{2}$) and longer breakout times [see equations (\ref{eq:Lbl^j S estiamte}) and (\ref{eq:Lbl^j F estiamte})].
For the photospheric velocities, we found that if measured [by measuring the combination of the luminosity and the temperature] it is possible to discriminate between the different jet models, and even measure the mass of the escaped cocoon [see Figure \ref{fig:NWF L v T} (middle)].
In terms of observed temperatures, we found that the cocoon is the hottest in our failed jet model (due to its relatively smaller photospheric radii), with temperatures in the order of $\sim 10^6-10^7$ K in the first $\sim 10$ s [see Figure \ref{fig:NWF L v T} (bottom)].

In terms of magnitudes, we confirmed that the UV band is the best domain to observe the cocoon emission [see Figure \ref{fig:NWF mag} (top); compared to optical (middle) and infrared bands (bottom)].
We explained that the cocoon emission can be detected with current (Swift UVOT) and upcoming (ULTRASAT) UV facilities, for sources with distances $\lesssim 100$ Mpc (for our failed and wide jet models, much less for our narrow jet model), if observations are carried out in the first 1000 s after the merger (before the KN becomes too bright).

We also found, for the first time, that even in our conservative parameter space, the cocoon's cooling emission can be bright in soft X-rays in the first few seconds (after the merger), in particular for the failed jet model.
We explained that this is detectable with MAXI and Swift XRT (if the source is caught in the FOV).
We also explained that, with more appropriate parameters, early cocoon emission might explain a subclass of XRFs (\citealt{2002ApJ...571L..31Y}; Hamidani \& Ioka in preparation) and MAXI Unidentified Short Soft Transient (MUSST; e.g., \citealt{2016GCN.20206....1D}; \citealt{2015ATel.7960....1M}; \citealt{2015GCN.17772....1H}; \citealt{2014GCN.16686....1U}; etc.).

To conclude, with the new generation of GW detectors (the upcoming LIGO O4; also with ET, CE, and LISA), we estimate that the cocoon emission is detectable in future GW170817-like events if early localization is achieved.
And with its observational features (luminosity, temperatures, and photospheric velocity) understood (with our analytic model), the cocoon emission can potentially be used to better understand NS mergers, \textit{s}GRBs, and KNe (together with the other EM counterparts: prompt emission, KN emission, and afterglow emission);
practically, the cocoon emission can be used to indirectly measure the escaped cocoon's mass and relate it to the mass of the dynamical ejecta (as explained in Section \ref{sec:U4 ph NWF}), 
infer the type of jet (as explained in Section \ref{sec:Mab NWF}),
and indirectly trace r-process nucleosynthesis and the abundance of heavy elements in the cocoon [e.g., from the opacity (e.g., \citealt{2022ApJ...934..117B}); or from spectral features (e.g., \citealt{2022arXiv220604232D})].

Furthermore, we predict that the cocoon is bright in soft X-ray in the first few seconds after the merger, especially for failed \textit{s}GRB jets.
Therefore, we argue that the cocoon, as a hybrid jet-ejecta outflow offers some advantages at probing NS mergers in Universe compared to \textit{s}GRBs' prompt emission, thanks to, e.g., the much larger opening angles of its emission ($\sim 20^\circ-30^\circ$), 
and its much longer timescales ($\sim 10$ s).
Also, the cocoon emission can theoretically be detected even if the jet is failed (no prompt emission and no afterglows emission).

\section*{Acknowledgements}
\addcontentsline{toc}{section}{Acknowledgements}
    We thank
    Amir Levinson, 
    Banerjee Smaranika, 
    Bing Zhang,
    Bing Theodore Zhang,
    Kazumi Kashiyama, 
    Kazuya Takahashi,
    Kenji Toma, 
    Kenta Kiuchi, 
    Kohta Murase, 
    Koutarou Kyutoku, 
    Kyohei Kawaguchi, 
    Masaomi Tanaka, 
    Masaru Shibata, 
    Pawan Kumar,
    Shigeo S. Kimura,
    Shota Kisaka,
    Shuta Tanaka,
    Suzuki Akihiro,
    Tomoki Wada, 
    Tsvi Piran, 
    Wataru Ishizaki,
    and Yudai Suwa,
    for their fruitful discussions and comments. 
    
    We thank the participants and the organizers of the workshops with the identification number YITP-T-19-04, YITP-W-18-11 and YITP-T-18-06, for their generous support and helpful comments. 
    
    Numerical computations were achieved thanks to the following: Cray XC50 of the Center for Computational Astrophysics at the National Astronomical Observatory of Japan, and Cray XC40 at the Yukawa Institute Computer Facility.
   
    This work was partly supported by JSPS KAKENHI nos. 20H01901, 20H01904, 20H00158, 18H01215, 17H06357, 17H06362, 22H00130 (KI). 

\section{Data availability}
The data underlying this article will be shared on reasonable request to the corresponding author.



\bibliographystyle{mnras}
\bibliography{04.2-mnras} 

\begin{thebibliography}{}
\makeatletter
\relax
\def\mn@urlcharsother{\let\do\@makeother \do\$\do\&\do\#\do\^\do\_\do\%\do\~}
\def\mn@doi{\begingroup\mn@urlcharsother \@ifnextchar [ {\mn@doi@}
  {\mn@doi@[]}}
\def\mn@doi@[#1]#2{\def\@tempa{#1}\ifx\@tempa\@empty \href
  {http://dx.doi.org/#2} {doi:#2}\else \href {http://dx.doi.org/#2} {#1}\fi
  \endgroup}
\def\mn@eprint#1#2{\mn@eprint@#1:#2::\@nil}
\def\mn@eprint@arXiv#1{\href {http://arxiv.org/abs/#1} {{\tt arXiv:#1}}}
\def\mn@eprint@dblp#1{\href {http://dblp.uni-trier.de/rec/bibtex/#1.xml}
  {dblp:#1}}
\def\mn@eprint@#1:#2:#3:#4\@nil{\def\@tempa {#1}\def\@tempb {#2}\def\@tempc
  {#3}\ifx \@tempc \@empty \let \@tempc \@tempb \let \@tempb \@tempa \fi \ifx
  \@tempb \@empty \def\@tempb {arXiv}\fi \@ifundefined
  {mn@eprint@\@tempb}{\@tempb:\@tempc}{\expandafter \expandafter \csname
  mn@eprint@\@tempb\endcsname \expandafter{\@tempc}}}

\bibitem[\protect\citeauthoryear{{Abbott} et~al.,}{{Abbott}
  et~al.}{2017a}]{2017PhRvL.119p1101A}
{Abbott} B.~P.,  et~al., 2017a, \mn@doi [Physical Review Letters]
  {10.1103/PhysRevLett.119.161101}, \href
  {https://ui.adsabs.harvard.edu/abs/2017PhRvL.119p1101A} {119, 161101}

\bibitem[\protect\citeauthoryear{{Abbott} et~al.,}{{Abbott}
  et~al.}{2017b}]{2017ApJ...848L..13A}
{Abbott} B.~P.,  et~al., 2017b, \mn@doi [\apjl] {10.3847/2041-8213/aa920c},
  \href {https://ui.adsabs.harvard.edu/abs/2017ApJ...848L..13A} {848, L13}

\bibitem[\protect\citeauthoryear{{Abbott} et~al.,}{{Abbott}
  et~al.}{2017c}]{2017ApJ...850L..39A}
{Abbott} B.~P.,  et~al., 2017c, \mn@doi [\apjl] {10.3847/2041-8213/aa9478},
  \href {https://ui.adsabs.harvard.edu/abs/2017ApJ...850L..39A} {850, L39}

\bibitem[\protect\citeauthoryear{{Abbott} et~al.,}{{Abbott}
  et~al.}{2020}]{2020ApJ...892L...3A}
{Abbott} B.~P.,  et~al., 2020, \mn@doi [\apjl] {10.3847/2041-8213/ab75f5},
  \href {https://ui.adsabs.harvard.edu/abs/2020ApJ...892L...3A} {892, L3}

\bibitem[\protect\citeauthoryear{{Abramowicz}, {Novikov}  \&
  {Paczynski}}{{Abramowicz} et~al.}{1991}]{1991ApJ...369..175A}
{Abramowicz} M.~A.,  {Novikov} I.~D.,   {Paczynski} B.,  1991, \mn@doi [\apj]
  {10.1086/169748}, \href
  {https://ui.adsabs.harvard.edu/abs/1991ApJ...369..175A} {369, 175}

\bibitem[\protect\citeauthoryear{Arcavi et~al.}{Arcavi
  et~al.}{2017}]{Arcavi:2017vbi}
Arcavi I.,  et~al., 2017, \mn@doi [Astrophys. J.] {10.3847/2041-8213/aa910f},
  848, L33

\bibitem[\protect\citeauthoryear{{Arnett}}{{Arnett}}{1980}]{1980ApJ...237..541A}
{Arnett} W.~D.,  1980, \mn@doi [\apj] {10.1086/157898}, \href
  {https://ui.adsabs.harvard.edu/abs/1980ApJ...237..541A} {237, 541}

\bibitem[\protect\citeauthoryear{{Balasubramanian} et~al.,}{{Balasubramanian}
  et~al.}{2022}]{2022arXiv220514788B}
{Balasubramanian} A.,  et~al., 2022, arXiv e-prints, \href
  {https://ui.adsabs.harvard.edu/abs/2022arXiv220514788B} {p. arXiv:2205.14788}

\bibitem[\protect\citeauthoryear{{Banerjee}, {Tanaka}, {Kawaguchi}, {Kato}  \&
  {Gaigalas}}{{Banerjee} et~al.}{2020}]{2020ApJ...901...29B}
{Banerjee} S.,  {Tanaka} M.,  {Kawaguchi} K.,  {Kato} D.,   {Gaigalas} G.,
  2020, \mn@doi [\apj] {10.3847/1538-4357/abae61}, \href
  {https://ui.adsabs.harvard.edu/abs/2020ApJ...901...29B} {901, 29}

\bibitem[\protect\citeauthoryear{{Banerjee}, {Tanaka}, {Kato}, {Gaigalas},
  {Kawaguchi}  \& {Domoto}}{{Banerjee} et~al.}{2022}]{2022ApJ...934..117B}
{Banerjee} S.,  {Tanaka} M.,  {Kato} D.,  {Gaigalas} G.,  {Kawaguchi} K.,
  {Domoto} N.,  2022, \mn@doi [\apj] {10.3847/1538-4357/ac7565}, \href
  {https://ui.adsabs.harvard.edu/abs/2022ApJ...934..117B} {934, 117}

\bibitem[\protect\citeauthoryear{{Banerjee}, {Tanaka}, {Kato}  \&
  {Gaigalas}}{{Banerjee} et~al.}{2023}]{2023arXiv230405810B}
{Banerjee} S.,  {Tanaka} M.,  {Kato} D.,   {Gaigalas} G.,  2023, \mn@doi [arXiv
  e-prints] {10.48550/arXiv.2304.05810}, \href
  {https://ui.adsabs.harvard.edu/abs/2023arXiv230405810B} {p. arXiv:2304.05810}

\bibitem[\protect\citeauthoryear{{Bauswein}, {Goriely}  \& {Janka}}{{Bauswein}
  et~al.}{2013}]{2013ApJ...773...78B}
{Bauswein} A.,  {Goriely} S.,   {Janka} H.~T.,  2013, \mn@doi [\apj]
  {10.1088/0004-637X/773/1/78}, \href
  {https://ui.adsabs.harvard.edu/abs/2013ApJ...773...78B} {773, 78}

\bibitem[\protect\citeauthoryear{{Begelman} \& {Cioffi}}{{Begelman} \&
  {Cioffi}}{1989}]{1989ApJ...345L..21B}
{Begelman} M.~C.,  {Cioffi} D.~F.,  1989, \mn@doi [\apjl] {10.1086/185542},
  \href {https://ui.adsabs.harvard.edu/abs/1989ApJ...345L..21B} {345, L21}

\bibitem[\protect\citeauthoryear{{Blandford} \& {Rees}}{{Blandford} \&
  {Rees}}{1974}]{1974MNRAS.169..395B}
{Blandford} R.~D.,  {Rees} M.~J.,  1974, \mn@doi [\mnras]
  {10.1093/mnras/169.3.395}, \href
  {https://ui.adsabs.harvard.edu/abs/1974MNRAS.169..395B} {169, 395}

\bibitem[\protect\citeauthoryear{{Bromberg}, {Nakar}, {Piran}  \&
  {Sari}}{{Bromberg} et~al.}{2011}]{2011ApJ...740..100B}
{Bromberg} O.,  {Nakar} E.,  {Piran} T.,   {Sari} R.,  2011, \mn@doi [\apj]
  {10.1088/0004-637X/740/2/100}, \href
  {http://adsabs.harvard.edu/abs/2011ApJ...740..100B} {740, 100}

\bibitem[\protect\citeauthoryear{{Burdge} et~al.,}{{Burdge}
  et~al.}{2020}]{2020ApJ...905...32B}
{Burdge} K.~B.,  et~al., 2020, \mn@doi [\apj] {10.3847/1538-4357/abc261}, \href
  {https://ui.adsabs.harvard.edu/abs/2020ApJ...905...32B} {905, 32}

\bibitem[\protect\citeauthoryear{{Burrows} et~al.,}{{Burrows}
  et~al.}{2000}]{2000SPIE.4140...64B}
{Burrows} D.~N.,  et~al., 2000, in {Flanagan} K.~A.,  {Siegmund} O.~H.,  eds,
  Society of Photo-Optical Instrumentation Engineers (SPIE) Conference Series
  Vol. 4140, X-Ray and Gamma-Ray Instrumentation for Astronomy XI. pp 64--75,
  \mn@doi{10.1117/12.409158}

\bibitem[\protect\citeauthoryear{{Chevalier} \& {Soker}}{{Chevalier} \&
  {Soker}}{1989}]{1989ApJ...341..867C}
{Chevalier} R.~A.,  {Soker} N.,  1989, \mn@doi [\apj] {10.1086/167545}, \href
  {https://ui.adsabs.harvard.edu/abs/1989ApJ...341..867C} {341, 867}

\bibitem[\protect\citeauthoryear{Chornock et~al.}{Chornock
  et~al.}{2017}]{Chornock:2017sdf}
Chornock R.,  et~al., 2017, \mn@doi [Astrophys. J.] {10.3847/2041-8213/aa905c},
  848, L19

\bibitem[\protect\citeauthoryear{Coulter et~al.}{Coulter
  et~al.}{2017}]{Coulter:2017wya}
Coulter D.~A.,  et~al., 2017, \mn@doi [Science] {10.1126/science.aap9811}

\bibitem[\protect\citeauthoryear{{D'Ai} et~al.,}{{D'Ai}
  et~al.}{2016}]{2016GCN.20206....1D}
{D'Ai} A.,  et~al., 2016, GRB Coordinates Network, \href
  {https://ui.adsabs.harvard.edu/abs/2016GCN.20206....1D} {20206, 1}

\bibitem[\protect\citeauthoryear{{De Colle}, {Kumar}  \& {Hoeflich}}{{De Colle}
  et~al.}{2022}]{2022MNRAS.512.3627D}
{De Colle} F.,  {Kumar} P.,   {Hoeflich} P.,  2022, \mn@doi [\mnras]
  {10.1093/mnras/stac742}, \href
  {https://ui.adsabs.harvard.edu/abs/2022MNRAS.512.3627D} {512, 3627}

\bibitem[\protect\citeauthoryear{{D{\'{\i}}az} et~al.,}{{D{\'{\i}}az}
  et~al.}{2017}]{2017ApJ...848L..29D}
{D{\'{\i}}az} M.~C.,  et~al., 2017, \mn@doi [\apjl] {10.3847/2041-8213/aa9060},
  \href {https://ui.adsabs.harvard.edu/abs/2017ApJ...848L..29D} {848, L29}

\bibitem[\protect\citeauthoryear{{Domoto}, {Tanaka}, {Kato}, {Kawaguchi},
  {Hotokezaka}  \& {Wanajo}}{{Domoto} et~al.}{2022}]{2022arXiv220604232D}
{Domoto} N.,  {Tanaka} M.,  {Kato} D.,  {Kawaguchi} K.,  {Hotokezaka} K.,
  {Wanajo} S.,  2022, arXiv e-prints, \href
  {https://ui.adsabs.harvard.edu/abs/2022arXiv220604232D} {p. arXiv:2206.04232}

\bibitem[\protect\citeauthoryear{{Drout} et~al.,}{{Drout}
  et~al.}{2017}]{2017Sci...358.1570D}
{Drout} M.~R.,  et~al., 2017, \mn@doi [Science] {10.1126/science.aaq0049},
  \href {https://ui.adsabs.harvard.edu/abs/2017Sci...358.1570D} {358, 1570}

\bibitem[\protect\citeauthoryear{{Duffell}, {Quataert}  \&
  {MacFadyen}}{{Duffell} et~al.}{2015}]{2015ApJ...813...64D}
{Duffell} P.~C.,  {Quataert} E.,   {MacFadyen} A.~I.,  2015, \mn@doi [\apj]
  {10.1088/0004-637X/813/1/64}, \href
  {https://ui.adsabs.harvard.edu/abs/2015ApJ...813...64D} {813, 64}

\bibitem[\protect\citeauthoryear{{Duffell}, {Quataert}, {Kasen}  \&
  {Klion}}{{Duffell} et~al.}{2018}]{2018ApJ...866....3D}
{Duffell} P.~C.,  {Quataert} E.,  {Kasen} D.,   {Klion} H.,  2018, \mn@doi
  [\apj] {10.3847/1538-4357/aae084}, \href
  {https://ui.adsabs.harvard.edu/abs/2018ApJ...866....3D} {866, 3}

\bibitem[\protect\citeauthoryear{{Eichler}, {Livio}, {Piran}  \&
  {Schramm}}{{Eichler} et~al.}{1989}]{1989Natur.340..126E}
{Eichler} D.,  {Livio} M.,  {Piran} T.,   {Schramm} D.~N.,  1989, \mn@doi
  [\nat] {10.1038/340126a0}, \href
  {https://ui.adsabs.harvard.edu/abs/1989Natur.340..126E} {340, 126}

\bibitem[\protect\citeauthoryear{{Goodman}}{{Goodman}}{1986}]{1986ApJ...308L..47G}
{Goodman} J.,  1986, \mn@doi [\apjl] {10.1086/184741}, \href
  {https://ui.adsabs.harvard.edu/abs/1986ApJ...308L..47G} {308, L47}

\bibitem[\protect\citeauthoryear{{Gottlieb}, {Nakar}  \& {Piran}}{{Gottlieb}
  et~al.}{2018}]{2018MNRAS.473..576G}
{Gottlieb} O.,  {Nakar} E.,   {Piran} T.,  2018, \mn@doi [\mnras]
  {10.1093/mnras/stx2357}, \href
  {https://ui.adsabs.harvard.edu/abs/2018MNRAS.473..576G} {473, 576}

\bibitem[\protect\citeauthoryear{{Gottlieb}, {Bromberg}, {Singh}  \&
  {Nakar}}{{Gottlieb} et~al.}{2020}]{2020MNRAS.498.3320G}
{Gottlieb} O.,  {Bromberg} O.,  {Singh} C.~B.,   {Nakar} E.,  2020, \mn@doi
  [\mnras] {10.1093/mnras/staa2567}, \href
  {https://ui.adsabs.harvard.edu/abs/2020MNRAS.498.3320G} {498, 3320}

\bibitem[\protect\citeauthoryear{{Gottlieb}, {Nakar}  \& {Bromberg}}{{Gottlieb}
  et~al.}{2021}]{2021MNRAS.500.3511G}
{Gottlieb} O.,  {Nakar} E.,   {Bromberg} O.,  2021, \mn@doi [\mnras]
  {10.1093/mnras/staa3501}, \href
  {https://ui.adsabs.harvard.edu/abs/2021MNRAS.500.3511G} {500, 3511}

\bibitem[\protect\citeauthoryear{{Hamidani} \& {Ioka}}{{Hamidani} \&
  {Ioka}}{2021}]{2021MNRAS.500..627H}
{Hamidani} H.,  {Ioka} K.,  2021, \mn@doi [\mnras] {10.1093/mnras/staa3276},
  \href {https://ui.adsabs.harvard.edu/abs/2021MNRAS.500..627H} {500, 627}

\bibitem[\protect\citeauthoryear{{Hamidani} \& {Ioka}}{{Hamidani} \&
  {Ioka}}{2023}]{2023MNRAS.520.1111H}
{Hamidani} H.,  {Ioka} K.,  2023, \mn@doi [\mnras] {10.1093/mnras/stad041},
  \href {https://ui.adsabs.harvard.edu/abs/2023MNRAS.520.1111H} {520, 1111}

\bibitem[\protect\citeauthoryear{{Hamidani}, {Takahashi}, {Umeda}  \&
  {Okita}}{{Hamidani} et~al.}{2017}]{2017MNRAS.469.2361H}
{Hamidani} H.,  {Takahashi} K.,  {Umeda} H.,   {Okita} S.,  2017, \mn@doi
  [\mnras] {10.1093/mnras/stx987}, \href
  {https://ui.adsabs.harvard.edu/abs/2017MNRAS.469.2361H} {469, 2361}

\bibitem[\protect\citeauthoryear{{Hamidani}, {Kiuchi}  \& {Ioka}}{{Hamidani}
  et~al.}{2020}]{2020MNRAS.491.3192H}
{Hamidani} H.,  {Kiuchi} K.,   {Ioka} K.,  2020, \mn@doi [\mnras]
  {10.1093/mnras/stz3231}, \href
  {https://ui.adsabs.harvard.edu/abs/2020MNRAS.491.3192H} {491, 3192}

\bibitem[\protect\citeauthoryear{{Harrison}, {Gottlieb}  \& {Nakar}}{{Harrison}
  et~al.}{2018}]{2018MNRAS.477.2128H}
{Harrison} R.,  {Gottlieb} O.,   {Nakar} E.,  2018, \mn@doi [\mnras]
  {10.1093/mnras/sty760}, \href
  {https://ui.adsabs.harvard.edu/abs/2018MNRAS.477.2128H} {477, 2128}

\bibitem[\protect\citeauthoryear{{Hjorth} et~al.,}{{Hjorth}
  et~al.}{2003}]{2003Natur.423..847H}
{Hjorth} J.,  et~al., 2003, \mn@doi [\nat] {10.1038/nature01750}, \href
  {http://adsabs.harvard.edu/abs/2003Natur.423..847H} {423, 847}

\bibitem[\protect\citeauthoryear{{Honda} et~al.,}{{Honda}
  et~al.}{2015}]{2015GCN.17772....1H}
{Honda} F.,  et~al., 2015, GRB Coordinates Network, \href
  {https://ui.adsabs.harvard.edu/abs/2015GCN.17772....1H} {17772, 1}

\bibitem[\protect\citeauthoryear{{Hotokezaka}, {Kiuchi}, {Kyutoku}, {Okawa},
  {Sekiguchi}, {Shibata}  \& {Taniguchi}}{{Hotokezaka}
  et~al.}{2013}]{2013PhRvD..87b4001H}
{Hotokezaka} K.,  {Kiuchi} K.,  {Kyutoku} K.,  {Okawa} H.,  {Sekiguchi} Y.-i.,
  {Shibata} M.,   {Taniguchi} K.,  2013, \mn@doi [\prd]
  {10.1103/PhysRevD.87.024001}, \href
  {https://ui.adsabs.harvard.edu/abs/2013PhRvD..87b4001H} {87, 024001}

\bibitem[\protect\citeauthoryear{{Hotokezaka}, {Sari}  \& {Piran}}{{Hotokezaka}
  et~al.}{2017}]{2017MNRAS.468...91H}
{Hotokezaka} K.,  {Sari} R.,   {Piran} T.,  2017, \mn@doi [\mnras]
  {10.1093/mnras/stx411}, \href
  {https://ui.adsabs.harvard.edu/abs/2017MNRAS.468...91H} {468, 91}

\bibitem[\protect\citeauthoryear{{Ioka} \& {Nakamura}}{{Ioka} \&
  {Nakamura}}{2018}]{2018PTEP.2018d3E02I}
{Ioka} K.,  {Nakamura} T.,  2018, \mn@doi [Progress of Theoretical and
  Experimental Physics] {10.1093/ptep/pty036}, \href
  {https://ui.adsabs.harvard.edu/abs/2018PTEP.2018d3E02I} {2018, 043E02}

\bibitem[\protect\citeauthoryear{{Ioka} \& {Nakamura}}{{Ioka} \&
  {Nakamura}}{2019}]{2019MNRAS.487.4884I}
{Ioka} K.,  {Nakamura} T.,  2019, \mn@doi [\mnras] {10.1093/mnras/stz1650},
  \href {https://ui.adsabs.harvard.edu/abs/2019MNRAS.487.4884I} {487, 4884}

\bibitem[\protect\citeauthoryear{{Ioka}, {Levinson}  \& {Nakar}}{{Ioka}
  et~al.}{2019}]{2019MNRAS.484.3502I}
{Ioka} K.,  {Levinson} A.,   {Nakar} E.,  2019, \mn@doi [\mnras]
  {10.1093/mnras/stz270}, \href
  {https://ui.adsabs.harvard.edu/abs/2019MNRAS.484.3502I} {484, 3502}

\bibitem[\protect\citeauthoryear{{Ishizaki}, {Kiuchi}, {Ioka}  \&
  {Wanajo}}{{Ishizaki} et~al.}{2021}]{2021ApJ...922..185I}
{Ishizaki} W.,  {Kiuchi} K.,  {Ioka} K.,   {Wanajo} S.,  2021, \mn@doi [\apj]
  {10.3847/1538-4357/ac23d9}, \href
  {https://ui.adsabs.harvard.edu/abs/2021ApJ...922..185I} {922, 185}

\bibitem[\protect\citeauthoryear{{Ito}, {Levinson}  \& {Nakar}}{{Ito}
  et~al.}{2020}]{2020MNRAS.499.4961I}
{Ito} H.,  {Levinson} A.,   {Nakar} E.,  2020, \mn@doi [\mnras]
  {10.1093/mnras/staa3125}, \href
  {https://ui.adsabs.harvard.edu/abs/2020MNRAS.499.4961I} {499, 4961}

\bibitem[\protect\citeauthoryear{{Ivezi{\'c}} et~al.,}{{Ivezi{\'c}}
  et~al.}{2019}]{2019ApJ...873..111I}
{Ivezi{\'c}} {\v{Z}}.,  et~al., 2019, \mn@doi [\apj]
  {10.3847/1538-4357/ab042c}, \href
  {https://ui.adsabs.harvard.edu/abs/2019ApJ...873..111I} {873, 111}

\bibitem[\protect\citeauthoryear{{Iwamoto} et~al.,}{{Iwamoto}
  et~al.}{1998}]{1998Natur.395..672I}
{Iwamoto} K.,  et~al., 1998, \mn@doi [\nat] {10.1038/27155}, \href
  {http://adsabs.harvard.edu/abs/1998Natur.395..672I} {395, 672}

\bibitem[\protect\citeauthoryear{{Just}, {Bauswein}, {Ardevol Pulpillo},
  {Goriely}  \& {Janka}}{{Just} et~al.}{2015}]{2015MNRAS.448..541J}
{Just} O.,  {Bauswein} A.,  {Ardevol Pulpillo} R.,  {Goriely} S.,   {Janka}
  H.-T.,  2015, \mn@doi [\mnras] {10.1093/mnras/stv009}, \href
  {https://ui.adsabs.harvard.edu/abs/2015MNRAS.448..541J} {448, 541}

\bibitem[\protect\citeauthoryear{{Kashiyama} \& {Quataert}}{{Kashiyama} \&
  {Quataert}}{2015}]{2015MNRAS.451.2656K}
{Kashiyama} K.,  {Quataert} E.,  2015, \mn@doi [\mnras]
  {10.1093/mnras/stv1164}, \href
  {https://ui.adsabs.harvard.edu/abs/2015MNRAS.451.2656K} {451, 2656}

\bibitem[\protect\citeauthoryear{{Kasliwal} et~al.,}{{Kasliwal}
  et~al.}{2017}]{2017Sci...358.1559K}
{Kasliwal} M.~M.,  et~al., 2017, \mn@doi [Science] {10.1126/science.aap9455},
  \href {https://ui.adsabs.harvard.edu/abs/2017Sci...358.1559K} {358, 1559}

\bibitem[\protect\citeauthoryear{Kilpatrick et~al.}{Kilpatrick
  et~al.}{2017}]{Kilpatrick:2017mhz}
Kilpatrick C.~D.,  et~al., 2017, \mn@doi [Science] {10.1126/science.aaq0073},
  358, 1583

\bibitem[\protect\citeauthoryear{{Kimura}, {Murase}, {Ioka}, {Kisaka}, {Fang}
  \& {M{\'e}sz{\'a}ros}}{{Kimura} et~al.}{2019}]{2019ApJ...887L..16K}
{Kimura} S.~S.,  {Murase} K.,  {Ioka} K.,  {Kisaka} S.,  {Fang} K.,
  {M{\'e}sz{\'a}ros} P.,  2019, \mn@doi [\apjl] {10.3847/2041-8213/ab59e1},
  \href {https://ui.adsabs.harvard.edu/abs/2019ApJ...887L..16K} {887, L16}

\bibitem[\protect\citeauthoryear{{Kisaka}, {Ioka}  \& {Takami}}{{Kisaka}
  et~al.}{2015}]{2015ApJ...802..119K}
{Kisaka} S.,  {Ioka} K.,   {Takami} H.,  2015, \mn@doi [\apj]
  {10.1088/0004-637X/802/2/119}, \href
  {https://ui.adsabs.harvard.edu/abs/2015ApJ...802..119K} {802, 119}

\bibitem[\protect\citeauthoryear{{Kisaka}, {Ioka}  \& {Sakamoto}}{{Kisaka}
  et~al.}{2017}]{2017ApJ...846..142K}
{Kisaka} S.,  {Ioka} K.,   {Sakamoto} T.,  2017, \mn@doi [\apj]
  {10.3847/1538-4357/aa8775}, \href
  {https://ui.adsabs.harvard.edu/abs/2017ApJ...846..142K} {846, 142}

\bibitem[\protect\citeauthoryear{{Klion}, {Duffell}, {Kasen}  \&
  {Quataert}}{{Klion} et~al.}{2021}]{2021MNRAS.502..865K}
{Klion} H.,  {Duffell} P.~C.,  {Kasen} D.,   {Quataert} E.,  2021, \mn@doi
  [\mnras] {10.1093/mnras/stab042}, \href
  {https://ui.adsabs.harvard.edu/abs/2021MNRAS.502..865K} {502, 865}

\bibitem[\protect\citeauthoryear{{Kouveliotou}, {Meegan}, {Fishman}, {Bhat},
  {Briggs}, {Koshut}, {Paciesas}  \& {Pendleton}}{{Kouveliotou}
  et~al.}{1993}]{1993ApJ...413L.101K}
{Kouveliotou} C.,  {Meegan} C.~A.,  {Fishman} G.~J.,  {Bhat} N.~P.,  {Briggs}
  M.~S.,  {Koshut} T.~M.,  {Paciesas} W.~S.,   {Pendleton} G.~N.,  1993,
  \mn@doi [\apjl] {10.1086/186969}, \href
  {http://adsabs.harvard.edu/abs/1993ApJ...413L.101K} {413, L101}

\bibitem[\protect\citeauthoryear{{Krolik} \& {Pier}}{{Krolik} \&
  {Pier}}{1991}]{1991ApJ...373..277K}
{Krolik} J.~H.,  {Pier} E.~A.,  1991, \mn@doi [\apj] {10.1086/170048}, \href
  {https://ui.adsabs.harvard.edu/abs/1991ApJ...373..277K} {373, 277}

\bibitem[\protect\citeauthoryear{{Kulkarni}}{{Kulkarni}}{2005}]{2005astro.ph.10256K}
{Kulkarni} S.~R.,  2005, arXiv e-prints, \href
  {https://ui.adsabs.harvard.edu/abs/2005astro.ph.10256K} {pp
  astro--ph/0510256}

\bibitem[\protect\citeauthoryear{{LSST Science Collaboration} et~al.,}{{LSST
  Science Collaboration} et~al.}{2009}]{2009arXiv0912.0201L}
{LSST Science Collaboration} et~al., 2009, arXiv e-prints, \href
  {https://ui.adsabs.harvard.edu/abs/2009arXiv0912.0201L} {p. arXiv:0912.0201}

\bibitem[\protect\citeauthoryear{{Lazzati}, {Morsony}  \& {Begelman}}{{Lazzati}
  et~al.}{2009}]{2009ApJ...700L..47L}
{Lazzati} D.,  {Morsony} B.~J.,   {Begelman} M.~C.,  2009, \mn@doi [\apjl]
  {10.1088/0004-637X/700/1/L47}, \href
  {https://ui.adsabs.harvard.edu/abs/2009ApJ...700L..47L} {700, L47}

\bibitem[\protect\citeauthoryear{{Lazzati}, {L{\'o}pez-C{\'a}mara},
  {Cantiello}, {Morsony}, {Perna}  \& {Workman}}{{Lazzati}
  et~al.}{2017}]{2017ApJ...848L...6L}
{Lazzati} D.,  {L{\'o}pez-C{\'a}mara} D.,  {Cantiello} M.,  {Morsony} B.~J.,
  {Perna} R.,   {Workman} J.~C.,  2017, \mn@doi [\apjl]
  {10.3847/2041-8213/aa8f3d}, \href
  {https://ui.adsabs.harvard.edu/abs/2017ApJ...848L...6L} {848, L6}

\bibitem[\protect\citeauthoryear{{Li} \& {Paczy{\'n}ski}}{{Li} \&
  {Paczy{\'n}ski}}{1998}]{1998ApJ...507L..59L}
{Li} L.-X.,  {Paczy{\'n}ski} B.,  1998, \mn@doi [\apjl] {10.1086/311680}, \href
  {https://ui.adsabs.harvard.edu/abs/1998ApJ...507L..59L} {507, L59}

\bibitem[\protect\citeauthoryear{{MacFadyen} \& {Woosley}}{{MacFadyen} \&
  {Woosley}}{1999}]{1999ApJ...524..262M}
{MacFadyen} A.~I.,  {Woosley} S.~E.,  1999, \mn@doi [\apj] {10.1086/307790},
  \href {http://adsabs.harvard.edu/abs/1999ApJ...524..262M} {524, 262}

\bibitem[\protect\citeauthoryear{{Matsumoto} \& {Masada}}{{Matsumoto} \&
  {Masada}}{2019}]{2019MNRAS.490.4271M}
{Matsumoto} J.,  {Masada} Y.,  2019, \mn@doi [\mnras] {10.1093/mnras/stz2821},
  \href {https://ui.adsabs.harvard.edu/abs/2019MNRAS.490.4271M} {490, 4271}

\bibitem[\protect\citeauthoryear{{Matsuoka} et~al.,}{{Matsuoka}
  et~al.}{2009}]{2009PASJ...61..999M}
{Matsuoka} M.,  et~al., 2009, \mn@doi [\pasj] {10.1093/pasj/61.5.999}, \href
  {https://ui.adsabs.harvard.edu/abs/2009PASJ...61..999M} {61, 999}

\bibitem[\protect\citeauthoryear{{Matzner} \& {McKee}}{{Matzner} \&
  {McKee}}{1999}]{1999ApJ...510..379M}
{Matzner} C.~D.,  {McKee} C.~F.,  1999, \mn@doi [\apj] {10.1086/306571}, \href
  {https://ui.adsabs.harvard.edu/abs/1999ApJ...510..379M} {510, 379}

\bibitem[\protect\citeauthoryear{{Metzger} et~al.,}{{Metzger}
  et~al.}{2010}]{2010MNRAS.406.2650M}
{Metzger} B.~D.,  et~al., 2010, \mn@doi [\mnras]
  {10.1111/j.1365-2966.2010.16864.x}, \href
  {https://ui.adsabs.harvard.edu/abs/2010MNRAS.406.2650M} {406, 2650}

\bibitem[\protect\citeauthoryear{{Mizuta} \& {Ioka}}{{Mizuta} \&
  {Ioka}}{2013}]{2013ApJ...777..162M}
{Mizuta} A.,  {Ioka} K.,  2013, \mn@doi [\apj] {10.1088/0004-637X/777/2/162},
  \href {http://adsabs.harvard.edu/abs/2013ApJ...777..162M} {777, 162}

\bibitem[\protect\citeauthoryear{{Mizuta}, {Nagataki}  \& {Aoi}}{{Mizuta}
  et~al.}{2011}]{2011ApJ...732...26M}
{Mizuta} A.,  {Nagataki} S.,   {Aoi} J.,  2011, \mn@doi [\apj]
  {10.1088/0004-637X/732/1/26}, \href
  {https://ui.adsabs.harvard.edu/abs/2011ApJ...732...26M} {732, 26}

\bibitem[\protect\citeauthoryear{{Mooley} et~al.,}{{Mooley}
  et~al.}{2018}]{2018Natur.561..355M}
{Mooley} K.~P.,  et~al., 2018, \mn@doi [\nat] {10.1038/s41586-018-0486-3},
  \href {https://ui.adsabs.harvard.edu/abs/2018Natur.561..355M} {561, 355}

\bibitem[\protect\citeauthoryear{{Morokuma}, {Tominaga}, {Tanaka}, {Sarugaku}
  \& {Kawai}}{{Morokuma} et~al.}{2015}]{2015ATel.7960....1M}
{Morokuma} T.,  {Tominaga} N.,  {Tanaka} M.,  {Sarugaku} Y.,   {Kawai} N.,
  2015, The Astronomer's Telegram, \href
  {https://ui.adsabs.harvard.edu/abs/2015ATel.7960....1M} {7960, 1}

\bibitem[\protect\citeauthoryear{{Morsony}, {Lazzati}  \& {Begelman}}{{Morsony}
  et~al.}{2007}]{2007ApJ...665..569M}
{Morsony} B.~J.,  {Lazzati} D.,   {Begelman} M.~C.,  2007, \mn@doi [\apj]
  {10.1086/519483}, \href {http://adsabs.harvard.edu/abs/2007ApJ...665..569M}
  {665, 569}

\bibitem[\protect\citeauthoryear{{Murguia-Berthier}, {Montes}, {Ramirez-Ruiz},
  {De Colle}  \& {Lee}}{{Murguia-Berthier} et~al.}{2014}]{2014ApJ...788L...8M}
{Murguia-Berthier} A.,  {Montes} G.,  {Ramirez-Ruiz} E.,  {De Colle} F.,
  {Lee} W.~H.,  2014, \mn@doi [\apjl] {10.1088/2041-8205/788/1/L8}, \href
  {https://ui.adsabs.harvard.edu/abs/2014ApJ...788L...8M} {788, L8}

\bibitem[\protect\citeauthoryear{{Nagakura}, {Hotokezaka}, {Sekiguchi},
  {Shibata}  \& {Ioka}}{{Nagakura} et~al.}{2014}]{2014ApJ...784L..28N}
{Nagakura} H.,  {Hotokezaka} K.,  {Sekiguchi} Y.,  {Shibata} M.,   {Ioka} K.,
  2014, \mn@doi [\apjl] {10.1088/2041-8205/784/2/L28}, \href
  {https://ui.adsabs.harvard.edu/abs/2014ApJ...784L..28N} {784, L28}

\bibitem[\protect\citeauthoryear{{Nakar} \& {Piran}}{{Nakar} \&
  {Piran}}{2017}]{2017ApJ...834...28N}
{Nakar} E.,  {Piran} T.,  2017, \mn@doi [\apj] {10.3847/1538-4357/834/1/28},
  \href {https://ui.adsabs.harvard.edu/abs/2017ApJ...834...28N} {834, 28}

\bibitem[\protect\citeauthoryear{{Nakar} \& {Sari}}{{Nakar} \&
  {Sari}}{2010}]{2010ApJ...725..904N}
{Nakar} E.,  {Sari} R.,  2010, \mn@doi [\apj] {10.1088/0004-637X/725/1/904},
  \href {https://ui.adsabs.harvard.edu/abs/2010ApJ...725..904N} {725, 904}

\bibitem[\protect\citeauthoryear{{Nakar} \& {Sari}}{{Nakar} \&
  {Sari}}{2012}]{2012ApJ...747...88N}
{Nakar} E.,  {Sari} R.,  2012, \mn@doi [\apj] {10.1088/0004-637X/747/2/88},
  \href {https://ui.adsabs.harvard.edu/abs/2012ApJ...747...88N} {747, 88}

\bibitem[\protect\citeauthoryear{{Nathanail}, {Gill}, {Porth}, {Fromm}  \&
  {Rezzolla}}{{Nathanail} et~al.}{2020}]{2020MNRAS.495.3780N}
{Nathanail} A.,  {Gill} R.,  {Porth} O.,  {Fromm} C.~M.,   {Rezzolla} L.,
  2020, \mn@doi [\mnras] {10.1093/mnras/staa1454}, \href
  {https://ui.adsabs.harvard.edu/abs/2020MNRAS.495.3780N} {495, 3780}

\bibitem[\protect\citeauthoryear{{Nathanail}, {Gill}, {Porth}, {Fromm}  \&
  {Rezzolla}}{{Nathanail} et~al.}{2021}]{2021MNRAS.502.1843N}
{Nathanail} A.,  {Gill} R.,  {Porth} O.,  {Fromm} C.~M.,   {Rezzolla} L.,
  2021, \mn@doi [\mnras] {10.1093/mnras/stab115}, \href
  {https://ui.adsabs.harvard.edu/abs/2021MNRAS.502.1843N} {502, 1843}

\bibitem[\protect\citeauthoryear{{Nativi}, {Bulla}, {Rosswog}, {Lundman},
  {Kowal}, {Gizzi}, {Lamb}  \& {Perego}}{{Nativi}
  et~al.}{2021}]{2021MNRAS.500.1772N}
{Nativi} L.,  {Bulla} M.,  {Rosswog} S.,  {Lundman} C.,  {Kowal} G.,  {Gizzi}
  D.,  {Lamb} G.~P.,   {Perego} A.,  2021, \mn@doi [\mnras]
  {10.1093/mnras/staa3337}, \href
  {https://ui.adsabs.harvard.edu/abs/2021MNRAS.500.1772N} {500, 1772}

\bibitem[\protect\citeauthoryear{{Nava}, {Sironi}, {Ghisellini}, {Celotti}  \&
  {Ghirlanda}}{{Nava} et~al.}{2013}]{2013MNRAS.433.2107N}
{Nava} L.,  {Sironi} L.,  {Ghisellini} G.,  {Celotti} A.,   {Ghirlanda} G.,
  2013, \mn@doi [\mnras] {10.1093/mnras/stt872}, \href
  {https://ui.adsabs.harvard.edu/abs/2013MNRAS.433.2107N} {433, 2107}

\bibitem[\protect\citeauthoryear{Nicholl et~al.}{Nicholl
  et~al.}{2017}]{Nicholl:2017ahq}
Nicholl M.,  et~al., 2017, \mn@doi [Astrophys. J.] {10.3847/2041-8213/aa9029},
  848, L18

\bibitem[\protect\citeauthoryear{{O'Connor} \& {Troja}}{{O'Connor} \&
  {Troja}}{2022}]{2022GCN.32065....1O}
{O'Connor} B.,  {Troja} E.,  2022, GRB Coordinates Network, \href
  {https://ui.adsabs.harvard.edu/abs/2022GCN.32065....1O} {32065, 1}

\bibitem[\protect\citeauthoryear{{Oke}}{{Oke}}{1974}]{1974ApJS...27...21O}
{Oke} J.~B.,  1974, \mn@doi [\apjs] {10.1086/190287}, \href
  {https://ui.adsabs.harvard.edu/abs/1974ApJS...27...21O} {27, 21}

\bibitem[\protect\citeauthoryear{{Paczynski}}{{Paczynski}}{1986}]{1986ApJ...308L..43P}
{Paczynski} B.,  1986, \mn@doi [\apjl] {10.1086/184740}, \href
  {https://ui.adsabs.harvard.edu/abs/1986ApJ...308L..43P} {308, L43}

\bibitem[\protect\citeauthoryear{{Paczy{\'n}ski}}{{Paczy{\'n}ski}}{1998}]{1998ApJ...494L..45P}
{Paczy{\'n}ski} B.,  1998, \mn@doi [\apjl] {10.1086/311148}, \href
  {https://ui.adsabs.harvard.edu/abs/1998ApJ...494L..45P} {494, L45}

\bibitem[\protect\citeauthoryear{Pian et~al.}{Pian et~al.}{2017}]{Pian:2017gtc}
Pian E.,  et~al., 2017, \mn@doi [Nature] {10.1038/nature24298}, 551, 67

\bibitem[\protect\citeauthoryear{{Piro} \& {Kollmeier}}{{Piro} \&
  {Kollmeier}}{2018}]{2018ApJ...855..103P}
{Piro} A.~L.,  {Kollmeier} J.~A.,  2018, \mn@doi [\apj]
  {10.3847/1538-4357/aaaab3}, \href
  {https://ui.adsabs.harvard.edu/abs/2018ApJ...855..103P} {855, 103}

\bibitem[\protect\citeauthoryear{{Piro}, {Haynie}  \& {Yao}}{{Piro}
  et~al.}{2021}]{2021ApJ...909..209P}
{Piro} A.~L.,  {Haynie} A.,   {Yao} Y.,  2021, \mn@doi [\apj]
  {10.3847/1538-4357/abe2b1}, \href
  {https://ui.adsabs.harvard.edu/abs/2021ApJ...909..209P} {909, 209}

\bibitem[\protect\citeauthoryear{{Preau}, {Ioka}  \&
  {M{\'e}sz{\'a}ros}}{{Preau} et~al.}{2021}]{2021MNRAS.503.2499P}
{Preau} E.,  {Ioka} K.,   {M{\'e}sz{\'a}ros} P.,  2021, \mn@doi [\mnras]
  {10.1093/mnras/stab652}, \href
  {https://ui.adsabs.harvard.edu/abs/2021MNRAS.503.2499P} {503, 2499}

\bibitem[\protect\citeauthoryear{{Radice}, {Galeazzi}, {Lippuner}, {Roberts},
  {Ott}  \& {Rezzolla}}{{Radice} et~al.}{2016}]{2016MNRAS.460.3255R}
{Radice} D.,  {Galeazzi} F.,  {Lippuner} J.,  {Roberts} L.~F.,  {Ott} C.~D.,
  {Rezzolla} L.,  2016, \mn@doi [\mnras] {10.1093/mnras/stw1227}, \href
  {https://ui.adsabs.harvard.edu/abs/2016MNRAS.460.3255R} {460, 3255}

\bibitem[\protect\citeauthoryear{{Radice}, {Perego}, {Hotokezaka}, {Fromm},
  {Bernuzzi}  \& {Roberts}}{{Radice} et~al.}{2018}]{2018ApJ...869..130R}
{Radice} D.,  {Perego} A.,  {Hotokezaka} K.,  {Fromm} S.~A.,  {Bernuzzi} S.,
  {Roberts} L.~F.,  2018, \mn@doi [\apj] {10.3847/1538-4357/aaf054}, \href
  {https://ui.adsabs.harvard.edu/abs/2018ApJ...869..130R} {869, 130}

\bibitem[\protect\citeauthoryear{{Ramirez-Ruiz} \&
  {Lloyd-Ronning}}{{Ramirez-Ruiz} \&
  {Lloyd-Ronning}}{2002}]{2002NewA....7..197R}
{Ramirez-Ruiz} E.,  {Lloyd-Ronning} N.~M.,  2002, \mn@doi [\na]
  {10.1016/S1384-1076(02)00106-9}, \href
  {http://adsabs.harvard.edu/abs/2002NewA....7..197R} {7, 197}

\bibitem[\protect\citeauthoryear{{Roming} et~al.,}{{Roming}
  et~al.}{2005}]{2005SSRv..120...95R}
{Roming} P. W.~A.,  et~al., 2005, \mn@doi [\ssr] {10.1007/s11214-005-5095-4},
  \href {https://ui.adsabs.harvard.edu/abs/2005SSRv..120...95R} {120, 95}

\bibitem[\protect\citeauthoryear{{Rouco Escorial} et~al.,}{{Rouco Escorial}
  et~al.}{2022}]{2022arXiv221005695R}
{Rouco Escorial} A.,  et~al., 2022, arXiv e-prints, \href
  {https://ui.adsabs.harvard.edu/abs/2022arXiv221005695R} {p. arXiv:2210.05695}

\bibitem[\protect\citeauthoryear{{Ruderman}}{{Ruderman}}{1975}]{1975NYASA.262..164R}
{Ruderman} M.,  1975, in {Bergman} P.~G.,  {Fenyves} E.~J.,   {Motz} L.,  eds,
  Vol. 262, Seventh Texas Symposium on Relativistic Astrophysics. pp 164--180,
  \mn@doi{10.1111/j.1749-6632.1975.tb31430.x}

\bibitem[\protect\citeauthoryear{{Rybicki} \& {Lightman}}{{Rybicki} \&
  {Lightman}}{1979}]{1979rpa..book.....R}
{Rybicki} G.~B.,  {Lightman} A.~P.,  1979, {Radiative processes in
  astrophysics}

\bibitem[\protect\citeauthoryear{{Sagiv} et~al.,}{{Sagiv}
  et~al.}{2014}]{2014AJ....147...79S}
{Sagiv} I.,  et~al., 2014, \mn@doi [\aj] {10.1088/0004-6256/147/4/79}, \href
  {https://ui.adsabs.harvard.edu/abs/2014AJ....147...79S} {147, 79}

\bibitem[\protect\citeauthoryear{{Savaglio}, {Glazebrook}  \& {Le
  Borgne}}{{Savaglio} et~al.}{2009}]{2009ApJ...691..182S}
{Savaglio} S.,  {Glazebrook} K.,   {Le Borgne} D.,  2009, \mn@doi [\apj]
  {10.1088/0004-637X/691/1/182}, \href
  {http://adsabs.harvard.edu/abs/2009ApJ...691..182S} {691, 182}

\bibitem[\protect\citeauthoryear{{Scheuer}}{{Scheuer}}{1974}]{1974MNRAS.166..513S}
{Scheuer} P.~A.~G.,  1974, \mn@doi [\mnras] {10.1093/mnras/166.3.513}, \href
  {https://ui.adsabs.harvard.edu/abs/1974MNRAS.166..513S} {166, 513}

\bibitem[\protect\citeauthoryear{{Schmidt}}{{Schmidt}}{1978}]{1978Natur.271..525S}
{Schmidt} W.~K.~H.,  1978, \mn@doi [\nat] {10.1038/271525a0}, \href
  {https://ui.adsabs.harvard.edu/abs/1978Natur.271..525S} {271, 525}

\bibitem[\protect\citeauthoryear{{Serino} et~al.,}{{Serino}
  et~al.}{2014}]{2014PASJ...66...87S}
{Serino} M.,  et~al., 2014, \mn@doi [\pasj] {10.1093/pasj/psu063}, \href
  {https://ui.adsabs.harvard.edu/abs/2014PASJ...66...87S} {66, 87}

\bibitem[\protect\citeauthoryear{Shappee et~al.}{Shappee
  et~al.}{2017}]{Shappee:2017zly}
Shappee B.~J.,  et~al., 2017, \mn@doi [Science] {10.1126/science.aaq0186}, 358,
  1574

\bibitem[\protect\citeauthoryear{{Shibata}}{{Shibata}}{1999}]{1999PhRvD..60j4052S}
{Shibata} M.,  1999, \mn@doi [\prd] {10.1103/PhysRevD.60.104052}, \href
  {https://ui.adsabs.harvard.edu/abs/1999PhRvD..60j4052S} {60, 104052}

\bibitem[\protect\citeauthoryear{{Shibata} \& {Ury{\= u}}}{{Shibata} \& {Ury{\=
  u}}}{2000}]{2000PhRvD..61f4001S}
{Shibata} M.,  {Ury{\= u}} K.~{\= o}.,  2000, \mn@doi [\prd]
  {10.1103/PhysRevD.61.064001}, \href
  {https://ui.adsabs.harvard.edu/abs/2000PhRvD..61f4001S} {61, 064001}

\bibitem[\protect\citeauthoryear{Smartt et~al.}{Smartt
  et~al.}{2017}]{Smartt:2017fuw}
Smartt S.~J.,  et~al., 2017, \mn@doi [Nature] {10.1038/nature24303}, 551, 75

\bibitem[\protect\citeauthoryear{Soares-Santos et~al.}{Soares-Santos
  et~al.}{2017}]{Soares-Santos:2017lru}
Soares-Santos M.,  et~al., 2017, \mn@doi [Astrophys. J.]
  {10.3847/2041-8213/aa9059}, 848, L16

\bibitem[\protect\citeauthoryear{{Tanaka} et~al.,}{{Tanaka}
  et~al.}{2017}]{2017PASJ...69..102T}
{Tanaka} M.,  et~al., 2017, \mn@doi [\pasj] {10.1093/pasj/psx121}, \href
  {https://ui.adsabs.harvard.edu/abs/2017PASJ...69..102T} {69, 102}

\bibitem[\protect\citeauthoryear{{Tanaka}, {Kato}, {Gaigalas}  \&
  {Kawaguchi}}{{Tanaka} et~al.}{2020}]{2020MNRAS.496.1369T}
{Tanaka} M.,  {Kato} D.,  {Gaigalas} G.,   {Kawaguchi} K.,  2020, \mn@doi
  [\mnras] {10.1093/mnras/staa1576}, \href
  {https://ui.adsabs.harvard.edu/abs/2020MNRAS.496.1369T} {496, 1369}

\bibitem[\protect\citeauthoryear{{Uchida} et~al.,}{{Uchida}
  et~al.}{2014}]{2014GCN.16686....1U}
{Uchida} D.,  et~al., 2014, GRB Coordinates Network, \href
  {https://ui.adsabs.harvard.edu/abs/2014GCN.16686....1U} {16686, 1}

\bibitem[\protect\citeauthoryear{Utsumi et~al.}{Utsumi
  et~al.}{2017}]{Utsumi:2017cti}
Utsumi Y.,  et~al., 2017, \mn@doi [Publ. Astron. Soc. Jap.]
  {10.1093/pasj/psx118}, 69, 101

\bibitem[\protect\citeauthoryear{Valenti et~al.,}{Valenti
  et~al.}{2017}]{Valenti:2017ngx}
Valenti S.,  et~al., 2017, \mn@doi [Astrophys. J.] {10.3847/2041-8213/aa8edf},
  848, L24

\bibitem[\protect\citeauthoryear{{Wanajo}, {Sekiguchi}, {Nishimura}, {Kiuchi},
  {Kyutoku}  \& {Shibata}}{{Wanajo} et~al.}{2014}]{2014ApJ...789L..39W}
{Wanajo} S.,  {Sekiguchi} Y.,  {Nishimura} N.,  {Kiuchi} K.,  {Kyutoku} K.,
  {Shibata} M.,  2014, \mn@doi [\apjl] {10.1088/2041-8205/789/2/L39}, \href
  {https://ui.adsabs.harvard.edu/abs/2014ApJ...789L..39W} {789, L39}

\bibitem[\protect\citeauthoryear{{Woosley}}{{Woosley}}{1993}]{1993ApJ...405..273W}
{Woosley} S.~E.,  1993, \mn@doi [\apj] {10.1086/172359}, \href
  {http://adsabs.harvard.edu/abs/1993ApJ...405..273W} {405, 273}

\bibitem[\protect\citeauthoryear{{Yamazaki}, {Ioka}  \& {Nakamura}}{{Yamazaki}
  et~al.}{2002}]{2002ApJ...571L..31Y}
{Yamazaki} R.,  {Ioka} K.,   {Nakamura} T.,  2002, \mn@doi [\apjl]
  {10.1086/341225}, \href
  {https://ui.adsabs.harvard.edu/abs/2002ApJ...571L..31Y} {571, L31}

\bibitem[\protect\citeauthoryear{{Yonetoku} et~al.,}{{Yonetoku}
  et~al.}{2020}]{2020SPIE11444E..2ZY}
{Yonetoku} D.,  et~al., 2020, in Society of Photo-Optical Instrumentation
  Engineers (SPIE) Conference Series. p. 114442Z, \mn@doi{10.1117/12.2560603}

\bibitem[\protect\citeauthoryear{{Yuan}, {Zhang}, {Chen}  \& {Ling}}{{Yuan}
  et~al.}{2022}]{2022arXiv220909763Y}
{Yuan} W.,  {Zhang} C.,  {Chen} Y.,   {Ling} Z.,  2022, \mn@doi [arXiv
  e-prints] {10.48550/arXiv.2209.09763}, \href
  {https://ui.adsabs.harvard.edu/abs/2022arXiv220909763Y} {p. arXiv:2209.09763}

\bibitem[\protect\citeauthoryear{{Zhu} et~al.,}{{Zhu}
  et~al.}{2023}]{2023ApJ...942...88Z}
{Zhu} J.-P.,  et~al., 2023, \mn@doi [\apj] {10.3847/1538-4357/aca527}, \href
  {https://ui.adsabs.harvard.edu/abs/2023ApJ...942...88Z} {942, 88}

\makeatother
\end{thebibliography}



\appendix

\section{The cocoon in numerical simulations}
\label{ap:cocoon}
\subsection{Identification of the cocoon fluid in numerical simulations}
As explained in \cite{2023MNRAS.520.1111H}, the following set of requirements has been used to categorize a certain fluid element as a ``cocoon" fluid element:
\begin{equation}
          \Gamma_{inf} < 10,\ 
          \Gamma < 5,\ 
          |\beta_\theta| > 0,\ 
          \rho > \rho_{CSM},
        \label{eq:cc condition}
\end{equation}
where $\Gamma_{} =(1-\beta^2)^{-1/2}$ is the Lorentz factor, $\beta$ is the velocity, $\Gamma_{inf}$ is the maximally achievable Lorentz factor at infinity (see Appendix \ref{ap:Bernoulli equation} for the definition of $\Gamma_{inf}$), $\beta_{\theta}$ is the angular velocity (in polar coordinates with $\theta$ as the angle measured from the polar axis, which is also the jet axis), and $\rho_{CSM}$ is the CSM density.

Here, the choice of $\Gamma_{inf}=10$, as the maximum Lorentz factor of the cocoon, is so that $\theta \sim 1/\Gamma_{}$ is of the order of typical GRB-jet opening angles. This is a very simplistic separations between the jet and the cocoon.

\subsection{Properties of the post-breakout cocoon}
\label{ap:post-breakout cocoon}
\subsubsection{Bernoulli equation and $\beta_{inf}$}
\label{ap:Bernoulli equation}
The maximum Lorentz factor is estimated by using Bernoulli equation
\begin{eqnarray}
    \Gamma_{inf} \approx h \Gamma ,
    \label{eq:Bernoulli Gamma}
\end{eqnarray}
where $h$ is the enthalpy (see \citealt{2023MNRAS.520.1111H} for more details about this approximation).
The corresponding velocity is $\beta_{inf}=\sqrt{1-1/\Gamma_{inf}^{2}}$.

\subsubsection{Energy distribution}
\label{ap:dEc sim}
\begin{figure}
    \centering
    \includegraphics[width=0.99\linewidth]{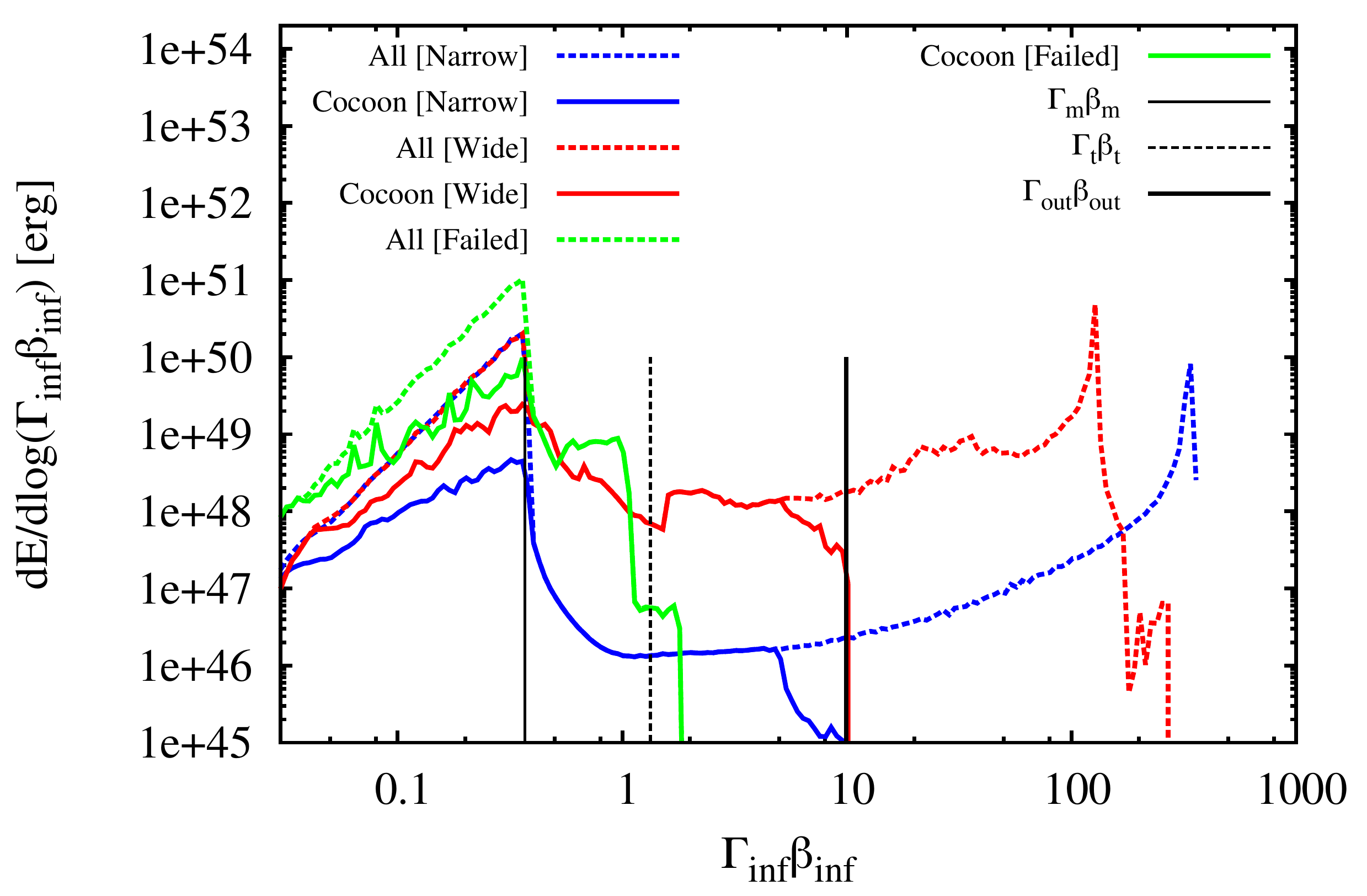}
  \caption{Total energy (kinetic + internal) distribution over logarithmic bins of the maximum four-velocity $\Gamma_{inf}\beta_{inf}$, for the three different models (narrow, wide and failed; in blue, red, and green; respectively) in the free-expansion phase ($t_1$) [see footnote \ref{foot:dE univ}].
  Dashed colored lines show all the outflow in the computational domain (jet, cocoon, and ejecta).
  Solid colored lines show only the cocoon component (respectively).
  There are three illustrative black vertical lines; 
  the thin solid line shows the edge of the ejecta ($\Gamma_m\beta_m$),
  the dashed line shows the separation between non-relativistic and relativistic cocoons ($\Gamma_t\beta_t$), 
  and the thick solid line shows the maximum velocity of the cocoon ($\Gamma_{out}\beta_{out}$) [see equations (\ref{eq:beta_t R NR cases}) and (\ref{eq:beta_t})].
  }
  \label{fig:dEc sim} 
\end{figure}

Figure \ref{fig:dEc sim} shows the distribution of energy (kinetic + internal) as a function of the maximum four-velocity $\Gamma_{inf}\beta_{inf}$ throughout the computational domain (highlighting the cocoon), at the approximate start of the free-expansion phase $t=t_1$ ($t_1 -t_0\sim 3$ s for successful jet models, and $t_1 -t_0 \sim 10$ s for the failed jet model; see \citealt{2023MNRAS.520.1111H})\footnote{\label{foot:dE univ}As the free-expansion phase is reached, no energy exchange is expected between different expanding shells.
As fluid elements are expected to converge to their maximum velocities $\Gamma_{inf}\beta_{inf}$, this distribution is not subject to change significantly.}.
This data was extracted by equally dividing the maximum four-velocity (logarithmically) making $d(\Gamma_{inf}\beta_{inf})$ numerical bins, and integrating the fluid energy throughout the computational domain over these bins.

The nature of the cocoon as an intermediate component, between the non-relativistic ejecta ($\beta_{inf}\le\beta_{m}$) and the highly relativistic jet ($\Gamma_{inf}>10$), is apparent. 
Also, the two parts of the cocoon, the non-relativistic cocoon and the relativistic cocoon [see equation (\ref{eq:beta_t R NR cases})], 
can be seen having distinguishable energetic properties across $\beta_t$ (also see Section \ref{sec:den sim}).

From Figure \ref{fig:dEc sim} it can be seen that the jet's energy is spread widely ($\Gamma_{inf}\beta_{inf} \sim 10- 100$) overlapping the relativistic cocoon at its lower end.
This spread is physical, accounting for the shocked jet, while the peak at the extreme right accounts for the unshocked jet.
However, as shown in \cite{2013ApJ...777..162M}, at the spatial boundary between the jet and the cocoon, numerical diffusion is inevitable, causing the formation of artificially baryon-polluted portions at the spatial edge of the jet (in contact with the cocoon).
Therefore, 
this numerical artifact may have contributed to enhance the cocoon's energy (internal energy in particular) with such polluted jets (also see \citealt{2021MNRAS.503.2499P} for neutron diffusion).
However, compared to 3D simulations where hydrodynamical jets are much less stable (Rayleigh-Taylor instability) and cocoon energies are enhanced (\citealt{2019MNRAS.490.4271M}; \citealt{2021MNRAS.500.3511G}), we estimate that our 2D simulations give a conservative evaluation of the cocoon energy (roughly within a factor two).
Hence, despite the numerical challenges, our numerical results should reflect the overall energetic nature of the cocoon conservatively.

\subsubsection{Velocity profile}
\label{ap:v sim}
\begin{figure}
    \centering
    \includegraphics[width=0.99\linewidth]{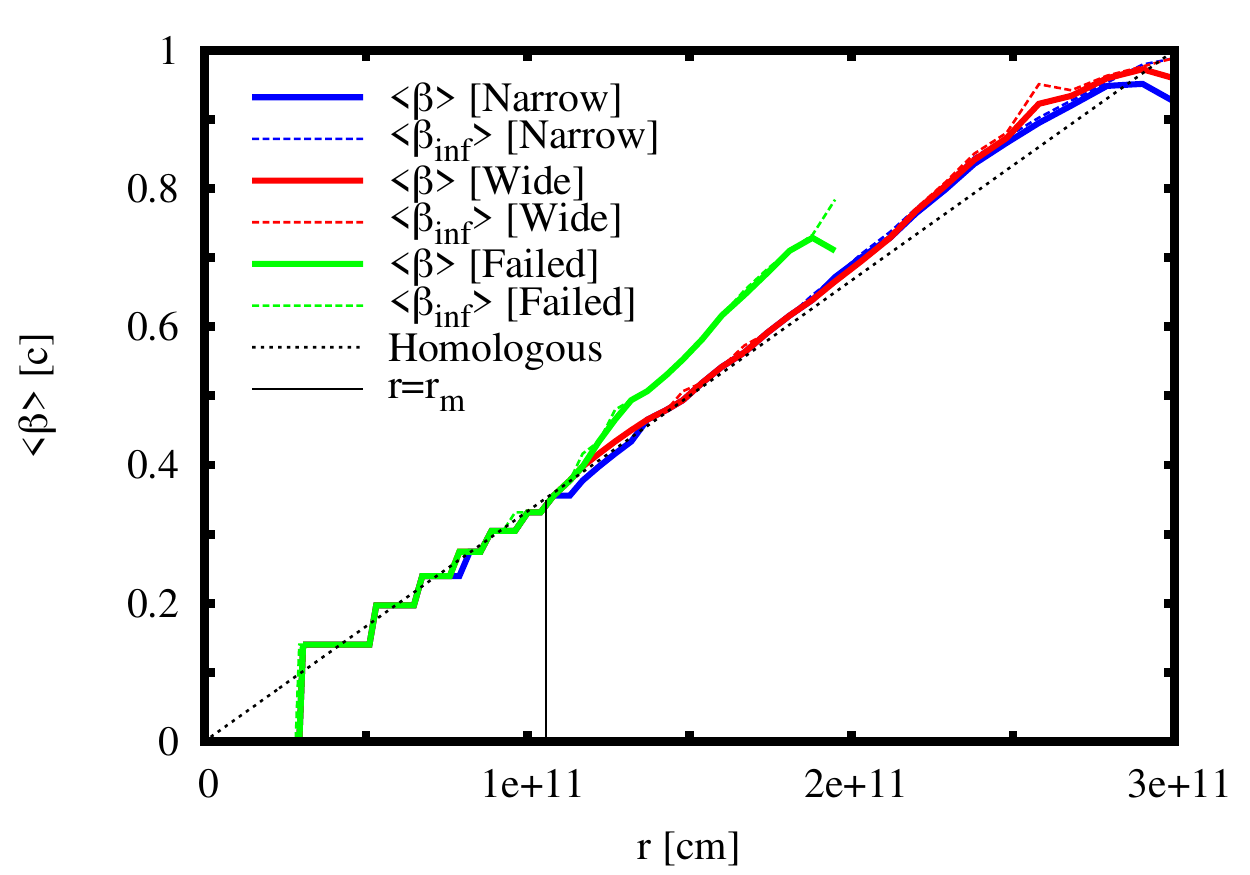} 
  \caption{Profiles of the cocoon velocity $\beta$ (solid colored lines), and the cocoon maximum velocity $\beta_{inf}$ (dashed colored lines), after averaging over logarithmically spaced radial bins [see equations (\ref{eq:gamma av}) and (\ref{eq:gamma inf av})]. 
  Three jet models [narrow (blue), wide (red), and failed (green)] are shown at the end of each simulation (all at $t-t_0=10$ s; in the laboratory frame).
  The edge of the ejecta is shown with a black vertical line [$r_m = c(t-t_m)\beta_m$, and $\beta_m =0.35$].
  The expected velocity profile for the case of a homologous expansion is shown (black dashed line), indicating that the cocoon can be approximated using $\beta_{inf}\approx\beta$ and $r\propto \beta$ [see equation (\ref{eq:av homologous})].
  }
  \label{fig:v sim} 
\end{figure}

In Figure \ref{fig:v sim}, the velocity profile of the cocoon in numerical simulations is shown (colored solid lines).
These velocities were calculated following several steps.
First, the computational radius $r$ was divided (following a uniform logarithmic scale) into bins $dr$.
Then, hydrodynamical properties of the cocoon (mass, volume, energy, etc.) were integrated for each bin.
The average Lorentz factor for a given bin is found using
\begin{eqnarray}
    \langle{\Gamma}\rangle =\frac{\int_r^{r+dr} dE_{c,k}}{\int_r^{r+dr} dM_c c^2} +1,
    \label{eq:gamma av}
\end{eqnarray}
where $dE_{c,k}$ is the kinetic energy of a given cocoon fluid element (located between $r$ and $r+dr$), and $dM_c$ is its mass.
Finally, average velocities are then found using $\langle{\beta}\rangle =\sqrt{1-1/\langle{\Gamma^2}\rangle}$.

For the maximum velocity $\beta_{inf}$ (dashed colored lines), the same procedure was followed, and the maximum Lorentz factor was found using
\begin{eqnarray}
    \langle{\Gamma_{inf}}\rangle =\frac{\int_r^{r+dr} dE_{c}}{\int_r^{r+dr} dM_c c^2} +1,
    \label{eq:gamma inf av}
\end{eqnarray}
where $dE_c$ is the total energy of a given cocoon fluid element (excluding the rest-mass energy).
Maximum velocities (averaged) are found as $\langle{\beta_{inf}}\rangle =\sqrt{1-1/\langle{\Gamma_{inf}^2}\rangle}$.

The profiles of these velocities show that, as $t\gg t_b$: i) cocoon fluid elements reach their maximum velocities $\beta\sim \beta_{inf}$; and ii) the cocoon expansion can be approximated as homologous
\begin{eqnarray}
\langle{\beta}\rangle \equiv \langle{\beta_{inf}}\rangle \approx r/[c(t-t_m)],
\label{eq:av homologous}
\end{eqnarray}
where $c$ is the speed of light.
Therefore, at late times, $\beta_{inf}$, $\beta$, and $r$ can be used interchangeably [see equation (\ref{eq:r=cbt})].

Since $\beta$ (and $\Gamma$) does not vary with time (i.e., free-expansion; unlike $r$), in the following, it will be used as a coordinate system so that each expanding cocoon shell is labeled by its velocity $\beta$ (or by $\Gamma$).

\section{Finding the time evolution of the photosphere}
\label{ap:photosphere}
In the following, we introduce reasonable approximations at the early and late time limits, so that the time evolution of $\Gamma_{ph}\beta_{ph}$ can be approximated to simple power-law functions.

\subsection{Relativistic cocoon}
We determine the time evolution by taking $l=0$, and by neglecting the term $\Gamma_{out}$ in equation (\ref{eq:tobs R}) [since $\Gamma_{ph}^{-6}\gg\Gamma_{out}^{-6}$], as
\begin{eqnarray}
\Gamma_{ph}(t_{obs})\approx \Gamma_t\left\{\frac{t_{obs}}{t_{obs}(\beta_{ph}=\beta_{t})}\right\}^{-\frac{1}{3}}.
\label{eq:Gamma ph R simple}
\end{eqnarray}
Using $\Gamma \approx [2(1-\beta)]^{-1/2}$ [see equation (\ref{eq:U4 R NR cases})], the photospheric velocity is $\beta_{ph}(t_{obs})\approx 1-(1-\beta_{t})\left\{\frac{t_{obs}}{t_{obs}(\beta_{ph}=\beta_{t})}\right\}^{2/3} $, where $t_{obs}(\beta_{ph}=\beta_{t})\equiv t_{obs}(\Gamma_{ph}=\Gamma_{t})$ can be found from equation (\ref{eq:tobs R}) using $\Gamma=\Gamma_t$ and $\tau_r=1$.

\subsection{Non-relativistic cocoon}
\label{ap:beta ph NR}

\subsubsection{Failed jet case}
\label{ap:ph failed}
In the failed jet case, without a relativistic cocoon component, 
the non-relativistic cocoon is observable from early times to late times.
Hence, there are two limits to consider.
First, at the early time limit, 
\begin{eqnarray}
\Gamma_{ph}\beta_{ph}\sim \Gamma_{t}\beta_t\sim \text{Const.}\propto t_{obs}^0,
\end{eqnarray}
as the photosphere is initially stuck (in the Lagrangian coordinate) at the outer edge of the non-relativistic cocoon (see $\Gamma_{ph}\beta_{ph}$ for the failed jet model in Figure \ref{fig:NWF L v T}).
Second, at the late time limit, i.e., the non-relativistic limit as $\beta_{ph}\sim \beta_m$, one has $\Gamma_{ph}\sim 1$ and $\Gamma_{ph}\beta_{ph}\sim \beta_{ph}$ [see equation (\ref{eq:U4 R NR cases})].
Using equations (\ref{eq:tobs NR}) and (\ref{eq:tau ph}) gives $t_{obs}(\beta_{ph})\propto {(1-\beta_{ph})}{\beta_{ph}^{-\frac{m-1}{2}}}$, so that
$t_{obs}(\beta_{ph})\propto {\beta_{ph}^{-\frac{m-1}{2}}}$ (for $m\approx 8\gg 1$) at this limit.
Hence,
\begin{eqnarray}
\beta_{ph}(t_{obs})\approx \beta_m\left\{\frac{t_{obs}}{t_{obs}(\beta_{ph}=\beta_m)}\right\}^{-\frac{2}{7}},
\label{eq:beta ph NR simple}
\end{eqnarray}
where $t_{obs}(\beta_{ph}=\beta_{m})$ can be found from equations (\ref{eq:tobs NR}) with $\beta=\beta_m$ and $\tau_{nr}=1$.

\subsubsection{Successful jet case}
\label{ap:ph successful}
In this case, as the relativistic cocoon covers the non-relativistic cocoon at early times, the non-relativistic cocoon comes into play only at late times.
Thus the temporal evolution of the phosphoric velocity is the same as equation (\ref{eq:beta ph NR simple}).


\section{Derivation of bolometric isotropic luminosities}
\label{ap:Lbl}
The luminosity is determined by the time derivative of the internal energy.
The total time derivative of the internal energy consists of two parts:
\begin{equation}
\begin{split}
    &\frac{d E_{c,i}(<\Gamma_{d}\beta_{d},t)}{dt}=\\&\frac{\partial E_{c,i}(<\Gamma_{d}\beta_{d},t)}{\partial t}
    +\frac{\partial E_{c,i}(<\Gamma_{d}\beta_{d},t)}{\partial (\Gamma_{d}\beta_{d})}
    \frac{\partial (\Gamma_{d}\beta_{d})}{\partial t}.
\end{split}
\end{equation}
The first term on the right hand side is evaluated with a diffusion radius fixed in the Lagrangian mass coordinate
(i.e., $\Gamma_{d}\beta_{d}=\rm{Const.}$).
Therefore, this term does not contribute to the luminosity because the emission comes only from the exterior of the diffusion surface (which is considered as sharp in our approximation; see Section \ref{sec:cc emission basics}).
This term represents the adiabatic cooling 
${\partial E_{c,i}(<\Gamma_{d}\beta_{d},t)}/{\partial t} = -E_{c,i}/t$ of internal energy [see equation (\ref{eq:Adiabatic 1/t})]
and the r-process heating. 
The second term does correspond to the luminosity because this term arises from the (inward) motion of the diffusion surface (in terms of velocity $\Gamma\beta$).
Therefore, under our approximation of sharp diffusion radius, the second term accounts for the photon radiation (radiated power or luminosity; although we are not taking into account observed timescales and beaming effects yet) from the internal energy stored in the cocoon (this argument can be extended to the relativistic limit; see footnote \ref{foot:L=L'}):

\begin{equation}
L = -\frac{\partial E_{c,i}(<\Gamma_d\beta_d,t)}{\partial (\Gamma_d \beta_d)}\frac{\partial (\Gamma_d \beta_d) }{\partial t}.
\label{eq:partial L first}
\end{equation}
Using $E_{c,i}(t) = E_{c,i}(<\Gamma_d\beta_d,t) + E_{c,i}(>\Gamma_d\beta_d,t)$ [with $E_{c,i}(t)$ only varying with $t$; see Section \ref{sec:Eint}], 
we have $-\frac{\partial E_{c,i}(<\Gamma_d\beta_d,t)}{\partial (\Gamma_d \beta_d)} =\frac{\partial E_{c,i}(>\Gamma_d\beta_d,t)}{\partial (\Gamma_d\beta_d)}$ and an alternative expression as:
\begin{equation}
L = \frac{\partial E_{c,i}(>\Gamma_d\beta_d,t)}{\partial (\Gamma_d\beta_d)}\frac{\partial (\Gamma_d \beta_d) }{\partial t}.
\label{eq:partial L}
\end{equation}

\subsection{Luminosity from jet-shock heating $L_{bl}^j$}
\label{ap:Lbl j}
As presented in equation (\ref{eq:Lbl tot}) the term $L_{bl}^{j}(t_{obs})$ is used to refer to the cooling emission, powered by the internal energy produced during jet propagation after taking into account adiabatic cooling.

Accounting for the difference between the observed time and the laboratory time $\frac{d t}{d t_{obs}}$, and for the beaming of the emission $\frac{4\pi}{\Omega_{EM}}$, the observed isotropic luminosity can be found as [using equations (\ref{eq:U4 R NR cases}) and (\ref{eq:partial L})]
\begin{equation}
\begin{split}
L_{bl}^j(t_{obs})=
\begin{cases}
\frac{\partial E_{c,i,r}^j(>\Gamma_d,t)}{\partial \Gamma_d }\frac{\partial \Gamma_d }{\partial t}\times\frac{dt_{}}{dt_{obs}}\times\frac{4\pi}{\Omega_{EM}} \\ \text{(Rela.  cocoon)},\\
\\
\frac{\partial E_{c,i,nr}^j(>\beta_d,t)}{\partial  \beta_d}\frac{\partial \beta_d }{\partial t}\times\frac{dt_{}}{dt_{obs}}\times\frac{4\pi}{\Omega_{EM}} \\
\text{(N. Rela.  cocoon)}.
\end{cases}
\end{split}
\label{eq:Lbl j}
\end{equation}
The terms $\partial \Gamma_{d}/\partial t$,
$\partial \beta_{d}/\partial t$ and $dt/dt_{obs}$ will be deduced in Appendix \ref{ap:deriv}.
The solid angle of the emission coming from the shell moving with $\beta_d$ can be found as
\begin{equation}
    \Omega_{EM} \approx 4\pi(1-\cos{\theta_{EM}}),
    \label{eq:Omega_EM}
\end{equation}
where the opening angle of the emission is
\begin{equation}
\theta_{EM} \approx \max{\{\arcsin[1/\Gamma(\beta_d)],\theta_{c}^{es}\}}.
    \label{eq:theta_EM}
\end{equation}
Note that the term $\frac{4\pi}{\Omega_{EM}}$ varies in two phases.
Early on, for $\Gamma_d>1/\theta_c^{es}$, this term is constant as $\sim \frac{4\pi}{\Omega}$ [see equations (\ref{eq:Omega}) and (\ref{eq:Omega_EM})].
At later times, once $\Gamma_d<1/\theta_c^{es}$, this term decreases roughly as $\propto \Gamma_d^2$ (approximating $\sin\theta_{EM}\sim \theta_{EM}$), and at the non-relativistic limit, it converges to $1$ (as $\Gamma_d=1$).

\subsection{Luminosity from r-process heating $L_{bl}^{rp}(t_{obs})$}
\label{ap:Lbl rp}
As presented in equation (\ref{eq:Lbl tot}), the second contributor to luminosity is r-process heating ($L_{bl}^{rp}$). 
For r-process heating, there are two terms to consider:
\begin{eqnarray}
L_{bl}^{rp}(t_{obs})
=  L_{bl}^{rp}(<\Gamma_d\beta_d,t_{obs}) + L_{bl}^{rp}(>\Gamma_d\beta_d,t_{obs}).
\label{eq:Lbl rp}
\end{eqnarray}
The first term $L_{bl}^{rp}(<\Gamma_d\beta_d,t_{obs})$ accounts for the heat that has been stored since the merger up until the time $t$, and is released as the diffusion shell moves inward (similarly to $L_{bl}^{j}$ although the source of internal energy is different).
Following the same argument as in Appendix \ref{ap:Lbl j} [see equation (\ref{eq:Lbl j})], this luminosity can be found as
\begin{equation}
\begin{split}
L_{bl}^{rp}(<\Gamma_d\beta_d,t_{obs})=
\begin{cases}
\frac{\partial E_{c,i,r}^{rp}(>\Gamma_d,t)}{\partial (\Gamma_d )}\frac{\partial (\Gamma_d) }{\partial t}\times\frac{dt_{}}{dt_{obs}}\times\frac{4\pi}{\Omega_{EM}} \\ \text{(Rela.  cocoon)},\\
\\
\frac{\partial E_{c,i,nr}^{rp}(>\beta_d,t)}{\partial ( \beta_d)}\frac{\partial ( \beta_d) }{\partial t}\times\frac{dt_{}}{dt_{obs}}\times\frac{4\pi}{\Omega_{EM}} \\
\text{(N. Rela.  cocoon)}.
\end{cases}
\end{split}
\label{eq:Lbl rp in}    
\end{equation}
The second term $L_{bl}^{rp}(>\Gamma_d\beta_d,t_{obs})$ accounts for the internal energy deposited in the diffusive region of the cocoon.
In reality, in the outer regions of the diffusion radius ($\beta\gtrsim \beta_d$) with $\tau(\beta) < 1$, the r-process deposited energy is not fully thermalized. However, as the r-process luminosity is very subdominant in the relativistic cocoon part (by about two orders of magnitude; relative to $L_{bl}^{j}$), and as the density profile is very steep in the non-relativistic cocoon part ($m\gg 1$), in both cases, the simplification that it is thermalized is reasonable\footnote{The fraction of this non-fully thermalized r-process-powered luminosity [defined in equation (\ref{eq:Lbl rp out})] can be found [from equations (\ref{eq:Lbl R rp S}) and (\ref{eq:Lbl NR rp S})] as accounting for, at most, $\sim 1/2$, and $\sim (4-k)/(2-k)\sim 1/4$ of the total r-process-powered luminosity $L_{bl}^{rp}$ of the relativistic and non-relativistic (at later times and for $m\gg 1$) cocoons, respectively.}. This luminosity can be found as:
\begin{equation}
\begin{split}
L_{bl}^{rp}(>\Gamma_d\beta_d,t_{obs})=
\begin{cases}
\dot{E}_{c,i,r}^{rp}(>\Gamma_d,t_{obs})\times \frac{dt}{dt_{obs}}\times\frac{4\pi}{\Omega_{EM}} \\ \text{(Rela.  cocoon)},\\
\\
\dot{E}_{c,i,nr}^{rp}(>\beta_d,t_{obs})\times \frac{dt}{dt_{obs}}\times\frac{4\pi}{\Omega_{EM}} \\
\text{(N. Rela.  cocoon)}.
\end{cases}
\end{split}
\label{eq:Lbl rp out}
\end{equation}
The terms $\partial \Gamma_{d}/\partial t$,
$\partial \beta_{d}/\partial t$ and $dt/dt_{obs}$ will be deduced in Appendix \ref{ap:deriv}.

As a remark, $\theta_{EM}$ has been calculated considering relativistic beaming at $\Gamma_d\beta_d$; this slightly underestimates the cocoon emission originating from r-process heating as the beaming in outer (and faster) regions ($\Gamma\beta>\Gamma_d\beta_d$) has slightly been underestimated (without significant effects on results).

Also, note that, due to the steep density profile of the non-relativistic cocoon (most of the mass is located in the inner region around $\beta_m$) and the small mass of the relativistic cocoon, these luminosities [equations (\ref{eq:Lbl rp in}) and (\ref{eq:Lbl rp out})] are very faint at early times.
Therefore, it is reasonable to find them using late time dependencies.

\subsection{Approximated time and $\Gamma_d\beta_d$ dependencies}
\label{ap:deriv}
Here we evaluate the terms 
$\frac{\partial (\Gamma_d\beta_d)}{\partial t}$ and  $\frac{d t_{}}{d t_{obs}}$ for later use [e.g., in equations (\ref{eq:Lbl j}), (\ref{eq:Lbl rp in}) and (\ref{eq:Lbl rp out})].

In the successful jet case, the relativistic cocoon is observable at early times and the non-relativistic cocoon is observable at late times.
First, considering the relativistic cocoon, equation (\ref{eq:t obs lab}), combined with equations (\ref{eq:tobs R}) [neglecting the $\Gamma_{out}$ term] and (\ref{eq:tau_d R}), one can find time dependencies as $t_{obs}\propto\Gamma^{-3}$ and $t_{}\propto\Gamma^{-1}$ for $\Gamma>\Gamma_t$ (taking $l=0$).
Second, at later times, taking the non-relativistic limit
$\beta_d\sim \beta_m$ and $\Gamma_d\sim 1$,
one can find
$t\propto \beta_d^{-\frac{m-2}{2}}$ and $t_{obs}\propto (1-\beta_d)t$ (for $m\gg 1$)
from equations (\ref{eq:t obs lab}), (\ref{eq:tau NR}), and (\ref{eq:tau_d late NR}).

In the failed jet case, the relativistic cocoon component is absent, and there are two limits as explained in Appendix \ref{ap:beta ph NR}.

First, at early times, the approximation $\beta_d\sim \beta_t\sim \text{Const.}$,  (i.e., $1-\beta_d\sim 1-\beta_t \sim \text{Const.}$) can be used, while $\beta_t-\beta_d \neq \text{Const.}$ is the only fast changing term (only in the failed jet case).
Implementing this approximation in equation (\ref{eq:tau NR}) [with equations (\ref{eq:t obs lab}) and (\ref{eq:tau_d late NR})], one can approximately find that $t_{obs}\propto t\propto (\beta_t-\beta_d) \propto \ln (\beta_t/\beta_d)$,
where the last step relies on Taylor Series;
we used the series $\ln(x)=\sum_{n=1}^{\infty}\left[\frac{(-1)^{n+1}}{n}(x-1)^n\right]$, with $x=\beta_t/\beta_d$, which is valid for $0<{x}<2$, as it is the case here [$0<(\beta_t/\beta_d)<2$], and we only considered the first term ($n=1$).

Second, at late times, the approximations $\beta_d\sim \beta_m$ and $\Gamma_d\sim 1$ give $t\propto \beta_d^{-\frac{m-2}{2}}$ and $t_{obs}\propto (1-\beta_d)t$ (with $m\gg 1$).

With the above $t$, $t_{obs}$, and $\Gamma_d$ (or $\beta_d$) dependencies, the term $\frac{d t_{}}{d t_{obs}}$ can be deduced for both the jet cases as 

\begin{equation}
\begin{split}
&\frac{dt}{dt_{obs}}  
\begin{cases}
\text{Successful jet case:}\\
\approx\frac{1}{3}\frac{t}{t_{obs}} &[\Gamma_d>\Gamma_t] \\
\text{(Rela. cocoon)} ,\\
\\
\approx \frac{\frac{t}{t_{obs}}}{1+\frac{\beta_d}{1-\beta_d}\left(\frac{2}{m-2}\right)^{}}
      \sim  \frac{t}{t_{obs}} &[\text{$\beta_d\gtrsim \beta_m$}] \\
      \text{(N. Rela.  cocoon)} .\\
\end{cases}
\\
\\
&\frac{dt}{dt_{obs}}  
\begin{cases}
\text{Failed jet case:}\\
\approx \frac{t}{t_{obs}} &\left[\beta_d\lesssim \beta_t \right] \\
\text{(N. Rela. cocoon)} ,\\
\\
\approx \frac{\frac{t}{t_{obs}}}{1+\frac{\beta_d}{1-\beta_d}\left(\frac{2}{m-2}\right)^{}}
      \sim  \frac{t}{t_{obs}} &[\text{$\beta_d\gtrsim \beta_m$}] \\
      \text{(N. Rela.  cocoon)} .\\
\end{cases} 
\end{split}
    \label{eq:dt dtobs cases}
\end{equation}
Similarly, using the above $t$ and $\Gamma_d$ (or $\beta_d$) dependencies, the term $\frac{\partial (\Gamma_d\beta_d)}{\partial t}$ is found for both the jet cases as
\begin{equation}
\begin{split}
&\frac{\partial (\Gamma_d\beta_d)}{\partial t} 
\begin{cases}
\text{Successful jet case:}\\
\approx  -\frac{\Gamma_d}{t} &[\Gamma_d>\Gamma_t] \\
\text{(Rela. cocoon)} ,\\
\\
\approx-\frac{\beta_d}{t}\left(\frac{2}{m-2}\right) &[\text{$\beta_d\gtrsim \beta_m$}]\\
\text{(N. Rela.  cocoon)} .\\
\end{cases}
\\
\\
&\frac{\partial (\Gamma_d\beta_d)}{\partial t} 
\begin{cases}
\text{Failed jet case:}\\
\approx \frac{\ln(\beta_t/\beta_d)}{t}\frac{\partial \beta_d}{\partial \ln(\beta_t/\beta_d)} &\left[\beta_d\lesssim \beta_t \right] \\
\text{(N. Rela. cocoon)} ,\\
\\
\approx-\frac{\beta_d}{t}\left(\frac{2}{m-2}\right) &[\text{$\beta_d\gtrsim \beta_m$}]\\
\text{(N. Rela.  cocoon)} .\\
\end{cases} 
\end{split}
        \label{eq:partial Gamma_d beta_d cases}
\end{equation}

Hence, with these derivations [equations (\ref{eq:dt dtobs cases}) and (\ref{eq:partial Gamma_d beta_d cases})] it is possible to analytically solve equations (\ref{eq:Lbl j}), (\ref{eq:Lbl rp in}) and (\ref{eq:Lbl rp out}); 
and find the bolometric luminosities for successful and failed jet cases.
The final results are presented in Section \ref{sec:Lbl}.

\section{The KN model}
\label{ap:KN}
In order to estimate the early KN emission [as shown in Figures \ref{fig:Lbl}, \ref{fig:NWF L v T}, and \ref{fig:NWF mag}], we followed the r-process model by \cite{2015ApJ...802..119K}.
The temperature and bolometric luminosity are estimated using equations (A.16) and (A.17) (for the thin diffusion phase) and equations (A.19) and (A.20) [for the thick diffusion phase; the start of this phase is determined by equation (A.18) and the diffusion radius is determined by equation (23)].
We take the following ejecta parameters: $v_{min}=0.1c$, $v_{min}=0.346c$, $M_{ej}=0.025M_{\odot}$, $\kappa =3.33$ cm$^{2}$ g$^{-1}$, $\dot{\varepsilon}_0=2 \times10^{10}$ erg s$^{-1}$ g$^{-1}$ (heating rate at 1 day), $\alpha=1.3$, and $\beta=3.5$.

\section{Glossary of the mathematical symbols}
\label{ap:Glossary}
\begin{table*}
\begin{minipage}{155mm}
\caption{Glossary of the mathematical symbols used throughout the paper.}
\begin{tabular}{lll}
    \hline
    \hline
\multicolumn{1}{c}{Symbols} & \multicolumn{1}{c}{Definition} & \multicolumn{1}{c}{References}\\
    \hline
    \hline
        \multicolumn{3}{c}{Time and the timeline}  \\
        \hline
        $t$ ($t'$) & Time in the laboratory (comoving) frame\\
        $t_{obs}$ & Time in the observer's frame & Equation (\ref{eq:t obs lab})\\
        $t_m$ & Merger time & Figure \ref{fig:key1}\\
        $t_0$ & Jet launch time & Figure \ref{fig:key1}\\
        $t_b$ & Jet breakout time & Figure \ref{fig:key1}\\
        $t_e$ & Prompt jet's turn-off time & Section \ref{sec:2}\\
        $t_1$ & Start of the free-expansion phase & Figure \ref{fig:key1}\\
        \hline
        \multicolumn{3}{c}{Parameters of the jet and the ejecta}  \\
        \hline
        $L_j$ & Jet luminosity (one side) & Table \ref{tab:1} \\
        $\theta_0$ & Maximum opening angle of the jet & Table \ref{tab:1} \\
        $ L_{iso,0} $ & Isotropic equivalent luminosity of the jet & Table \ref{tab:1}\\
        $M_e$ & Mass of the ejecta (relevant to the jet propagation) & Table \ref{tab:1}\\  
        $n$ & Index of the power-law density profile for the ejecta & Table \ref{tab:1}\\
        $\beta_m$ & Maximum velocity of the ejecta & Table \ref{tab:1}\\
        \hline
        \multicolumn{3}{c}{Cocoon quantities in numerical simulations}  \\
        \hline
        $r$ & Radial distance from the central engine frame & Appendix \ref{ap:cocoon}\\
        $\Gamma$ ($\beta$) & Lorentz factor (velocity) & Appendix \ref{ap:cocoon}\\
        $\Gamma_{inf}$ ($\beta_{inf}$) & Maximally attained Lorentz factor (velocity) at infinity & Appendix \ref{ap:Bernoulli equation}\\
        $V_c$ & Cocoon volume & Section \ref{sec:Numerical profiles of the cocoon}\\
        $M_c$ & Cocoon mass & Section \ref{sec:den sim}\\
        $\langle{\rho_c}\rangle$ & Average cocoon density (as a function of $\Gamma_{inf}\beta_{inf}$) & Equation (\ref{eq:den av})\\
        $E_{c,i}$ & Cocoon internal energy & Section \ref{sec:Eint sim}\\
        $P$ & Gas pressure in the comoving frame & Section \ref{sec:Eint sim}\\
        $\Gamma_a$ & Adiabatic index & Section \ref{sec:Eint sim}\\
        $dE_{c,i}/dV_c$ & Average internal energy density in the cocoon (as a function of $\Gamma_{inf}\beta_{inf}$) & Equation (\ref{eq:Eint average})\\
        \hline
        \multicolumn{3}{c}{Parameters defining the escaped cocoon}  \\
        \hline
        $\beta_t$ ($\Gamma_t$) & Maximum velocity (Lorentz factor) of the non-relativistic cocoon & Section \ref{sec:escaped R NR}\\
        $\beta_{out}$ ($\Gamma_{out}$) & Maximum velocity (Lorentz factor) of the relativistic cocoon & Section \ref{sec:escaped R NR}\\
        $m$ & Index of the power-law density profile for the non-relativistic cocoon & Equation (\ref{eq:dm/dv cases})\\
        $l$ & Index of the power-law density profile for the relativistic cocoon & Equation (\ref{eq:dm/dv cases})\\
        $\theta_c^{es}$ & Opening angle of the escaped cocoon & Section \ref{sec:theta es 1}\\
        $M_c^{es}$ & Total mass of the escaped cocoon & Section \ref{sec:Analytic density profile}\\
        $M_{c,nr}^{es}$ & Mass of the escaped non-relativistic cocoon & Section \ref{sec:Analytic density profile}\\
        $M_{c,r}^{es}$ & Mass of the escaped relativistic cocoon & Section \ref{sec:Analytic density profile}\\
        $E_{c,i}^{es}$ & Total internal energy in the escaped cocoon  & Equation (\ref{eq:Lbl^j S estiamte})\\
        $E_{c,i,nr}^{es}$ & Internal energy in the non-relativistic cocoon  & Equation (\ref{eq:Lbl^j S estiamte})\\
        $E_{c,i,r}^{es}$ & Internal energy in the relativistic cocoon  & Equation (\ref{eq:Lbl^j S estiamte})\\
        \hline
        \multicolumn{3}{c}{Quantities used to define the escaped cocoon's mass and internal energy density}  \\
        \hline
        $\Omega$ & Solid angle of the escaped cocoon & Equation (\ref{eq:Omega})\\
        $\rho_t$ & Cocoon density at $\beta_t$ & Equation (\ref{eq:rho_t v2})\\
        $\rho_{c,nr}$ & Density in the non-relativistic cocoon & Section \ref{sec:Analytic density profile}\\
        $\rho_{c,r}$ & Density in the relativistic cocoon & Section \ref{sec:Analytic density profile}\\
        $e_{t}^j$ & Internal energy (from jet-shock heating) density in the cocoon at $\beta_t$ & Section \ref{sec:Eint jet}\\
        $e_{c,i,nr}^j$ & Internal energy (from jet-shock heating) density in the non-relativistic cocoon & Section \ref{sec:Eint jet}\\
        $e_{c,i,r}^j$ & Internal energy (from jet-shock heating) density in the relativistic cocoon & Section \ref{sec:Eint jet}\\
        $E_{c,i}^{j}$ & Total internal energy from jet-shock heating in the escaped cocoon  & Section \ref{sec:Eint jet}\\
        $E_{c,i,nr}^{j}$ & Internal energy from jet-shock heating in the non-relativistic cocoon  & Section \ref{sec:Eint jet}\\
        $E_{c,i,r}^{j}$ & Internal energy from jet-shock heating in the relativistic cocoon  & Section \ref{sec:Eint jet}\\
        $\dot{\varepsilon}$ & The r-process heating rate per rest-mass & Equation (\ref{eq:Epsilon dot})\\
        $\dot{\varepsilon}_0$ & The initial r-process heating rate per rest-mass at $t_h'$ & Section \ref{sec:Edep rp}\\
        $k$ & Power-law index in the time decay of r-process radioactive heating & Section \ref{sec:Edep rp}\\
        $\dot{E}_{c,i}^{rp}$ & Total r-process internal energy deposition rate in the escaped cocoon & Equation (\ref{eq:rp dep cases})\\
        $\dot{E}_{c,i,nr}^{rp}$ & R-process internal energy deposition rate in the non-relativistic cocoon & Equation
        (\ref{eq:Edep cum R+NR})\\
        $\dot{E}_{c,i,r}^{rp}$ & R-process internal energy deposition rate in the relativistic cocoon & Equation
        (\ref{eq:Edep cum R+NR})\\
        ${E}_{c,i}^{rp}$ & R-process internal energy available in the escaped cocoon & Equation
        (\ref{eq:E conserved rp+adiabatic})\\
        ${E}_{c,i,nr}^{rp}$ & R-process internal energy available in the non-relativistic cocoon & Equation
        (\ref{eq:Erp cum R+NR})\\
        ${E}_{c,i,r}^{rp}$ & R-process internal energy available in the relativistic cocoon & Equation
        (\ref{eq:Erp cum R+NR})\\
        \hline
\end{tabular}
\end{minipage}
\label{tab:ap}
\end{table*}

\begin{table*}
\begin{minipage}{155mm}
\caption{Glossary of the mathematical symbols used throughout the paper. (Continued)}
\begin{tabular}{lll}
    \hline
    \hline
    \multicolumn{1}{c}{Symbols} & \multicolumn{1}{c}{Definition} & \multicolumn{1}{c}{References}\\
    \hline
    \hline
        \multicolumn{3}{c}{Photon scattering and diffusion}  \\
        \hline
        $\theta_v$ & The angle between the observer's LOS and the polar axis & Section \ref{sec:Optical depth}\\
        $\Theta$ & The angle between the observer's LOS and the fluid's velocity vector $\vec{\beta}$ & Section \ref{sec:Optical depth}\\
        $z$ & Distance from the central engine along the observer's LOS & Equation (\ref{eq:tau def})\\
        $Z_{obs}=R_{obs}$ & Distance between the central engine and the observer & Section \ref{sec:Optical depth}\\
        $\kappa$ & Opacity & Equation (\ref{eq:tau def})\\
        $\tau$ & Optical depth across the escaped cocoon & Equation (\ref{eq:tau def})\\
        $\tau_{r}$ & Optical depth across the relativistic part of the cocoon & Equation (\ref{eq:tau R})\\
        $\tau_{nr}$ & Optical depth across the non-relativistic part of the cocoon & Equation (\ref{eq:tau NR})\\
        $\Gamma_d$ ($\beta_d$) & Lorentz factor (velocity) at the sharp diffusion shell & Section \ref{sec:photon diff}\\
        $N$ & Number of photon scatterings & Equation (\ref{eq:N cases})\\
        $\Delta \beta'$ & The relative velocity between the diffusion shell and the edge of the cocoon & Section \ref{sec:random walk}\\
        $\beta_x$ & The velocity at which inner and outer optical depths (from $\beta_d$ to $\beta_t$) are equal & Equation (\ref{eq:betax})\\
        $\tau_{ph}$ & Optical depth at the photosphere & Equation (\ref{eq:tau ph})\\
        $r_{ph}$ & Photospheric radius & Equation (\ref{eq:rph})\\
        $\Gamma_{ph}$ ($\beta_{ph}$) & Lorentz factor (velocity) at the photosphere & Section \ref{sec:Photospheric velocity}\\
        \hline
        \multicolumn{3}{c}{Luminosity, temperature, and magnitude}  \\
        \hline
        $L$ & Luminosity & Equation (\ref{eq:partial L})\\
        $L_{bl}$ & Total bolometric luminosity & Equation (\ref{eq:Lbl tot})\\
        $L_{bl}^j$ & Bolometric luminosity from jet-shock heating & Equation (\ref{eq:Lbl j})\\
        $L_{bl,r}^j$ & Bolometric luminosity from jet-shock heating in the relativistic cocoon & Equation (\ref{eq:Lbl R j S})\\
        $L_{bl,nr}^j$ & Bolometric luminosity from jet-shock heating in the non-relativistic cocoon & Equations (\ref{eq:Lbl NR j S}) and (\ref{eq:Lbl NR j F})\\
        $L_{bl}^{rp}$ & Bolometric luminosity from r-process heating & Equation (\ref{eq:Lbl rp})\\
        $L_{bl,r}^{rp}$ & Bolometric luminosity from r-process heating in the relativistic cocoon & Equation (\ref{eq:Lbl R rp S})\\
        $L_{bl,nr}^{rp}$ & Bolometric luminosity from r-process heating in the non-relativistic cocoon & Equation (\ref{eq:Lbl NR rp S})\\
        $\theta_{EM}$ ($\Omega_{EM}$) & Opening (solid) angle of the photon emission & Equations (\ref{eq:theta_EM}) and (\ref{eq:Omega_EM})\\ 
        $p$ & Power-law index for the bolometric luminosity's light curve & Equations (\ref{eq:Lbl^j S estiamte}) and (\ref{eq:Lbl^j F estiamte})\\
        $\epsilon$ ($\epsilon'$) & Internal energy density in the laboratory (comoving frame) & Section \ref{sec:Tobs}\\
        $T_{obs}$ ($T'$) & Observed (comoving) temperature & Equation (\ref{eq:Teff})\\
        $F_\nu$ & Observed blackbody flux density at the frequency $\nu$ & Equation (\ref{eq:Fnu})\\
        $m_{AB}$ ($M_{AB}$) & Apparent (absolute) magnitude & Equations (\ref{eq:mag ap}) and (\ref{eq:mag ab})\\

\hline
\hline
\end{tabular}
\end{minipage}
\end{table*}

\bsp	
\label{lastpage}
\end{document}